\documentclass[12pt]{article}

\usepackage{pst-feyn, pst-node, pst-arrow}
\usepackage{feynmp}
\usepackage{tikz-feynman}
\tikzfeynmanset{compat=1.1.0}

\usepackage{tocloft}
\usepackage{CJKutf8}
\usepackage{amsfonts}
\usepackage{amsmath}
\usepackage{amssymb}
\usepackage{array}
\usepackage{bigints}
\usepackage{bm}
\usepackage{booktabs}
\usepackage[nosort]{cite}
\usepackage{color}
\usepackage{dsfont}
\usepackage{float}
\usepackage{framed}
\usepackage{graphicx}
\usepackage{indentfirst}
\usepackage{mathrsfs}
\usepackage{multirow}
\usepackage{pdflscape}
\usepackage{setspace}
\usepackage{subdepth}
\usepackage{subfig}
\usepackage{titlesec}
\usepackage{wrapfig}
\usepackage[all]{xy}
\usepackage{young}
\usepackage[vcentermath]{youngtab}
\usepackage{relsize}
\usepackage{stackengine}
\usepackage{verbatim}
\usepackage{slashed}
\usepackage[utf8]{inputenc}
\usepackage{bbm}

\usepackage{hyperref}
\hypersetup{colorlinks=true}
\hypersetup{linkcolor=black}
\hypersetup{citecolor=black}
\hypersetup{urlcolor=black}

\numberwithin{equation}{section}

\usepackage[left=2.5cm,right=2.5cm,top=2.5cm,bottom=3cm]{geometry}
\fontdimen2\font=1.2\fontdimen2{\jot}{5pt}

\titleformat{\section}{\large\bfseries}{\thesection.}{4pt}{}
\titlespacing{\section}{0pt}{20pt}{6pt}

\titleformat{\subsection}{\normalfont\bfseries}{\thesubsection.}{4pt}{}
\titlespacing{\subsection}{0pt}{15pt}{6pt}

\titleformat{\subsubsection}{\normalfont\itshape}{\thesubsubsection.}{4pt}{}
\titlespacing{\subsubsection}{0pt}{15pt}{6pt}

\titleformat{\paragraph}{\normalfont\itshape}{\theparagraph.}{4pt}{}
\titlespacing{\paragraph}{0pt}{15pt}{6pt}

\newcommand{\bea}{\begin{eqnarray}}
\newcommand{\eea}{\end{eqnarray}}
\newcommand{\beq}{\begin{equation}}
\newcommand{\eeq}{\end{equation}}

\def\<{\langle}
\def\>{\rangle}

\def\nn{\nonumber}

\def\cO {{\mathcal O}}

\DeclareFontShape{OT1}{cmr}{mx}{n}%
{<->cmr10}{}
\newcommand{\mytitlefont}{\fontseries{mx}\selectfont}
\DeclareMathAlphabet{\titlemath}{OT1}{cmr}{mx}{n}

\begin{document}

\begin{titlepage}
% \begin{flushright} \small
% UUITP-17/19
%  \end{flushright}

\begin{center}
			
~\\[0.7cm]
			
{\fontsize{20pt}{0pt} \mytitlefont Conformal 3-point correlators in momentum space, method of subgraphs and the $1/N$ expansion}
			
~\\[0.4cm]
%\begin{CJK*}{UTF8}{gbsn}
Zhijin Li

{\it Shing-Tung Yau Center and School of Physics, Southeast University, Nanjing 210096, China}~\\[0.2cm]

%\end{CJK*}			
\end{center}

\vskip0.5cm
			
Conformal 3-point correlators of conserved currents play important roles in numerous directions. These correlators are fixed by conformal symmetry up to a few parameters, which are  known only at leading order in perturbative expansions. The major challenges come from the multi-loop Feynman integrals with three external momenta. In this work, we employ the method of subgraphs to compute the subleading order corrections to the conformal current 3-point correlators in the large $N$ expansion.  We show that the method of subgraphs generates diagrammatic expansions for the conformal 3-point correlators, and that it is closely related to the operator product expansions in momentum space. We verify the subgraph expansions of conserved current 3-point correlators using exact results in 3D.
We demonstrate that multi-loop 3-point Feynman integrals can be significantly simplified by taking the subgraph expansions. Due to constraints from conformal symmetry, it suffices to keep only the first few terms in the subgraph expansions to completely fix the subleading order corrections. 
We apply this method to compute the $1/N$ corrections to current correlators $\langle JJJ\rangle$ in the critical $O(N)$ vector model and the Gross-Neveu-Yukawa model. We also compute the $1/N$  corrections to the coefficients in the current-current-scalar correlators $\langle JJ\sigma_{T}\rangle$ and $\langle JJ\sigma\rangle$ in the critical $O(N)$ vector model. We compare the perturbative results with the bootstrap data and discuss their application to conductivity near the quantum critical point.

% \vfill 
% \begin{flushleft} 
% November 2022
%  \end{flushleft}

\end{titlepage}

% TABLE OF CONTENTS
	
\setcounter{tocdepth}{3}
\renewcommand{\cfttoctitlefont}{\large\bfseries}
\renewcommand{\cftsecaftersnum}{.}
\renewcommand{\cftsubsecaftersnum}{.}
\renewcommand{\cftsubsubsecaftersnum}{.}
\renewcommand{\cftdotsep}{6}
\renewcommand\contentsname{\centerline{Contents}}
	
\tableofcontents

% MAIN TEXT
\newpage
\section{Introduction}

The 3-point correlators of conserved current operators in conformal field theories (CFTs) have been systematically studied in the fundamental works \cite{Osborn:1993cr, Erdmenger:1996yc}. Here the conserved current operators of interest include both the conserved spin 1 vector field operator $J_\mu^a$ and the stress tensor operator $T_{\mu\nu}$. The results in \cite{Osborn:1993cr,Erdmenger:1996yc} show that in general dimensions, the conformal 3-point correlation functions can be  fixed by conformal symmetry up to a few free parameters, and the numbers of independent free parameters are further reduced by the conservation conditions of the conserved currents. These parameters are the key ingredients in the studies of the conformal current 3-point correlators. The parameters in the correlators $\langle TJJ\rangle$, $\langle TTT\rangle$ are subject to the constraints from unitarity and causality, namely the conformal collider bounds \cite{Hofman:2008ar,Buchel:2009sk,Chowdhury:2012km,Hofman:2016awc}, and they play important roles in the studies of the strongly coupled systems in condensed matter physics \cite{Chowdhury:2012km,Witczak-Krempa:2013nua,Katz:2014rla,Witczak-Krempa:2015pia,Lucas:2016fju,Lucas:2017dqa} and quantum gravity through holography \cite{Camanho:2014apa,Afkhami-Jeddi:2016ntf}. The conformal current 3-point correlators have also been applied to study the cosmological correlators \cite{Maldacena:2002vr, Maldacena:2011nz, Antoniadis:2011ib, Bzowski:2012ih, Arkani-Hamed:2018kmz,Baumann:2020dch}. For instance, the stress tensor 3-point correlator corresponds to the non-Gaussianity of gravitational waves generated during  the early stage of cosmology \cite{Maldacena:2011nz}.

We will be interested in the conserved current 3-point correlators $\langle J^a_\mu J^b_\nu \cO\rangle $ and $\langle T_{\mu\nu}T_{\rho\sigma} \cO\rangle $, in which the operators $\cO$ are the leading terms in the operator product expansions (OPEs):
\begin{align}
    J_\mu^a &\times J_\nu^b  \sim \mathbf{1} \,\delta^{ab} +\cO_S \, \delta^{ab} + T_{\mu\nu} \,\delta^{ab}+\cdots, \\
    T_{\mu\nu} &\times T_{\rho\sigma} \sim \mathbf{1} + \cO_S+ T_{\mu\sigma} +\cdots.
\end{align}
The above OPEs can be used to determine the charge conductivity and viscosity near quantum critical points at finite temperature  \cite{Witczak-Krempa:2013nua,Katz:2014rla,Witczak-Krempa:2015pia,Lucas:2016fju,Lucas:2017dqa}, for which the dominat contributions are from the vacuum state, relevant scalars $\cO_S$ and the stress tensor $T_{\mu\nu}$.  A recent conformal bootstrap study has shown that the conserved current operators can help to improve the bootstrap results to remarkably high precision \cite{Chang:2024whx}. It is expected that a similar bootstrap implementation associated with the conserved currents can be used to improve the bootstrap results for the strongly coupled conformal gauge theories \cite{Li:2020bnb}, in particular the infrared (IR) fixed point of the three-dimensional quantum electrodynamics (QED$_3$) coupled to $N_f=4$ flavors electrons \cite{Albayrak:2021xtd}, and the perturbative results on the current 3-point correlators can provide useful input data for the bootstrap computations, reminiscent of the special roles played by the $1/N$ expansions of the QED$_3$ spectrum \cite{Pufu:2013vpa,Chester:2016ref,Chester:2017vdh}.

The goal of this work is to develop large $N$ diagrammatic techniques to compute the $1/N$ corrections to the parameters in the conserved current 3-point correlators. The method can be applied to general  conformal field theories. Here we will focus on the critical $O(N)$ vector model and the Gross-Neveu-Yukawa (GNY) model. The critical $O(N)$ vector model has been extensively studied in the past decades using $1/N$ perturbative expansions, and plenty of CFT data for this theory have been obtained, see 
\cite{Henriksson:2022rnm} for a comprehensive review. Nevertheless, most of the perturbative studies focused on the spectrum \cite{Ma:1974qh,Okabe:1978mp,Kazakov:1979ik,Vasilev19811nEC,Derkachov:1997ch,Gracey:2002qa} and the central charges \cite{Petkou:1995vu,Huh:2013vga,Huh:2014eea,Diab:2016spb} as free parameters in the conserved current  2-point correlators; \footnote{The $1/N$ corrections to the scalar three-point correlators have been computed recently in \cite{Goykhman:2019kcj,Chai:2021uhv}.} while the perturbative results on the conserved current 3-point correlators were  computed only at the leading order in position space \cite{Osborn:1993cr,Erdmenger:1996yc} and in momentum space \cite{Chowdhury:2012km, Maldacena:2011nz}. These results are given by the free field theories. 
The main challenge is that the Feynman loop integrals for the 3-point correlators are more difficult to evaluate than those for the 2-point correlators. In fact, although the conformal 3-point correlators are fixed by conformal symmetry up to free parameters and have compact forms in position space, their counterparts in momentum space can only be written in complicated expressions in general dimensions.\footnote{In 3D the conformal conserved current 3-point correlators in momentum space are given by conformal integrals, which can be written in compact forms.} Since the Feynman loop integrals are usually evaluated in momentum space,  one cannot directly apply the conformal 3-point correlation functions in position space \cite{Osborn:1993cr,Erdmenger:1996yc} to the Feynman loop integrals. 

The conformal 3-point correlators in momentum space, especially those for the conserved current operators have been extensively studied in the past decade \cite{Bzowski:2013sza,Coriano:2013jba,Bzowski:2014qja,Bzowski:2015pba,Bzowski:2017poo,Gillioz:2019lgs,Bzowski:2019kwd,Bzowski:2020kfw,Coriano:2020ees,Gillioz:2021sce, Marotta:2022jrp}, motivated by their fundamental applications in holographic cosmology and other directions. It has been shown that the conformal 3-point correlators in momentum space can be expressed as integrals of triple Bessel $K$ functions \cite{Bzowski:2013sza} or, equivalently, as Appell functions \cite{Coriano:2013jba}. These formulas provide the leading order results in the perturbative expansions of the 3-point correlators, and the Feynman loop integrals for the subleading order corrections are considerably more challenging. In this work, we apply the method of subgraphs \cite{Smirnov:1994tg, Smirnov:2002pj, Smirnov:2012gma,Smirnov:2021dkb} to decompose the original multi-loop 3-point Feynman diagrams into a series of simpler Feynman integrals. Due to the constraints from conformal symmetry, the subleading order corrections to the 3-point correlators can be completely fixed by the first few terms in the subgraph expansions, therefore it significantly simplifies the Feynman loop integrals for the $1/N$ corrections.

The method of subgraphs provides a graph-theoretical expansion for multi-scale Feynman loop integral in terms of its subgraphs. 
The Feynman loop integrals with multiple scales appear ubiquitously in QFTs, while in general they are hard to evaluate analytically. A natural idea is to expand the original loop integral with respect to a small/large scale parameter $p$. The subtlety is that such expansion does not commute with the integration and the preconditions for the $p$-expansion of the integrand are not satisfied in certain regions of the integration domain. This problem can be systematically resolved using the method of subgraphs or the method of regions \cite{Smirnov:1994tg,Smirnov:2002pj,Smirnov:2012gma, Smirnov:2021dkb,Beneke:1997zp} . The method of regions is physically motivated: one divides the integration domain into different regions $\mathcal{D}_i$, within which the integrand can be Taylor expanded with respect to parameters that are considered small in $\mathcal{D}_i$. This method has been widely applied in Feynman loop computations, see \cite{Gardi:2022khw} for new development. Nevertheless, a strict proof of the method of regions is not available yet. 

For the Feynman integrals in Euclidean spacetime, or those with off-shell momenta and masses in Minkowski spacetime, the expansion of the original integral can be represented in a graph-theoretical way, i.e., by its subgraphs. Here the role of the subgraphs is to provide graphical labels for the small/large internal momenta in the Taylor expansion of the integrand.  The method of subgraphs has been proved for general  Feynman diagrams in Euclidean space, see Appendix B.2 in \cite{Smirnov:2002pj}. Coincidentally, the Feynman loop integrals for the CFT data, including the conformal 3-point correlators are evaluated in Euclidean spacetime, therefore the method of subgraphs can be applied as a solid approach to simplify the loop computations. Another remarkable advantage of applying the method of subgraphs to CFTs is that the conformal 3-point correlators have already been {\it almost} fixed by conformal symmetry. As a result, the $1/N$ corrections to the free parameters can be completely solved from the first few terms in the subgraph expansion. This is however different from the Feynman loop integrals in general QFTs, in which the method of subgraphs provides an asymptotic expansion and requires taking the subgraph expansions to high orders to obtain sufficiently precise data for the original Feynman integral \cite{Smirnov:2023bxf}.

\vspace{3mm}
{\bf \noindent A short summary of the results.} 
The gist of this work is to show that the method of subgraphs, together with the constraints from conformal symmetry provides an efficient approach to compute the higher order corrections of the conformal 3-point correlators. Our results can be summarized into three parts.

\vspace{3mm}
{\bf a.} In Section \ref{sec2} we apply the method of subgraphs to study the general scalar conformal 3-point correlators in momentum space and demonstrate that it provides a graph-theoretical expansion for the complicated correlation functions. A particularly important property is that the subgraph expansion automatically captures the singularities in the OPE limit in the momentum space. We compare the subgraph expansions for the scalar 3-point functions $\langle\cO_1\cO_2\cO_3\rangle$ with the solutions from the conformal Ward identities, and find perfect agreement. In Section \ref{free3pt} we  present the subgraph expansions of the conserved current 3-point correlators $\langle JJJ\rangle$, $\langle TJJ\rangle$ and $\langle TTT\rangle$, and verify the results using the exact correlation functions in 3D.  

\vspace{3mm}
{\bf b.} We compute the $1/N$ corrections to the conserved current 3-point correlators $\langle JJJ\rangle$ in the critical $O(N)$ vector model and GNY model. In both theories the 3-point correlators can be expanded in terms of the free field theories
\begin{equation}
    \langle JJJ\rangle=n_s \langle JJJ\rangle_{\mathrm{free ~scalar}}+n_f\langle JJJ\rangle_{\mathrm{free ~fermion}}. \label{ferbos}
    \end{equation}
The parameters $n_s$ and $n_f$, up to the order $O(1/N)$, are given by
    \begin{align}
    n_s =1-\frac{2^{D+2} (3 D-4) \sin \left(\frac{\pi  D}{2}\right) \Gamma \left(\frac{D-1}{2}\right)}{\pi ^{3/2} (D-2)^2 D^2 \Gamma \left(\frac{D}{2}-2\right)N}&, ~~
    n_f=-\frac{2^{D+3} \sin \left(\frac{\pi  D}{2}\right) \Gamma \left(\frac{D-1}{2}\right)}{\pi ^{3/2} (D-2)^2 D^2 \Gamma \left(\frac{D}{2}-2\right)N}, 
\end{align}
for the critical $O(N)$ vector model and 
\begin{align}
    n_s =\frac{2^{D+2} \sin \left(\frac{\pi  D}{2}\right) \Gamma \left(\frac{D-1}{2}\right)}{\pi ^{3/2} D^2 \Gamma \left(\frac{D}{2}-1\right)N}, ~~~
    n_f=1+\frac{2 (D+1) \sin \left(\frac{\pi  D}{2}\right) \Gamma (D-1)}{\pi  \Gamma \left(\frac{D}{2}+1\right)^2 N},
\end{align}
for the GNY model. Details of the computations and the special role of the method of subgraphs are presented in Section \ref{sec4}.

\vspace{3mm}
{\bf c.} We compute the $1/N$ corrections to the correlators $\langle JJ\cO\rangle$ in the critical $O(N)$ vector model motivated by their applications in the conductivity at finite temperature. We focus on the lowest $O(N)$ traceless symmetric scalar $\sigma_T$ and the $O(N)$ singlet scalar $\sigma$. Up to the subleading order, the results are
\begin{align}
    \lambda_{JJ\sigma_T}=D-2+\frac{((D-12) D+16) (\cos (\pi  D)-1) \Gamma \left(2-\frac{D}{2}\right) \Gamma (D-2)}{\pi ^2 D \Gamma \left(\frac{D}{2}+1\right)N},
\end{align}
and 
\begin{equation}
    \left. \lambda_{JJ\sigma}\right|_{D=3}=\frac{4}{\pi \sqrt{N}}-\frac{368}{9 \pi ^3\sqrt{N}N}.
\end{equation}
The normalization for above OPE coefficients is introduced in (\ref{JJO}), which follows \cite{Reehorst:2019pzi}. Details of the $1/N$ loop computations and renormalization are explained in Section \ref{sec5}.

% In this work, we will firstly compute the subgraph expansions of the conserved current 3-point correlators based on the free boson and free fermion theories in general dimensions, which provide the leading order results for the conserved current 3-point correlators and can be explicitly verified based on the 3D exact results, as well as the solutions obtained from the conformal Ward identities in momentum space. The results will be presented in section 3. They provide the leading order result for the conserved current 3-point correlators in momentum space in the large $N$ expansion, and only the free parameters receive higher order corrections. in section 4 we apply the method of subgraphs to compute the order $O(1/N)$ corrections to the $\langle J^aJ^bJ^c\rangle$ in both the $O(N)$ vector and Gross-Neveu models, which further demonstrates how the subgraph expansions can drastically simplify the Feynman loop integrals for the 3-point CFT data. In section 5 we compute the current 3-point correlators $\langle J J S \rangle$ where $S$ is a scalar constructing different irreducible representations of the global symmetry. These correlation functions provide necessary ingredients to study the finite temperature conductivity near quantum critical point and also the conformal bootstrap computations. 

\vspace{3mm}
We expect the method developed in this work can be employed to provide a perturbative study for the conserved current 3-point correlators in many physically interesting CFTs. We discuss some future works in Section \ref{sec6}.

\section{Method of subgraphs and conformal 3-point correlators in momentum space} \label{sec2}
Feynman diagrams are usually evaluated in momentum space. Therefore, to study conformal correlation functions using this method, one needs to know the momentum-space form of the correlators.
The conformal 2-point correlation functions in momentum space can be obtained using Fourier transformations, and the results are given in simple forms. However, for the conformal 3-point correlators,  the Fourier transformations of the correlation functions in position space lead to complicated expressions \cite{Bzowski:2013sza,Coriano:2013jba}. Besides, there are extra contributions from the contact terms of the correlation functions. In this section, we first review the results on the conformal 3-point correlators in momentum space obtained from the Fourier transformations and the conformal Ward identities \cite{Bzowski:2013sza,Coriano:2013jba,Coriano:2020ees}, then we propose that the method of subgraphs provides a useful graph-theoretical expansion for the conformal 3-point correlation function in momentum space.

\subsection{Conformal 3-point correlators in momentum space}
We use the scalar 3-point correlator to exhibit the complexity of the expressions for the conformal 3-point correlation functions in momentum space. The 3-point correlators of spinning operators can be decomposed into scalar correlation functions associated with specific tensor structures \cite{Bzowski:2013sza}. 

The conformal 3-point correlator in momentum space can be obtained through Fourier transformations of the correlators in position space. Consider
a 3-point correlator of scalars $\cO_i$ with scaling dimensions $\Delta_i$. In position space, the correlation function is given by
\begin{equation}
    \langle \cO_1(x_1)\cO_2(x_2)\cO_3(x_3)\rangle=\frac{\lambda_{\cO_1\cO_2\cO_3}}{x_{12}^{\Delta_1+\Delta_2-\Delta_3}x_{23}^{\Delta_2+\Delta_3-\Delta_1}x_{31}^{\Delta_1+\Delta_3-\Delta_2}}, \label{3pt}
\end{equation}
where $x_{ij}\equiv|x_i-x_j|$. We take the Fourier transformation of the above formula to get the same conformal 3-point correlation function in momentum space $\langle\cO_1(p_1)\cO_2(p_2)\cO_3(p_3)\rangle$, up to the contributions from contact terms.\footnote{The conformal 3-point correlation functions in position space like (\ref{3pt}) are valid when the three coordinates $x_i$ do not coincide. The contact terms appear when the coordinates $x_i$ coincide, and their Fourier transformations correspond to analytic terms of the external momenta \cite{Maldacena:2011nz}.}  
The Fourier transformation of the 3-point correlation function (\ref{3pt})  is proportional to the one-loop integral 
\begin{equation}
   \langle\cO_1(p_1)\cO_2(p_2)\cO_3(p_3)\rangle \propto \mathcal{I}(\nu_1,\nu_2,\nu_3)\equiv\int \frac{d^D k}{\left(2\pi\right)^D}\frac{1}{\left(k^2\right)^{\nu_3}\left((p_1-k)^2\right)^{\nu_2}\left((p_2+k)^2\right)^{\nu_1}}, \label{3kint}
\end{equation}
where we have implicitly assumed momentum conservation condition $p_3^\mu=-p_1^\mu-p_2^\mu$ and
$$\Delta_t=\Delta_1+\Delta_2+\Delta_3, ~~~ \nu_i=\frac{D-\Delta_t}{2}+\Delta_i.$$
The integral $\mathcal{I}(\nu_1,\nu_2,\nu_3)$ can be studied using recursive relations derived from either conformal symmetry constraints or integration-by-parts relations \cite{Coriano:2013jba, Davydychev:1992eww}. Using standard techniques for Feynman integrals, the scalar conformal 3-point correlation function in momentum space can be rewritten in an integral form \cite{Bzowski:2013sza}
\begin{align}
    \langle\cO_1(p_1)\cO_2(p_2)\cO_3(p_3)\rangle= &c_{\cO_1\cO_2\cO_3} \delta^d(p_1+p_2+p_3) 
    p_1^{\Delta_1-\frac{d}{2}}p_2^{\Delta_2-\frac{d}{2}}p_3^{\Delta_3-\frac{d}{2}} \nn \\
    &\times \int_0^\infty dx x^{\frac{d}{2}-1}K_{\Delta_1-\frac{d}{2}}(p_1x)K_{\Delta_2-\frac{d}{2}}(p_2x)K_{\Delta_3-\frac{d}{2}}(p_3x), \label{3pres}
\end{align}
where $c_{\cO_1\cO_2\cO_3}$ is a constant related to $\lambda_{\cO_1\cO_2\cO_3}$ in (\ref{3pt}) and $K_a(z)$ is the modified Bessel function of the second kind. For general parameters, the above triple-$K$ integral cannot be solved in a compact form. In fact, it may be divergent and a suitable regularization is needed, see \cite{Bzowski:2013sza} for more details.

% The conformal 3-point correlators can also be obtained directly in momentum space using conformal Ward identities, without resorting to position-space results \cite{Maldacena:2011nz,Bzowski:2013sza,Coriano:2013jba}. Specifically, Poincaré symmetry imposes the momentum conservation condition and dictates that the scalar 3-point functions depend solely on the magnitudes of the external momenta $p_i\equiv|p_i^\mu|$. 

The conformal 3-point correlators can be derived directly in momentum space by solving the conformal Ward identities, without referring to their position-space counterparts \cite{Maldacena:2011nz,Bzowski:2013sza,Coriano:2013jba}.
Poincaré invariance enforces momentum conservation and implies that the scalar 3-point function depends only on the magnitudes of the external momenta, $p_i \equiv |p_i^\mu|$.
The additional conformal Ward identities are derived from dilatation and special conformal transformations, which can be written in momentum space:
\begin{align}
    \left( 2D+\sum\limits_{i=1}^3\left(p_i\frac{\partial}{\partial p_i}-\Delta_i\right) \right)\langle\cO_1(p_1)\cO_2(p_2)\cO_3(p_3)\rangle=0, \label{dilatation}\\
    \sum\limits_{i=1}^3\left( \frac{\partial^2}{\partial p_i^2}+\frac{D+1-2\Delta_i}{p_i}\frac{\partial}{\partial p_i}\right) p_i^\mu\langle\cO_1(p_1)\cO_2(p_2)\cO_3(p_3)\rangle=0. \label{sct}
\end{align}
The dilatation symmetry (\ref{dilatation})  implies that the scalar 3-point function is a homogeneous function of the external momenta $p_i$ with degree $\Delta_t-2D$:
\begin{equation}
    \langle\cO_1(p_1)\cO_2(p_2)\cO_3(p_3)\rangle=p_3^{\Delta_t-2D}f(\frac{p_1}{p_3},\frac{p_2}{p_3}).
\end{equation}
Substituting this ansatz into the special conformal Ward identity (\ref{sct}) yields a system of partial differential equations for the function $f$.
The general solutions to these equations are expressed in terms of the generalized hypergeometric function of two variables, the Appell $F_4$ functions.
Among these, the physical solution corresponds precisely to the momentum-space expression given in (\ref{3pres}).

We will be particularly interested in the zero-momentum limit, where $$p_1^\mu=p^\mu,~ p_2^\mu=-p^\mu -q^\mu, ~ p_3^\mu=q^\mu,~~ q\equiv |q^\mu|\ll p\equiv |p^\mu|,$$
and the Bessel function $K_a(p_3 x)$ in (\ref{3pres}) admits the following series expansion $$K_a(p_3 x)= 2^{a-1}\Gamma(a) p_3^{-a} x^{-a}\left(1+O(p_3^2)\right)+2^{-a-1}\Gamma(-a) p_3^a x^{a}\left(1+O(p_3^2)\right).
$$
By applying the above expansion to (\ref{3pres}),
the triple-$K$ integral for the 3-point correlation function (\ref{3pres}) can be simplified to
\begin{align}
    \langle \cO_1(p)&\cO_2(-p-q)\cO_3(q)\rangle \propto \label{3ptpq}\\
    & p^{\Delta _t-2D}\,2^{\Delta_3-\frac{D}{2} -1} \Gamma \left(\Delta_3 -\frac{D}{2}\right)  R\left(D-\Delta_3, \Delta_1-\frac{D}{2}, \Delta_2-\frac{D}{2}\right)\left( 1+O(q^2)\right)+ \nn\\
     &p^{\Delta_t-2\Delta_3-D} q^{2 \Delta_3 -D}\,2^{\frac{D}{2}-\Delta_3 -1} \Gamma \left(\frac{D}{2}-\Delta_3 \right)  R\left(\Delta,\Delta_1-\frac{D}{2}, \Delta_2-\frac{D}{2}\right)\left( 1+O(q^2)\right), \nn
\end{align}
in which we have applied the double-$K$ integral \cite{book1992}
\begin{align}
    R(a,b,c)\equiv&\int\limits_0^\infty dx x^{a-1}K_b(x)K_c(x)= \\
    &\frac{2^{a-3} \Gamma \left(\frac{1}{2} (a-b-c)\right) \Gamma \left(\frac{1}{2} (a+b-c)\right) \Gamma \left(\frac{1}{2} (a-b+c)\right) \Gamma \left(\frac{1}{2} (a+b+c)\right)}{\Gamma (a)}. \nn
\end{align}
An important property in (\ref{3ptpq}) is that, for general  $\Delta_i$, the small $q$ expansion separates into two parts: a regular part whose leading term is independent of $q$ and a second part with leading term proportional to $q^{2 \Delta_3 -D}$. 
For $\Delta_3>D/2$ the entire correlation function is regular in the zero-momentum limit, while for $\Delta_3<D/2$ the 3-point function is dominated by a singular term proportional to $q^{2\Delta_3-D}$ in the zero-momentum limit. Such behavior of the conformal 3-point correlators has been studied in \cite{Lucas:2016fju,Lucas:2017dqa} to evaluate the dynamical response functions near quantum critical points. They also play an important role in analyzing the relation between the scale and conformal invariance \cite{Bzowski:2014qja,Dymarsky:2014zja}. 

The  small $q$ expansion of the scalar 3-point correlation function (\ref{3ptpq}) can also be obtained from the properties of the Appell function $F_4$ \cite{Coriano:2013jba}. 
In the following section, we show that the two terms in the zero-momentum limit can be reconstructed through the subgraph expansion of the one-loop integral (\ref{3kint}). The advantage of the subgraph expansion is that it provides a simplified diagrammatic representation of the asymptotic behaviors of the correlation functions, without using the mathematical properties of the complicated expressions for the original integrals.

\subsection{Method of subgraphs for the 2-scale loop integrals} \label{methodsbg}
The conformal 3-point correlation functions in momentum space (\ref{3kint}) depend on the one-loop integral $\mathcal{I}(\nu_1,\nu_2,\nu_3)$, which contains two scales $p_1 $ and $ p_2$. The Feynman loop diagrams with multiple scales are usually difficult to evaluate analytically.
For the multi-scale Feynman loop integrals in Euclidean space, the method of subgraphs provides an efficient approach to evaluate the original integral by decomposing it into its subgraphs with lower scales \cite{Smirnov:1994tg,Smirnov:2002pj,Smirnov:2012gma}. 
In this section, we briefly review the subgraph expansions of Feynman diagrams, and then apply the method of subgraphs to compute the conformal 3-point correlators in momentum space.

The multi-scale Feynman integrals in the limits with small or large external momenta have been studied for a long time \cite{Weinberg:1959nj}. This problem is subtle as the limit of the integral cannot be obtained simply by taking the same limit on the integrand. Take the one-loop integral $\mathcal{I}(1,1,1)$ for example:
\begin{equation}
    \mathcal{I}(1,1,1)=\int \frac{d^D k}{\left(2\pi\right)^D}\frac{1}{k^2(p+k)^2(k-q)^2}. \label{3DI111}
\end{equation}
This is an integral with two external momenta. In this work, the integrals with $n$ external momenta are dubbed $n$-scale integral. For $n=0$, the integral becomes scaleless and vanishes
\begin{equation}
    \int \frac{d^Dk}{(2\pi)^D}\frac{1}{\left(k^2\right)^a}=0. \label{scalelessint}
\end{equation}
The 2-point correlation functions are given by the 1-scale integrals, and their one-loop diagrams can be exactly solved through
\begin{equation}
    \int \frac{d^d k}{(2 \pi )^D }\frac{1}{\left(k^2\right)^a \left((k-p)^2\right)^b}=\frac{1}{(4 \pi )^{D/2}}\frac{\Gamma \left(\frac{D}{2}-a\right)   \Gamma \left(\frac{D}{2}-b\right) \Gamma \left(a+b-\frac{D}{2}\right) }{ \Gamma (a) \Gamma (b) \Gamma (D-a-b)}p^{D-2a-2b}. \label{2ptint}
\end{equation}
The integrals for the 2-point correlators with tensor structures can be reduced to the scalar integrals (\ref{2ptint}) through tensor reductions. 
The one-loop integrals will be extensively employed throughout this work. In contrast, there are no compact solutions for the one-loop 2-scale integrals like (\ref{3DI111}), and we take asymptotic expansions for these integrals.

We now show the subtlety in the small $q$ expansion of the 2-scale integrals. Consider the  Taylor expansion with small $q$ ($\mathcal{T}_q$) for the integrand in (\ref{3DI111}):
\begin{equation}
    \frac{1}{k^2(p+k)^2(k-q)^2}=\frac{1}{k^2(p+k)^2}\mathcal{T}_q\frac{1}{(k-q)^2}=\frac{1}{(k^2)^2(p+k)^2}+\frac{2k\cdot q}{(k^2)^3(p+k)^2}+\cdots. \label{expansion}
\end{equation}
The integral of each term in the expansion is regular in $q$
\begin{align}
    \int \frac{d^D k}{\left(2\pi\right)^D}\frac{1}{\left(k^2\right)^2(p+k)^2} &+\frac{2k\cdot q}{(k^2)^3(p+k)^2}= \frac{2^{-D} \pi ^{-\frac{D}{2}}  \Gamma \left(3-\frac{D}{2}\right) \Gamma \left(\frac{D}{2}-2\right) \Gamma \left(\frac{D}{2}-1\right)}{\Gamma (D-3)}p^{D-6}\nn \\
    &-\frac{2^{-D} \pi ^{-\frac{D}{2}} \Gamma \left(4-\frac{D}{2}\right) \Gamma \left(\frac{D}{2}-2\right) \Gamma \left(\frac{D}{2}-1\right)}{\Gamma (D-3)}p^{D-8} \, p\cdot q.\label{smallqintegrand}
\end{align}
With $D=3$, the two leading terms vanish.
Actually the integral (\ref{smallqintegrand}) is  IR divergent in 3D, and the right-hand side of (\ref{smallqintegrand}) gives the integral in dimensional regularization.
The result in (\ref{smallqintegrand}) can be compared with the integral (\ref{3DI111}) in 3D, which becomes a conformal integral and can be evaluated exactly:
\begin{equation}
    \left.\mathcal{I}(1,1,1)\right|_{D=3}=\frac{1}{8p q |p+q|}=\frac{1}{8p^2 q}-\frac{p\cdot q}{8p^4 q}+\cdots. \label{I3D}
\end{equation}
The original integral has a pole in the zero-momentum limit $q\rightarrow 0$, which is not shown in (\ref{smallqintegrand}). The small $q$ expansion of the integrand failed to capture the zero-momentum limit of the one-loop integral (\ref{3DI111}).
The reason is that the Taylor expansion (\ref{expansion})  is broken down in the integration domain $k<q$. In this region the integrand of $\mathcal{I}(1,1,1)$ can be alternatively expanded with small $k$:
\begin{equation}
    \frac{1}{k^2(p+k)^2(k-q)^2}=\frac{1}{k^2(k-q)^2}\mathcal{T}_k\frac{1}{(p+k)^2}=\frac{1}{p^2 k^2(k-q)^2}-\frac{2k\cdot p}{(p^2)^2k^2(k-q)^2}+\cdots, \label{expansion1}
\end{equation}
and its integral is given by
\begin{align}
\int \frac{d^D k}{\left(2\pi\right)^D} \frac{1}{p^2 k^2(k-q)^2} -\frac{2k\cdot p}{(p^2)^2k^2(k-q)^2} & = \\
\frac{2^{-D} \pi ^{-\frac{D}{2}}  \Gamma \left(2-\frac{D}{2}\right) \Gamma \left(\frac{D}{2}-1\right)^2}{\Gamma (D-2) p^2}  q^{D-4} &- \frac{2^{1-D} \pi ^{-\frac{D}{2}} \Gamma \left(2-\frac{D}{2}\right) \Gamma \left(\frac{D}{2}-1\right) \Gamma \left(\frac{D}{2}\right) p_{\mu }}{\Gamma (D-1) p^4} q^{D-4} q\cdot p , \nn
\end{align}
which produces the exact asymptotic behavior (\ref{I3D}) of $\mathcal{I}(1,1,1)$ in 3D.

This example demonstrates that the integration domains of the loop momenta are crucial for obtaining  proper asymptotic expansions of the Feynman integrals. The method of regions is proposed to resolve this problem \cite{Beneke:1997zp}.
It leads to asymptotic expansions for general Feynman integrals in certain limits of external momenta and masses. The procedure can be summarized as follows: first divide the integration domain of the internal momenta into different regions $\mathcal{D}_i$ associated with a set of parameters $\mathcal{C}_i$ which are considered small in $\mathcal{D}_i$, and take the Taylor expansion for the integrand $f$ in each region $\mathcal{D}_i$ with respect to the small parameters $\mathcal{C}_i$: $\mathcal{T}_{\mathcal{C}_i}f$; then integrate the new integrands $\mathcal{T}_{\mathcal{C}_i} f$ over the entire integration domain of the internal momenta. See \cite{Smirnov:2002pj} for detailed explanations of the method of regions.

The method of regions can be applied to general Feynman integrals, although a strict proof has not yet been achieved. However, for Feynman integrals in Euclidean space, the asymptotic expansion of the Feynman integrals can be formulated in a graph-theoretical language, namely the method of subgraphs \cite{Smirnov:1994tg,Smirnov:2002pj,Smirnov:2012gma}. The Feynman loop integrals for the CFT data are of the Euclidean type; therefore, their asymptotic expansions can be effectively studied using the method of subgraphs. 

Let us go back to the one-loop integral $\mathcal{I}(1,1,1)$. As explained after (\ref{smallqintegrand}), a naive  Taylor expansion of the integrand with small $q$ generates IR divergence. 
This problem can be cured by expanding the integrand with small $k$ (\ref{expansion1}). 
Consider the remainder of above Taylor expansions up to the order $n$, denoted $\mathcal{T}_{q/k}^n$:
\begin{equation}
    R^n \,\mathcal{I}(1,1,1)= \int \frac{d^D k}{\left(2\pi\right)^D} \frac{1}{k^2}\left(\left(1-\mathcal{T}^n_k\right)\frac{1}{(p+k)^2}\right)\left(\left(1-\mathcal{T}^n_q\right)\frac{1}{(k-q)^2}\right). \label{remainder}
\end{equation}
The second Taylor expansion $\left(1-\mathcal{T}^n_q\right)$ generates IR divergences for the integral, which is cured by the remainder from the first Taylor expansion $\left(1-\mathcal{T}^n_q\right)$, and vice versa for the integral in the UV limit.
The operators in the remainder (\ref{remainder}) can be interpreted graphically 
\begin{equation}
    R^n=(1-M^n_\gamma)(1-M^n_\Gamma),
\end{equation}
in which $M_\Gamma$ refers to the operation on the original Feynman diagram and only the momentum $q_\mu$ is considered small, while $M_\gamma$ refers to the operation on the subgraph given by the propagator $\frac{1}{(k-q)^2}$ and both the momenta $q_\mu, k_\mu$ are considered small. The original integral is given by
\begin{align}
    \mathcal{I}(1,1,1)&=(1-R^n)\mathcal{I} (1,1,1)+R^n\mathcal{I}(1,1,1) \\
    &=(M_\Gamma^n +M_\gamma^n+M_\Gamma^n\cdot M_\gamma^n+ R^n )\; \mathcal{I} (1,1,1).
\end{align}
The first two terms consist of the one-loop integrals with a single scale, which can be evaluated using the formula (\ref{2ptint}).
The third term generates a scaleless integral and vanishes (\ref{scalelessint}). Taking the order of the Taylor expansions to infinity $n\rightarrow \infty$, the remainder term vanishes as well, and the  original one-loop integral has an asymptotic expansion
\begin{equation}
    \mathcal{I}(1,1,1)=M_\Gamma \mathcal{I}(1,1,1) + M_\gamma \mathcal{I}(1,1,1).
\end{equation}
The operation $M_\Gamma$ ($M_\gamma$) corresponds to the Taylor expansion with respect to small momentum $q$ (momenta $q, k$) for the original Feynman diagram  $\Gamma$ (subgraph $\gamma$).

This procedure can be generalized to  higher-loop Feynman diagrams in Euclidean spacetime \cite{Smirnov:1994tg,Smirnov:2002pj}.  Consider an $L$-loop Feynman diagram $\Gamma$ in Euclidean space. Its external momenta are separated into two sets: the large and small external momenta $\{Q_1, Q_2, \dots, Q_i\}$, $\{q_1, q_2, \dots, q_j\}$. All the masses are tuned to zero to reach the IR fixed points. It can be shown that the remainder of the large $Q$ or small $q$ momenta expansion is given by the forest formula
\begin{equation}
    R^n=\sum\limits_S \prod\limits_{\gamma\in S} \left(-M_\gamma^{n(\gamma)}\right),  \label{remainderG}
\end{equation}
where $S$ denotes the set of non-overlapping subgraphs, i.e. the forests \cite{Zimmermann:1969jj}. The rules to determine $S$ will be explained later. The order of the Taylor expansion $n(\gamma)$ depends on the UV degree of divergence $\Lambda(\gamma)$ of the subgraph $\gamma$: $n(\gamma)=\Lambda(\gamma)+ n$, so that the remainder $R^n$ scales as $(q/Q)^n$ in the large $Q$ (small $q$) limit \cite{Smirnov:1990rz}. The subgraphs $\gamma$ in (\ref{remainderG}) are determined by two rules.
First, the subgraphs $\gamma\subseteq \Gamma$ contain all the vertices with large external momenta $Q$. Besides, the new graph is 1PI after identifying the external vertices of $\gamma$ with large momenta $Q$. Let $k_1, \dots, k_{h(\gamma)}$ denote  the loop momenta restricted in the subgraph $\gamma$, and $k_{h(\gamma)+1,\dots, k_L}$ denote the loop momenta which also flow through the complementary part of the subgraph $\Gamma\slash \gamma$. When taking the Taylor expansion for the subgraph $\gamma$, the internal momenta are separated into two scales: the momenta  $k_1, \dots, k_{h(\gamma)}$ are considered large, while the momenta $k_{k(\gamma)+1,\dots, k_L}$ are considered small. The operation $M_\gamma^{n(\gamma)}$ applies on the subgraph $\gamma$ while commutes with its complementary part $\Gamma\slash\gamma$:
\begin{equation}
    M_\gamma^{n(\gamma)}\mathcal{I}_\Gamma= \mathcal{I}_{\Gamma\slash \gamma}\circ M_\gamma^{n(\gamma)}\mathcal{I}_{\gamma}=\int \frac{dk_1}{(2\pi)^d}\dots \frac{dk_L}{(2\pi)^d} \mathcal{P}_{\Gamma\slash \gamma} \mathcal{T}_{q_1,\cdots, q_j, k_{h(\gamma)+1},\dots, k_L}^{n(\gamma)} \mathcal{P}_\gamma,
\end{equation}
where $\mathcal{P}_\gamma$ represents the partial integrand from the subgraph $\gamma$ and we use $\mathcal{I}_\gamma$ to denote its integral. 

The remainder form (\ref{remainderG}) leads to an identity for the original Feynman integral $\mathcal{I}_\Gamma$:
\begin{equation}
    \mathcal{I}_\Gamma=(1-R^n)\,\mathcal{I}_\Gamma+R^n \, \mathcal{I}_\Gamma=\sum \limits_\gamma \mathcal{I}_{\Gamma\slash \gamma}\circ M_\gamma^{n(\gamma)}\mathcal{I}_{\gamma}+R^{n}\mathcal{I}_\Gamma.
\end{equation}
Here when expanding the subtraction $1-R^n$, we have implicitly employed the fact that the multiple operations like $M_{\gamma_1} M_{\gamma_2} \mathcal{I}_\Gamma$ generate scaleless Feynman integrals which vanish automatically. Now take the infinity $n$ limit, above formula gives the asymptotic expansion of the original Feynman integral
\begin{equation}
    \mathcal{I}_\Gamma=\sum \limits_\gamma \mathcal{I}_{\Gamma\slash \gamma}\circ M_\gamma\mathcal{I}_{\gamma}. \label{subexpansion}
\end{equation}
The subgraph expansion can be applied to general Feynman loop diagrams in Euclidean spacetime with external momenta separated into two scales.

~

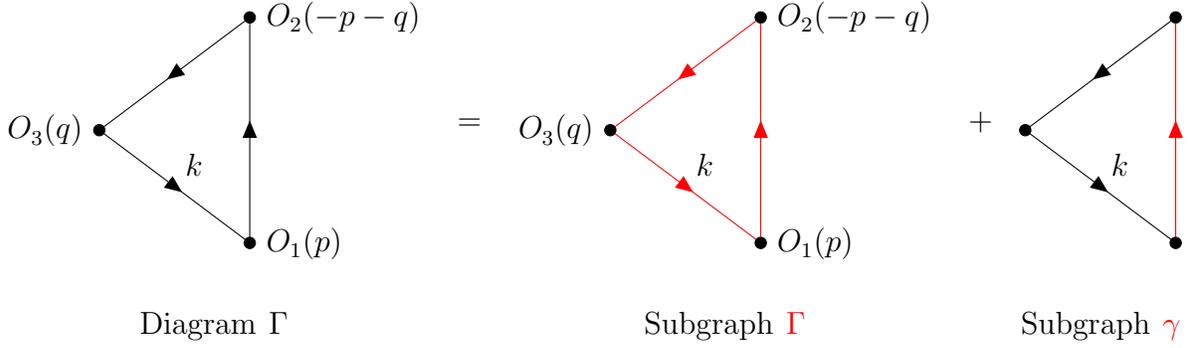
\begin{figure}
   \begin{equation*}
   \begin{array}{ccccc}
    \begin{tikzpicture}[baseline=(a.base)]
  \begin{feynman}
    \vertex [dot,label=right:$O_2(-p-    q)$] (c) at (2,3/2) {};
    \vertex[dot,label=right:$O_1(p)$] (b) at (2,-3/2) {};
    \vertex[dot,label=left:$O_3(q)$] (a)  at (0,0) {};

    \diagram* {
       (a) -- [fermion, edge label=\(k\)] (b) --[fermion] (c) --[fermion] (a)
    };
  \end{feynman}
\end{tikzpicture} &=&
\begin{tikzpicture}[baseline=(a.base)]
  \begin{feynman}
    \vertex [dot,label=right:$O_2(-p-    q)$] (c) at (2,3/2) {};
    \vertex[dot,label=right:$O_1(p)$] (b) at (2,-3/2) {};
    \vertex[dot,label=left:$O_3(q)$] (a)  at (0,0) {};

    \diagram* {
       (a) -- [fermion,red, edge label=\({\color{black}k}\)] (b) --[fermion,red] (c) --[fermion,red] (a)
    };
  \end{feynman}
\end{tikzpicture}
&+&
\begin{tikzpicture}[baseline=(a.base)]
  \begin{feynman}
    \vertex [dot] (c) at (2,3/2) {};
    \vertex[dot] (b) at (2,-3/2) {};
    \vertex[dot] (a)  at (0,0) {};

    \diagram* {
       (a) -- [fermion, edge label=\(k\)] (b) --[fermion,red] (c) --[fermion] (a)
    };
  \end{feynman}
\end{tikzpicture} \\
& & & & 
\\ \mathrm{Diagram~ }\displaystyle{\Gamma} & & \mathrm{Subgraph ~ } {\textstyle\color{red}\Gamma} & & \mathrm{Subgraph ~ } {\color{red}\gamma}
\end{array}
\end{equation*}
    \caption{Diagrammatic representation of $\mathcal{I}(\nu_1,\nu_2,\nu_3)$ and the two subgraphs in the expansion: $\Gamma$ and $\gamma$ marked in red color.}
    \label{fig:sss}
\end{figure}

Now we compute the subgraph expansion of the scalar 3-point correlator $\langle\cO_1(p)\cO_2(-p-q)\cO_3(q)\rangle$ in the limit $q\ll p$. The correlator is proportional to the integral $\mathcal{I}(\nu_1,\nu_2,\nu_3)$ 
\begin{equation}
    \mathcal{I}(\nu_1,\nu_2,\nu_3)=\int \frac{d^D k}{\left(2\pi\right)^D}\frac{1}{\left((p+k)^2\right)^{\nu_3}\left(k^2\right)^{\nu_2}\left((k-q)^2\right)^{\nu_1}}. \label{3DI}
\end{equation}
This is a generalization of the integral (\ref{3DI111}) to general external scalars. The diagrammatic representation of the integral $\mathcal{I}(\nu_1,\nu_2,\nu_3)$ is shown in Figure \ref{fig:sss}, in which the links should be considered as generalized scalar propagators: $1/(k^2)^{\nu_i}$. 
According to the rules for the method of subgraphs (\ref{subexpansion}), the subgraphs of the diagram $\mathcal{I}(\nu_1,\nu_2,\nu_3)$ consist of the 1PI subgraphs containing all the external vertices with large external momenta, i.e., $\cO_1(p)$ and $\cO_2(-p-q)$.
So there are two subgraphs for the original Feynman diagram $\Gamma$: $\Gamma$ itself and the link connecting $\cO_1(p)$ with $\cO_2(-p-q)$, see the graphs in red color in Figure \ref{fig:sss}.
We take Taylor expansions for the subgraphs with respect to the momenta of small scale. The external momentum $q$ is consider small for all the subgraphs.
The internal loop momentum $k$ is considered large for the subgraph $\Gamma$; while for the subgraph $\gamma$, the loop momentum $k$ also flows over $\Gamma\slash \gamma$ so is considered small.

Note in Figure \ref{fig:sss} we have ignored another putative subgraph $\Gamma/\gamma$, i.e., the two links connecting operators $\{O_2, O_3\}$ and $\{O_3, O_1\}$. The reason is that the integral $\mathcal{I}_{\Gamma/\gamma}$ for this subgraph, after taking Taylor expansion with respect to $k$ and $q$, becomes an integral with no scale and vanishes automatically\footnote{Here the internal momentum $k$ has been shifted to $k\rightarrow k-p$ comparing with (\ref{3DI}) so that the large external momentum $p$ only flows inside the new subgraph $\Gamma/\gamma$.}
\begin{equation}
    \mathcal{I}_{\Gamma/\gamma}=\int \frac{d^D k}{\left(2\pi\right)^D}\frac{1}{\left((k)^2\right)^{\nu_3}} \mathcal{T}_{k,q}
    \frac{1}{\left((k-p)^2\right)^{\nu_2}\left((k-p-q)^2\right)^{\nu_1}}=0.
\end{equation}
We use $\mathcal{I}_{\gamma,n}$ to denote the contribution from the subgraph $\gamma$ at the order $n$ in the Taylor expansion. The leading terms of the two subgraphs are given by
\begin{align}
   \mathcal{I}_{\Gamma,0}&= \int \frac{d^D k}{\left(2\pi\right)^D}\mathcal{T}_q^0 \frac{1}{\left((p+k)^2\right)^{\nu_3}\left(k^2\right)^{\nu_2}\left((k-q)^2\right)^{\nu_1}}=\int \frac{d^D k}{\left(2\pi\right)^D}\frac{1}{\left(k^2\right)^{\nu_1+\nu_2}\left((p+k)^2\right)^{\nu_3}}\nn\\
   &= \frac{2^{-D} \pi ^{-\frac{D}{2}} p^{-2 D+\Delta _t}\Gamma \left(\frac{1}{2} \left(\Delta _1+\Delta _2-\Delta _3\right)\right) \Gamma \left(D-\frac{\Delta _t}{2}\right) \Gamma \left(\Delta _3-\frac{D}{2}\right)}{\Gamma \left(D-\Delta _3\right) \Gamma \left(\frac{1}{2} \left(D-\Delta _1-\Delta _2+\Delta _3\right)\right) \Gamma \left(\frac{1}{2} \left(-D+\Delta _t\right)\right)}, \label{sssG}
   \end{align}
  and
   \begin{align}
   \mathcal{I}_{\gamma,0}&= \int \frac{d^D k}{\left(2\pi\right)^D} \frac{1}{\left(k^2\right)^{\nu_2}\left((k-q)^2\right)^{\nu_1}}\mathcal{T}_{q,k}^0\frac{1}{\left((p+k)^2\right)^{\nu_3}}= \int \frac{d^D k}{\left(2\pi\right)^D}\frac{1}{\left(p^2\right)^{\nu_3}\left(k^2\right)^{\nu_2}\left((k-q)^2\right)^{\nu_1}} \nn\\
   &=\frac{2^{-D} \pi ^{-\frac{D}{2}} p^{- D+\Delta _t-2\Delta _3} q^{2\Delta_3-D}\Gamma \left(\frac{1}{2} \left(\Delta _t-2\Delta _3\right)\right) \Gamma \left(\frac{1}{2} \left(\Delta_t-2\Delta _1\right)\right) \Gamma \left(\frac{D}{2}-\Delta _3\right)}{\Gamma \left(\Delta _3\right) \Gamma \left(\frac{1}{2} \left(D+\Delta _1-\Delta _2-\Delta _3\right)\right) \Gamma \left(\frac{1}{2} \left(D-\Delta _t+2\Delta _2\right)\right)}. \label{sssg}
\end{align}
The two leading terms $\mathcal{I}_{\Gamma,0}$ and $\mathcal{I}_{\gamma,0}$ are consistent with the results (\ref{3ptpq}) up to an overall normalization factor. Note the solution (\ref{3ptpq}) is obtained using asymptotic behavior and integral properties of the Bessel functions. In contrast, the above results can be directly evaluated using the diagrammatic subgraph expansions.
In the following sections of this work, we will use the method of subgraphs to compute the conformal 3-point correlators of conserved currents and higher loop Feynman integrals.

\subsection{Comments on the subgraph expansion and operator product expansion}
We provide physical interpretations for the subgraph expansions of the conformal 3-point correlator (\ref{sssG}), (\ref{sssg}). We show that the subgraph expansions automatically reproduce the singularities in the correlation functions in the OPE limit, and that the method of subgraphs provides a useful technique to study the OPE in momentum space.

In position space, the product of two operators can be expanded
\begin{equation}
\cO_1(x_1)\cO_2(x_2)=\sum\limits_i C_i(x,\partial_x)\cO_i(x_2), \label{OPEinx}
\end{equation}
in which we define $x^\mu\equiv x_{1}^\mu-x_2^\mu$ for brevity.
The above OPE is convergent in the regions where no other operators are inserted. In the short distance limit $x\rightarrow 0$, the $\cO_3$-dependent part in the OPE is dominated by
\begin{equation}
    \cO_1(x_1)\cO_2(x_2)\sim \frac{\lambda_{\cO_1\cO_2\cO_3}}{x_{12}^{\Delta_1+\Delta_2-\Delta_3}}\cO_3(x_2)+\cdots.
\end{equation}
Applying the above OPE in the conformal scalar 3-point correlator (\ref{3pt}), it leads to
\begin{equation}
    \langle \cO_1(x_1)\cO_2(x_2)\cO_3(x_3)\rangle\sim \frac{\lambda_{\cO_1\cO_2\cO_3}}{x^{\Delta_1+\Delta_2-\Delta_3}_{12}}\langle\cO_3(x_2)\cO_3(x_3)\rangle+\cdots. \label{OPEin3p}
\end{equation}
The OPE in momentum space is given by the Fourier transformation of (\ref{OPEinx}):
\begin{equation}
    \cO_1(p)\cO_2(0)\equiv \int d^Dx e^{ip\cdot x}\,\cO(x)\cO(0)\sim\sum\limits_i \tilde{C}_i(p,\partial_p) \cO_i(0). \label{OPEinLargep}
\end{equation}
The OPE limit in position space $x\rightarrow 0$ corresponds to the large momentum  limit $p\rightarrow \infty$  in momentum space.
One might naively expect that the correlation function in momentum space obtained from (\ref{OPEinLargep}) should be identical to the Fourier transformation of the 3-point correlation function on the right-hand side of (\ref{OPEin3p}). However, this is not correct.

The Fourier transformation of the OPE limit in position space (\ref{OPEin3p}) gives
\begin{equation}
    \langle \cO_1(p)\cO_2(-p-q)\cO_3(q)\rangle \propto p^{\Delta_1+\Delta_2-\Delta_3}q^{2\Delta_3-D}+\cdots,  \label{OPE3}
\end{equation}
which is exactly the subgraph contribution $\mathcal{I}_\gamma$ (\ref{sssg}). Recall that when evaluating $\mathcal{I}_\gamma$, the Taylor expansion $\mathcal{T}_{q,k}$ with respect to small $q$ and $k$, or equivalently the large $p$ limit has been applied. This is just the OPE limit $p\rightarrow \infty$ in momentum space. However, we have shown that in the subgraph expansion with large $p$, the whole conformal 3-point correlator also contains a $q$-regular term $\mathcal{I}_\Gamma$ contributed from the subgraph $\Gamma$. For the scalar $\cO_3$ with $\Delta_3>D/2$, the $q$-regular term is dominating and the naive OPE expansion (\ref{OPE3}) completely misses the asymptotic behavior of the correlator in this limit.

The key is that the OPE in position space (\ref{OPEinx}) is convergent only when no extra operator in between $\cO_1(x)$ and $\cO_2(0)$. However, for the OPE in momentum space (\ref{OPEinLargep}), one integrates over the whole spacetime $x\in\mathbb{R}^D$, and when applying it in the 3-point correlator with an extra operator $\cO(x_3)$, the integration domain covers the neighborhood of $\cO(x_3)$: $x_3\in\mathbb{U}_{x_3}$, where the operator $\cO_1(x_1)$ is more close to the third operator $\cO(x_3)$ and the OPE (\ref{OPEinx}) for $\cO_1(x_1)\times \cO_2(x_2)$ is not convergent any more. The contribution from the integration domain $\mathbb{U}_{x_3}$ corresponds to the large loop momentum $k$ which is comparable to the large external momentum $k\sim p$, and the integrand should be expanded with respect to large $k$ and $p$, or equivalently small $q$ denoted by $\mathcal{T}_q$ in (\ref{sssG}), this is precisely the contribution $\mathcal{I}_\Gamma$ from the subgraph $\Gamma$! The correlation functions with coincident coordinates, e.g. $x_2\rightarrow x_3$ are the so-called contact terms, and their Fourier transformations generate functions that are analytic in $q$ \cite{Maldacena:2011nz}. The extra term $\mathcal{I}_\Gamma$ in the OPE limit plays an important role in the studies of the relation between scale and conformal invariance \cite{Bzowski:2014qja,Dymarsky:2014zja} and also the conductivity at finite temperature \cite{Lucas:2016fju,Lucas:2017dqa}.

In short, the subgraph expansion of the conformal 3-point correlator provides a comprehensive description for the  asymptotic behavior of the correlator in the OPE limit in momentum space, which is known to be a subtle problem \cite{Bzowski:2014qja,Dymarsky:2014zja} as the integration domain in the Fourier transformation  covers the neighborhood of the third operator where the OPE stops converging. In general such kind of problems can be carefully treated using the method of regions, and for the integrals in Euclidean space, the results are given by the subgraphs of the original integral. For the conformal correlators in Lorentzian space, the method of subgraphs does not apply and one needs to go back to the method of regions to evaluate contributions from different integration domains.
We expect the method of subgraphs can play an important  role for the conformal 4-point correlators and the related studies, in particular the conformal bootstrap in momentum space, for which we will provide more discussions at the end of this paper.

\section{Subgraph expansions of conserved current 3-point correlators} \label{free3pt}

We compute the conformal 3-point correlators of the conserved currents in momentum space using the subgraph expansion. It has been shown in \cite{Osborn:1993cr,Erdmenger:1996yc} that the conformal current 3-point correlators can be decomposed into those derived from free field theories, and the correlation functions have been solved in compact forms in position space. In momentum space, the conformal conserved current 3-point correlation functions can also be solved from the conformal Ward identities \cite{Bzowski:2013sza}. Nevertheless, they consist of complicated expressions that are not convenient for further study. The method of subgraphs, as explained previously, provides a simple graphical expansion for the conformal 3-point correlation functions, which is useful for related studies. Specifically, the results in this section provide the bases to compute the subleading order corrections in the $1/N$ expansion.

We study the three conformal 3-point correlators of conserved currents: $\langle JJJ\rangle$, $\langle TJJ\rangle$ and $\langle TTT\rangle$. The independent tensor structures in these correlators have been classified in \cite{Osborn:1993cr, Erdmenger:1996yc}, which can be realized by free boson and free fermion theories. For the stress tensor 3-point correlator $\langle TTT\rangle$ in 4D, its tensor structures also contain those from the free vector field theory. In this work, we compute the conformal 3-point correlation functions in general dimensions. It has been known that the free vector field theory is not conformal in $D\neq 4$ \cite{El-Showk:2011xbs}. Therefore, we focus on the tensor structures realized by the free boson and free fermion theories. The method in this work can be directly applied to the 4D free vector theory.

% \begin{table}  
% \centering
% \caption{Independent  tensor structures and parameters in the conformal current 3-point correlators.}\label{table:JTcorrelators}
% \begin{tabular}{cccc}
%  \hline\hline  \\[-1em]
%   & \bf ~ Tensor structures  & ~ \bf  Parameters ~
%  & \bf  ~ Realized by ~  \\[.3em]
% %\mbox{Three}&\mbox{Four}&\mbox{Five}\\
% \hline \\[-1em]
% $\langle JJJ \rangle$  & {\bf 2}  & $C_J, ~\lambda_{JJJ}$ & free boson \& fermion     \\[.3em]
% \hline \\[-1em]
% $ \langle TJJ \rangle$  & {\bf 2} & $C_J, ~ \lambda_{TJJ}$ & free boson \& fermion    \\[.3em]
% \hline\\[-1em]
% $\langle TTT\rangle_{D=3}$  & {\bf 2} & $n_B, ~ n_F$ &  free boson \& fermion    \\[.3em]
% \hline \\[-1em]
% $\langle TTT\rangle_{D=4}$ & {\bf 3} &  $C_T, ~ a,~ c$ & free boson \& fermion \& vector      \\[.3em]
%  \hline\hline
% \end{tabular}  
% \end{table}

To compute the 3-point correlators in momentum space using Feynman diagrams, we need to specify the propagators and vertices of the external operators. Here we briefly introduce these ingredients for free scalars $\phi^i$ and free fermions $\psi^i$. A more detailed introduction to the interacting theories will be provided in Section \ref{largeNdiags}.

\begin{figure}
  \begin{equation*}
\begin{tikzpicture}[baseline=(a.base)]
  \begin{feynman}
    \vertex [label=right:$i$] (c) at (2,3/2) {};
    \vertex[label=right:$j$] (b) at (2,-3/2) {};
    \vertex[dot,label=left:$J
    ^a_{\mu}(p)$] (a)  at (0,0) {};

    \diagram* {
       (c) -- [fermion, edge label=\(k\)] (a) --[fermion, edge label=\(k+p\)] (b) 
    };
  \end{feynman}
\end{tikzpicture} =U(p,k)\left(t^a\right)^{i}_j,
~~
    \begin{tikzpicture}[baseline=(a.base)]
  \begin{feynman}
    \vertex [label=right:$i$] (c) at (2,3/2) {};
    \vertex[label=right:$j$] (b) at (2,-3/2) {};
    \vertex[dot,label=left:$T_{\mu\nu}(p)$] (a)  at (0,0) {};

    \diagram* {
       (c) -- [fermion, edge label=\(k\)] (a) --[fermion, edge label=\(k+p\)] (b) 
    };
  \end{feynman}
\end{tikzpicture} =V(p,k) \,\delta^{i}_j
\end{equation*}
    \caption{Vertices of the conserved currents $J^a_\mu(p)$ and $T_{\mu\nu}(p)$.}
    \label{fig:vertex}
\end{figure}
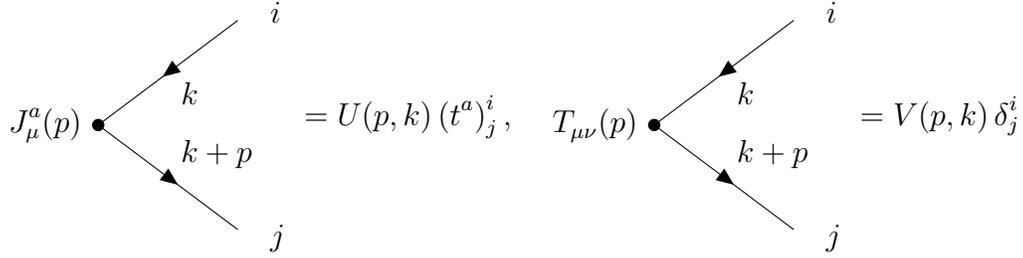

Consider the theories with $N$ real scalars or $N$  4-component Dirac fermions. Their Lagrangians $\mathcal{L}_{s/f}$ are
\begin{equation}
    \mathcal{L}_{s}=\frac{1}{2}\left( \partial\phi_0^i\right)^2, ~~~~ \mathcal{L}_{f}=-\bar{\psi}_{0i}\slashed{\partial}\psi_0^i,
\end{equation}
and the propagators of the free scalars $\phi_0^i$ and free fermions $\psi_0^i$ are given by
\begin{equation}
    \langle \phi_0^i(p) \phi^j_0(-p)\rangle=\frac{\delta^{ij}}{p^2}, ~~~ \langle \psi^i_0(p) \bar{\psi}_{0j}(-p)\rangle=\frac{\delta^{i}_j\slashed{p}}{p^2}. \label{prop}
\end{equation}
The conserved spin 1 current $J^a_\mu$ and the stress tensor operator $T_{\mu\nu}$ read
\begin{align}
     J^a_{\mu}(p)&=\frac{i}{2}\int\frac{d^Dk}{(2\pi)^D} \left(2k_\mu+p_\mu\right) \phi_{0}^i(-k)\left(t^a\right)^{ij} \phi_0^j(p+k), \label{Jop}\\
    T_{\mu\nu}(p)&=\int \frac{d^Dk}{(2\pi)^D} \left(k_\mu k_\nu+\frac{k_\mu p_\nu+p_\mu k_\nu}{2}+\frac{D-2}{4(D-1)}p_\mu p_\nu\right)\phi_{0}^i(k+p) \phi_0^i(-k)-\mathrm{trace}, \label{Top}
\end{align}
for the free scalar theory and
\begin{align}
    J^a_{\mu}(p)&=-\int\frac{d^Dk}{(2\pi)^D} \bar{\psi}_{0i}(-k)\left(t^a\right)^i_j \gamma_\mu \psi_0^j(p+k), \label{Jopf}\\
    T_{\mu\nu}(p)&=-\frac{i}{2}\int \frac{d^Dk}{(2\pi)^D} \bar{\psi}_{0i} \gamma_\mu \left(k_\nu+\frac{1}{2}p_\nu\right) \psi_0^i(p+k)+(\mu\leftrightarrow \nu)-\mathrm{trace}, \label{Topf}
\end{align}
for the free fermion theory. In the perturbative computations, the expressions for $J$ and $T$ correspond to the vertices shown in Figure \ref{fig:vertex}. The momentum-dependent factors in the vertices are given by
\begin{equation}
  U(p,k)=i(2k_\mu+p_\mu),   ~~ V(p,k)=2k_\mu k_\nu+k_\mu p_\nu+p_\mu k_\nu+\frac{D-2}{2(D-1)}p_\mu p_\nu-\mathrm{trace}, \label{vertexON}
\end{equation}
for the free boson theory and 
\begin{equation}
  U(p,k)=-\gamma_\mu,   ~~ V(p,k)=-\frac{i}{4}\left(2k_\mu+p_\mu \right)\gamma_\nu + (\mu\leftrightarrow\nu) -\mathrm{trace}, \label{vectexGNY}
\end{equation}
for the free fermion theory. 

\begin{figure}
   \begin{equation*}
   \begin{array}{ccccc}
    \begin{tikzpicture}[baseline=(a.base)]
  \begin{feynman}
    \vertex [dot,label=right:$J^c_{\rho}(-p-    q)$] (c) at (2,3/2) {};
    \vertex[dot,label=right:$J^b_{\nu}(p)$] (b) at (2,-3/2) {};
    \vertex[dot,label=left:$J^a_{\mu}(q)$] (a)  at (0,0) {};

    \diagram* {
       (a) -- [fermion, edge label=\(k\)] (b) --[fermion] (c) --[fermion] (a)
    };
  \end{feynman}
\end{tikzpicture} &=&
\begin{tikzpicture}[baseline=(a.base)]
  \begin{feynman}
    \vertex [dot,label=right:$J^c_{\rho}(-p-    q)$] (c) at (2,3/2) {};
    \vertex[dot,label=right:$J^b_{\nu}(p)$] (b) at (2,-3/2) {};
    \vertex[dot,label=left:$J^a_{\mu}(q)$] (a)  at (0,0) {};

    \diagram* {
       (a) -- [fermion,red, edge label={\color{black}\(k\)}] (b) --[fermion,red] (c) --[fermion,red] (a)
    };
  \end{feynman}
\end{tikzpicture}
&+& 
\begin{tikzpicture}[baseline=(a.base)]
  \begin{feynman}
    \vertex [dot] (c) at (2,3/2) {};
    \vertex[dot] (b) at (2,-3/2) {};
    \vertex[dot] (a)  at (0,0) {};

    \diagram* {
       (a) -- [fermion, edge label=\(k\)] (b) --[fermion,red] (c) --[fermion] (a)
    };
  \end{feynman}
\end{tikzpicture} \\
& & & &
\\ \mathrm{Diagram~ }\displaystyle{\Gamma} & & \mathrm{Subgraph ~ } {\textstyle\color{red}\Gamma} & & \mathrm{Subgraph ~ } {\color{red}\gamma}
\end{array}
\end{equation*}
    \caption{Feynman diagram for the current 3-point correlator $\langle J^a_\mu J^b_\nu J^c_\rho\rangle$ and its expansion with two subgraphs: $\Gamma$ and $\gamma$ marked in red color. The Feynman diagrams for the correlators $\langle TJJ \rangle$ and $\langle TTT \rangle$ are similar but with different vertices for the stress tensor. }
    \label{fig:JJJ}
\end{figure}
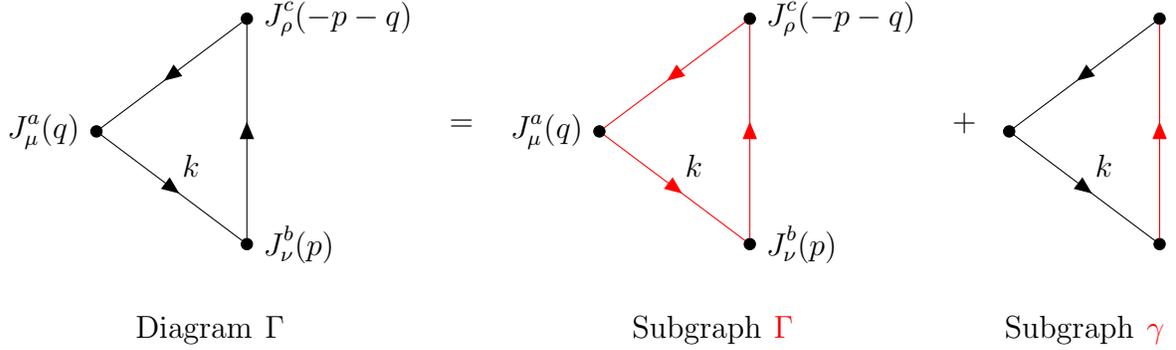

In free field theories, it is straightforward to compute the conformal 3-point correlation functions of the conserved currents $\langle JJJ \rangle$, $\langle TJJ \rangle$ and $\langle TTT \rangle$ using Wick contractions. The final results are given by the one-loop Feynman integrals. In Figure \ref{fig:JJJ} we show the Feynman diagram for the correlator $\langle JJJ\rangle$, and the Feynman diagrams for $\langle TJJ \rangle$ and $\langle TTT \rangle$ have the same geometry but different vertices.
The subgraph expansions of the conserved current 3-point correlators are similar to those of the scalar 3-point correlator: there are two subgraphs $\Gamma$ and $\gamma$, in which the integrands are expanded with respect to two sets of small momenta $\{q\}$ and $\{q, k\}$, respectively. Then the one-loop 2-scale Feynman integrals are reduced to one-loop 1-scale Feynman integrals, which can be evaluated analytically. 

In principle, one can take the Taylor expansions of the integrands for each subgraph  $\mathcal{T}_{q/q,k}$ to arbitrarily high orders.  The resulting integrals  are given by the series expansions of the small parameter $q/p$. For the Feynman diagrams in general QFTs, the expansions to high orders are needed to approximate the original integrals \cite{Smirnov:2023bxf}. However, in CFTs the 3-point correlation functions are fixed by conformal symmetry up to a few free parameters, and  the $1/N$ corrections are only applied to these free parameters. To extract these  corrections from the Feynman loop integrals, we just need to keep the first few terms in the Taylor expansion of each subgraph.

In the following parts, we present the subgraph expansions for the correlators $\langle JJJ \rangle$, $\langle TJJ \rangle$ and $\langle TTT \rangle$ to the subleading order, which will be sufficient to extract the subleading corrections in the large $N$ expansion. Higher order terms can be  computed using the same method. We use $\mathcal{I}_{\gamma,i}$ to denote the $i$-th order contribution from the subgraph $\gamma$.

\subsection{Conserved current 3-point correlator $\langle JJJ \rangle$} \label{JT3pt}
In the free scalar theory, the spin-1 current 3-point correlation function is given by
\begin{align}
    \mathcal{I}_{\Gamma,0}=-&\frac{2^{1-D} \pi ^{1-\frac{D}{2}} \left(p^2\right)^{\frac{D}{2}-3} \csc \left(\frac{\pi  D}{2}\right) \Gamma \left(\frac{D}{2}\right)}{\Gamma (D)} \label{JJJG0}\\
    &\times \left(p_{\mu } \left((D-4) p_{\nu } p_{\rho }-(D-2) p^2 \delta (\nu ,\rho )\right)+p^2 \left(p_{\rho } \delta (\mu ,\nu )+p_{\nu } \delta (\mu ,\rho )\right)\right), \nn\\
    \mathcal{I}_{\Gamma,1}=-& \frac{2^{1-2 D} \pi ^{\frac{3}{2}-\frac{D}{2}} \left(p^2\right)^{\frac{D}{2}-4} \csc \left(\frac{\pi  D}{2}\right)}{\Gamma \left(\frac{D+1}{2}\right)} \label{JJJG1}\\
    & \times\left(p^2 \left((D-4) p_{\mu } p_{\rho } q_{\nu }+(D-4) p_{\nu } \left(p_{\rho } q_{\mu }+p_{\mu } q_{\rho }\right)\right)+(D-6) (D-4) p_{\alpha } p_{\mu } p_{\nu } p_{\rho } q_{\alpha }\right) \nn \\
    &+ \delta\mathrm{-dependent ~ terms} \nn \\
     \mathcal{I}_{\gamma,0}=-& \frac{2^{1-D} \pi ^{1-\frac{D}{2}} \left(q^2\right)^{\frac{D}{2}-2} \csc \left(\frac{\pi  D}{2}\right) \Gamma \left(\frac{D}{2}\right)}{\left(p^2\right)^2 \Gamma (D)} \label{JJJg0}\\
     &\times \left(p^2 p_{\nu } \left(q^2 \delta (\mu ,\rho )-q_{\mu } q_{\rho }\right)-p_{\rho } \left(p^2 \left(q_{\mu } q_{\nu }-q^2 \delta (\mu ,\nu )\right)+q^2 p_{\mu } p_{\nu }\right)+p_{\alpha } p_{\nu } p_{\rho } q_{\alpha } q_{\mu }\right). \nn
\end{align}
In the free fermion theory, the correlation function is
\begin{align}
    \mathcal{I}_{\Gamma,0}=&\frac{2^{1-2 D} (D-2) \pi ^{\frac{3}{2}-\frac{D}{2}} \left(p^2\right)^{\frac{D}{2}-3} \csc \left(\frac{\pi  D}{2}\right)}{\Gamma \left(\frac{D+1}{2}\right)} \\
    &\times p_{\mu } \left((D-4) p_{\nu } p_{\rho }-(D-2) p^2 \delta (\nu ,\rho )\right)+p^2 \left(p_{\rho } \delta (\mu ,\nu )+p_{\nu } \delta (\mu ,\rho )\right), \nn\\
    \mathcal{I}_{\Gamma,1}=& \frac{4^{-D} \pi ^{\frac{3}{2}-\frac{D}{2}} \left(p^2\right)^{\frac{D}{2}-4} \csc \left(\frac{\pi  D}{2}\right)}{\Gamma \left(\frac{D+1}{2}\right)}(D-4) \\
    & \times \left(p_{\nu } \left((D-2) p_{\rho } \left((D-6) p_{\mu } p_{\sigma } q_{\sigma }+p^2 q_{\mu }\right)+(2 D-3) p^2 p_{\mu } q_{\rho }\right)-p^2 p_{\mu } p_{\rho } q_{\nu }\right) \nn \\
    &+ \delta\mathrm{-dependent ~ terms} \nn \\
     \mathcal{I}_{\gamma,0}=&\frac{2^{1-2 D} (D-2) \pi ^{\frac{3}{2}-\frac{D}{2}} \left(q^2\right)^{\frac{D}{2}-2} \csc \left(\frac{\pi  D}{2}\right)}{p^2 \Gamma \left(\frac{D+1}{2}\right)} \label{JJJfg1} \\
     &\times \left(\delta (\nu ,\rho ) \left(q^2 p_{\mu }-p_{\alpha } q_{\alpha } q_{\mu }\right)+p_{\rho } \left(q_{\mu } q_{\nu }-q^2 \delta (\mu ,\nu )\right)+p_{\nu } \left(q_{\mu } q_{\rho }-q^2 \delta (\mu ,\rho )\right)\right). \nn
\end{align}

\subsection{Conserved current 3-point correlator $\langle TJJ \rangle$}
The subgraph expansion of the conformal correlation function $\langle TJJ \rangle$ in the free scalar theory
is given by
\begin{align}
    \mathcal{I}_{\Gamma,0}=&\frac{2^{1-2 D} \pi ^{\frac{3}{2}-\frac{D}{2}} \left(p^2\right)^{\frac{D}{2}-3} \csc \left(\frac{\pi  D}{2}\right)}{D \Gamma \left(\frac{D+1}{2}\right)} \\
    &\times \left(D p_{\mu } p_{\nu } \left((D-4) p_{\rho } p_{\sigma }-(D-2) p^2 \delta (\rho ,\sigma )\right)-(D-4) p^2 p_{\rho } p_{\sigma } \delta (\mu ,\nu )+\right. \nn\\
    &~~~~\left. \left(p^2\right)^2 \left(D \delta (\mu ,\sigma ) \delta (\nu ,\rho )+D \delta (\mu ,\rho ) \delta (\nu ,\sigma )+(D-4) \delta (\mu ,\nu ) \delta (\rho ,\sigma )\right)\right) \nn, \\
    \mathcal{I}_{\Gamma,1}=& \frac{4^{-D} \pi ^{\frac{3}{2}-\frac{D}{2}} \left(p^2\right)^{\frac{D}{2}-4} \csc \left(\frac{\pi  D}{2}\right)}{D \Gamma \left(\frac{D+1}{2}\right)} (D-4) D p_{\rho } \\
    & \times\left((D-6) p_{\beta } p_{\mu } p_{\nu } p_{\sigma } q_{\beta }+p^2 \left(p_{\mu } p_{\sigma } q_{\nu }+p_{\nu } \left(p_{\sigma } q_{\mu }+2 p_{\mu } q_{\sigma }\right)\right)\right)+ \delta\mathrm{-dependent ~ terms}, \nn \\
     \mathcal{I}_{\gamma,0}=& \frac{4^{-D} \pi ^{\frac{3}{2}-\frac{D}{2}} \left(q^2\right)^{\frac{D-4}{2}} \csc \left(\frac{\pi  D}{2}\right)}{(D-1) (p^2)^2 \Gamma \left(\frac{D+3}{2}\right)}\\
     &\times \left(-\left((D-1) q^2 p_{\mu } q_{\nu } \left(p_{\sigma } q_{\rho }+p_{\rho } q_{\sigma }\right)\right)+2 (D-2) p_{\alpha } q_{\alpha } q_{\mu } q_{\nu } \left(p_{\sigma } q_{\rho }+p_{\rho } q_{\sigma }\right)+\right. \nn \\
     &~~~~\left.q_{\mu } \left(-2 q_{\nu } \left((D-2) p^2 q_{\rho } q_{\sigma }-2 q^2 p_{\rho } p_{\sigma }\right)-\left((D-1) q^2 p_{\nu } \left(p_{\sigma } q_{\rho }+p_{\rho } q_{\sigma }\right)\right)\right) \right) \nn\\
     &+ \delta\mathrm{-dependent ~ terms}. \nn
\end{align}
Note that the lowest non-vanishing term in $\mathcal{I}_{\gamma}$ appears at the order $O(q^{D})$, in contrast to the order $O(q^{D-2})$ for the correlator $\langle JJJ\rangle$.

In the free fermion theory, the correlation function $\langle TJJ \rangle$ is
\begin{align}
    \mathcal{I}_{\Gamma,0}=&-\frac{4^{-D} (D-2) \pi ^{\frac{3}{2}-\frac{D}{2}} \left(p^2\right)^{\frac{D-6}{2}} \csc \left(\frac{\pi  D}{2}\right)}{D \Gamma \left(\frac{D+1}{2}\right)} \\
    &\times \left(p^2 \left(2 (D-2) p^2 \delta (\mu ,\nu ) \delta (\rho ,\sigma )+D p_{\nu } p_{\sigma } \delta (\mu ,\rho )+p_{\rho } \left(D p_{\nu } \delta (\mu ,\sigma )-2 (D-2) p_{\sigma } \delta (\mu ,\nu )\right)\right)\right. \nn\\
    &\left. ~~~~+D p_{\mu } \left(2 p_{\nu } \left((D-4) p_{\rho } p_{\sigma }-(D-2) p^2 \delta (\rho ,\sigma )\right)+p^2 \left(p_{\sigma } \delta (\nu ,\rho )+p_{\rho } \delta (\nu ,\sigma )\right)\right)\right) \nn, \\
    \mathcal{I}_{\Gamma,1}=& -\frac{2^{-D-1} (D-2) \pi ^{1-\frac{D}{2}} z \left(p^2\right)^{\frac{D}{2}-4} \csc \left(\frac{\pi  D}{2}\right) \Gamma \left(\frac{D}{2}\right)}{\Gamma (D+1)} \\
    & \times\left((D-6) p_{\alpha } p_{\mu } p_{\nu } p_{\sigma } q_{\alpha }+p^2 \left(p_{\mu } p_{\sigma } q_{\nu }+p_{\nu } \left(p_{\sigma } q_{\mu }+2 p_{\mu } q_{\sigma }\right)\right)\right)+ \delta\mathrm{-dependent ~ terms}, \nn \\
     \mathcal{I}_{\gamma,0}=& \, O(q^{D+1}).
\end{align}
The subgraph $\gamma$ vanishes at the order $O(q^D)$ in the free fermion theory.

\subsection{Conserved current 3-point correlator $\langle TTT \rangle$}
The correlation function $\langle TTT\rangle$
has more complicated tensor structures. In the free scalar theory, it is given by
\begin{align}
    \mathcal{I}_{\Gamma,0}=&\frac{2^{-2 (D+1)} (D-2) \pi ^{\frac{3}{2}-\frac{D}{2}} \left(p^2\right)^{D/2} \csc \left(\frac{\pi  D}{2}\right)}{(D-1) p^6 \Gamma \left(\frac{D+3}{2}\right)}\left(2 (D-4) p_{\alpha } p_{\beta } p_{\mu } p_{\nu } p_{\rho } p_{\sigma }+\right. \\
    &\left. p_{\alpha } \left(-\left((D-1) p^2 p_{\mu } p_{\nu } \left(p_{\sigma } \delta (\beta ,\rho )+p_{\rho } \delta (\beta ,\sigma )\right)\right)-2 p^2 p_{\beta } \delta (\mu ,\nu )+p^2 p_{\mu } p_{\nu } \delta (\rho ,\sigma )\right) \right. \nn\\
    &\left. ~~~+p^2 p_{\mu } p_{\nu } \left(p_{\rho } \left(2 p_{\sigma } \delta (\alpha ,\beta )-(D-1) p_{\beta } \delta (\alpha ,\sigma )\right)-(D-1) p_{\beta } p_{\sigma } \delta (\alpha ,\rho )\right)\right) +\cdots \nn, \\
    \mathcal{I}_{\Gamma,1}=& \frac{4^{-D-1} (D-4) (D-2) \pi ^{\frac{3}{2}-\frac{D}{2}} \left(p^2\right)^{\frac{D-8}{2}} \csc \left(\frac{\pi  D}{2}\right) }{(D-1) \Gamma \left(\frac{D+3}{2}\right)} p_{\rho } p_{\sigma }\\
    & \times\left((D-6) p_{\alpha } p_{\beta } p_{\mu } p_{\nu } p_{\tau } q_{\tau }+p^2 \left(p_{\nu } \left(2 p_{\mu } \left(p_{\beta } q_{\alpha }+p_{\alpha } q_{\beta }\right)+p_{\alpha } p_{\beta } q_{\mu }\right)+p_{\alpha } p_{\beta } p_{\mu } q_{\nu }\right)\right) \nn\\
    &+ \delta\mathrm{-dependent ~ terms}, \nn \\
    \mathcal{I}_{\gamma,0}=& O(q^{D-1}).
\end{align}
For brevity, here  we have ignored terms with more than one $\delta$ functions in $\mathcal{I}_{\Gamma,0}$, and all the $\delta$-dependent terms in $\mathcal{I}_{\Gamma,1}$ and $\mathcal{I}_{\gamma,0}$. 

The correlation function in the free fermion theory is
\begin{align}
    \mathcal{I}_{\Gamma,0}=&-\frac{4^{-D} (D-4) (D-2) \pi ^{\frac{3}{2}-\frac{D}{2}} \left(p^2\right)^{\frac{D}{2}-3} \csc \left(\frac{\pi  D}{2}\right)}{\Gamma \left(\frac{D+3}{2}\right)}p_{\alpha } p_{\beta } p_{\mu } p_{\nu } p_{\rho } p_{\sigma } \\
    &+ \delta\mathrm{-dependent ~ terms}\nn, \\
    \mathcal{I}_{\Gamma,1}=& \frac{p_{\rho } p_{\sigma } \left(2^{-2 D-1} (D-4) (D-2) \pi ^{\frac{3}{2}-\frac{D}{2}} \left(p^2\right)^{\frac{D}{2}-4} \csc \left(\frac{\pi  D}{2}\right)\right)}{\Gamma \left(\frac{D+3}{2}\right)}\\
    & \times \left((D-6) p_{\alpha } p_{\beta } p_{\gamma } p_{\mu } p_{\nu } q_{\gamma }+p^2 \left(p_{\alpha } p_{\beta } p_{\nu } q_{\mu }+p_{\mu } \left(2 p_{\nu } \left(p_{\beta } q_{\alpha }+p_{\alpha } q_{\beta }\right)+p_{\alpha } p_{\beta } q_{\nu }\right)\right)\right) \nn\\
    &+ \delta\mathrm{-dependent ~ terms}, \nn \\
    \mathcal{I}_{\gamma,0}=& O(q^{D-1}).
\end{align}
The correlators with stress tensor $T_{\mu\nu}$ are found to have weaker singularity in $q$ within the subgraph $\gamma$.

\subsection{Compare the subgraph expansions with the 3D exact results}
We verify the subgraph expansions of the conformal current 3-point correlators. It is straightforward to compare the above results with the solutions of the conformal Ward identities in momentum space \cite{Bzowski:2013sza}. This has been done for the scalar 3-point correlators in (\ref{sssG}) and (\ref{sssg}).
Here we show another verification of the subgraph expansions based on the exact results in 3D, in which the Feynman diagrams generate  conformal integrals  and can be evaluated analytically for the conformal current 3-point correlators in free scalar and free fermion theories. 

Let us take the correlator $\langle JJJ \rangle$ in the 3D free boson theory as an example.
The correlation function is given by the Feynman diagram in Figure \ref{fig:JJJ} with vertices of the operator $J$, which corresponds to the integral
\begin{align}
    \langle J^a_\mu(q)J^b_\nu(p)J^c_\rho(-p-q)\rangle= \alpha_J \int \frac{d^3k}{(2\pi)^{6}}\frac{\left(2 k_{\mu }-q_{\mu }\right) \left(p_{\nu }+2 k_{\nu }\right) \left(2 k_{\rho }+p_{\rho }-q_{\rho }\right)}{k^2 (k+p)^2 (k-q)^2}, \label{JJJ3D}
\end{align}
where $\alpha_J=i^3\mathrm{tr}(t^at^bt^c)$ is the symmetry factor and will be assumed implicitly.
Using tensor reduction, above integral can be reduced to scalar conformal 3-point integrals or the 2-point integrals (\ref{2ptint}), both of which can be solved in compact forms.
The exact integral in (\ref{JJJ3D}) is given by
\begin{align}
    \mathcal{I}_{JJJ,\;\mathrm{3D}}=\;&\frac{p_{\mu } p_{\nu } p_{\rho } q\left(p^2+p q+(2 p+q)+2 q^2|p+q|+|p+q|^2\right)+\cdots}{8 p q|p+q|(p+q+|p+q|)^3} \nn \\
    & +\frac{\delta (\mu ,\nu ) \left(p_{\rho } \left(p+|p+q|\right)-q_{\rho } \left(q+|p+q|\right)\right)}{8 \left(p+q+|p+q|\right)^2} +\cdots
    % +\frac{\delta (\mu ,\rho ) \left(q_{\nu } \left(2 p+q+|p+q|\right)+p_{\nu } \left(p+|p+q|\right)\right)}{8 \left(p+q+|p+q|\right)^2} \nn\\
    % &-\frac{\delta (\nu ,\rho ) \left(p_{\mu } \left(p+2 q+|p+q|\right)+q_{\mu } \left(q+|p+q|\right)\right)}{8 \left(p+q+|p+q|\right)^2}.
\end{align}
In the zero-momentum limit $q\rightarrow 0$, above correlation function admits a small $q$ Taylor expansion, and indeed there are two parts which are respectively regular/singular in the small $q$ limit.
The results completely agree with the subgraph expansions (\ref{JJJG0}-\ref{JJJg0}). In particular, there is a $q$-singular sector in $\mathcal{I}_{JJJ,\;\mathrm{3D}}$ which contains a factor $1/q$:
\begin{equation}
  -\frac{p_{\nu } p_{\rho } q_{\mu }}{8 p q |p+q|} +\frac{p_{\nu } p_{\rho } q_{\mu }}{4 q |p+q| \left(p+q+|p+q|\right)} +\cdots.
\end{equation}
This sector corresponds to $\mathcal{I}_{\gamma,0}$ in the subgraph expansion (\ref{JJJg0}). 

We have verified the subgraph expansions of the correlators $\langle TJJ\rangle$ and $\langle TTT\rangle$
in both free scalar and free fermion theories using the 3D exact results, and found perfect agreements. 
Comparing with the conformal 3-point correlation functions solved from the conformal Ward identities, the subgraph expansions of the correlators provide diagrammatic representations for the complicated results, and are convenient for further studies of these correlators, e.g., the $   1/N$ corrections for the conformal 3-point correlators which we will compute in the following two sections.

\section{Method of subgraphs and the $1/N$ expansions: $\langle JJJ\rangle$} \label{sec4}
In this section, we compute the subleading order corrections to the conformal current 3-point correlators $\langle JJJ\rangle$ in the $1/N$ expansion. The correlation function $\langle JJJ\rangle$ is, in general, given by a linear combination of those from the free scalar and free fermion theories, which have already been computed in Section \ref{free3pt}. It is the coefficients in the linear combinations that receive $1/N$ corrections.

The higher order corrections to the conformal 3-point correlators involve 2-scale multi-loop Feynman diagrams, which are quite challenging to evaluate analytically. We use the method of subgraphs to simplify the loop computations. Here, we study the large $N$ expansion of the correlator $\langle JJJ\rangle$ in the $O(N)$ vector model and GNY model. We aim to illustrate the role of the method of subgraphs in computing loop corrections to higher-point correlators. This method can be directly applied to general CFTs.

\subsection{The large $N$ diagrammatic techniques} \label{largeNdiags}
The large $N$ perturbative expansion has been widely applied to study critical phenomena, see \cite{Moshe:2003xn} for a review. In this work, we follow the theoretical implementation of the large $N$ expansion developed in \cite{Diab:2016spb,Giombi:2016fct}, which has been employed to compute the $1/N$ corrections to the conformal current 2-point correlators $\langle JJ\rangle$ and $\langle TT\rangle$. 

The critical $O(N)$ vector model can be realized as the IR fixed point of the theory
\begin{equation}
  ~~~~~~~~~  \mathcal{L}=\frac{1}{2}\left(\partial\phi_0^i\right)^2+\lambda \left(\phi_0^i\phi_0^i\right)^2, ~~~~~ i=1,2,\dots, N.
\end{equation}
Using the Hubbard-Stratonovich transformation, the Lagrangian can be rewritten as
\begin{equation}
    \mathcal{L}=\frac{1}{2}\left(\partial\phi_0^i\right)^2+\frac{1}{\sqrt{N}}\sigma_0 \phi_0^i\phi_0^i-\frac{1}{4N\lambda}\sigma_0^2. \label{Lag1}
\end{equation}
The propagator of the bare scalar field $\phi_0$ has already been shown in (\ref{prop}).
The auxiliary field $\sigma_0$ receives quantum loop corrections through the interaction $\sigma_0\phi_0^i\phi_0^i$ and becomes dynamical. According to the normalization in (\ref{Lag1}), its effective propagator is \cite{Diab:2016spb}
\begin{equation}
    \langle \sigma_0(p)\sigma_0(-p)\rangle= \frac{C^0_\sigma}{\left(p^2\right)^{\frac{d}{2}-2+\epsilon}}, ~~~C^0_\sigma=2^{D+1} (4\pi)^{\frac{D-3}{2}}\Gamma\left(\frac{D-1}{2}\right)\sin(\frac{\pi D}{2}), \label{Lag2}
\end{equation}
in which the parameter $\epsilon$ denotes the shift of the scaling dimension of $\sigma_0$. It plays the role of a regulator in the loop computations and vanishes in the final results. The quadratic term $\sigma_0^2$ in (\ref{Lag2}) is irrelevant near the IR fixed point; therefore, the effective Lagrangian for the $O(N)$ critical point is
\begin{equation}
    \mathcal{L}_{IR}=\frac{1}{2}\left(\partial\phi_0^i\right)^2+\frac{1}{\sqrt{N}}\sigma_0 \phi_0^i\phi_0^i, \label{Lag3}
\end{equation}
Since we have introduced the regularization parameter $\epsilon$ for $\sigma_0$, its  dimension is modified accordingly, which makes the vertex in $\mathcal{L}_{IR}$ dimensionful. To make the vertex dimensionless, an arbitrary renormalization scale $\mu$ is introduced in the vertex
\begin{equation}
    \mathcal{L}_{\mathrm{vertex}}=\mu^\epsilon \frac{1}{\sqrt{N}}\sigma_0\phi_0^i\phi_0^i.
\end{equation}
The renormalization factor $\mu^\epsilon$ will be implicitly assumed throughout this work.

It is straightforward to apply renormalization within the minimal subtraction scheme. The renormalization factors $Z_{\phi}$ and $Z_{\sigma}$  are introduced for the bare operators $\phi_0$ and $\sigma_0$. Then the physical fields $\phi$ and $\sigma$ and given by
\begin{equation}
    \phi^i=Z_\phi^{\frac{1}{2}}\phi_0^i,~~\sigma=Z_\sigma^{\frac{1}{2}}\sigma_0,
\end{equation}
which absorb the divergences in the loop computations
\begin{equation}
    Z_\phi=1+\frac{1}{N}\frac{r_\phi}{\epsilon}+O(N^{-2}), ~~Z_\sigma=1+\frac{1}{N}\frac{r_\sigma}{\epsilon}+O(N^{-2}). \label{renormalizations}
\end{equation}
After renormalization, the 2-point correlators of the renormalized fields have the standard forms of the 2-point conformal correlation functions in momentum space
\begin{equation}
    \langle \phi(p)^i\phi(-p)^j\rangle=\delta^{ij}\frac{C_\phi}{\left(p^2\right)^{\frac{D}{2}-\Delta_\phi}},~~~ \langle \sigma(p)\sigma(-p)\rangle=  \frac{C_\sigma}{\left(p^2\right)^{\frac{D}{2}-\Delta_\sigma}}. \label{phisigpropagator}
\end{equation}
Both the scaling dimensions $\Delta_{\phi}$ and the coefficient $C_{\phi}$ have been computed to the order $O(1/N)$ in \cite{Diab:2016spb}. 
We will compute the subleading order corrections to the coefficient $C_\sigma$ and the anomalous dimension $\Delta_\sigma$ in Section \ref{section:JJSig}.
We will also be interested in the conserved currents $J$ and $T$, whose definitions have been given in (\ref{Jop}) and (\ref{Top}). It has been found that at the order $O(1/N)$, the renormalization factor of the spin-1 conserved current $J^a_\mu$ is trivial, $Z_J=1+O(N^{-2})$, while the stress tensor operator $T_{\mu\nu}$ needs a special renormalization factor $Z_T$
\begin{equation}
    T_{\mu\nu}^p=Z_T T_{\mu\nu}, ~~~ Z_T=1+\frac{1}{N}\left(\frac{r_T}{\epsilon}+s_T\right)+O(N^{-2}),
\end{equation}
and the coefficients $r_T, s_T$ are given by \cite{Diab:2016spb}
\begin{equation}
    r_T=\frac{2\Gamma(D-2)\sin(\pi D/2)}{\pi\Gamma(\frac{D}{2}-2)\Gamma(\frac{D}{2}+2)},~~~~ s_T=\frac{4\Gamma(D-2)\sin(\pi D/2)}{\pi\Gamma(\frac{D}{2}-1)\Gamma(\frac{D}{2}+2)}. \label{TRenormalization}
\end{equation}

We now combine the large $N$ diagrammatic approach and method of subgraphs to compute the $1/N$ corrections to the conformal current 3-point correlator $\langle JJJ\rangle$.

\subsection{$\langle JJJ\rangle$ in the $O(N)$ vector model}
Our strategy for compute the $1/N$ corrections to the conformal 3-point correlators can be briefly summarized as follows. We start with the effective Lagrangian (\ref{Lag3}), and write all the Feynman diagrams for the 3-point correlator $\langle \cO_1\cO_2\cO_3\rangle$  up to the order $O(1/N)$. The Feynman diagrams are evaluated using the method of subgraphs, and the results contain divergences regularized by the parameter $\epsilon$ in (\ref{Lag2}). Then we introduce the renormalization factors $Z_{\cO_i}$ to cancel the divergences.  After renormalization, the Feynman loop integrals give the ``physical" 3-point correlation function up to the order $O(1/N)$, which should be able to reproduce the correlation functions solved from the conformal Ward identities and also provide the subleading order corrections to the free parameters in the conformal 3-point correlators. 

\subsubsection{Subgraph expansions of the Feynman diagrams}
\begin{figure}
\begin{equation*}
\begin{array}{cccc}
  \begin{tikzpicture}[baseline=(a.base)]
  \begin{feynman}
    \vertex [dot,label=right:$J^c_{\rho}(-p-    q)$] (c) at (2,3/2) {};
    \vertex[dot,label=right:$J^b_{\nu}(p)$] (b) at (2,-3/2) {};
    \vertex[dot,label=left:$J^a_{\mu}(q)$] (a)  at (0,0) {};

    \diagram* {
       (a) -- [fermion] (b) --[fermion] (c) --[fermion] (a)
    };
  \end{feynman}
\end{tikzpicture}   & 
 \begin{tikzpicture}[baseline=(a.base)]
  \begin{feynman}
 \vertex [dot] (c) at (2,3/2) {};
    \vertex[dot] (b) at (2,-3/2) {};
    \vertex[dot] (a)  at (0,0) {};

    \vertex (d) at (6/4, 9/8);
    \vertex (e) at (2/4, 3/8);

    \diagram* {
       (a) -- [fermion] (b) --[fermion] (c) -- (d) --[fermion] (e) -- (a),
       (e) --[fermion, dashed, half right,looseness=1.7] (d);
    };
  \end{feynman}
\end{tikzpicture} & ~~~~~~~~\begin{tikzpicture}[baseline=(a.base)]
  \begin{feynman}
 \vertex [dot] (c) at (2,3/2) {};
    \vertex[dot] (b) at (2,-3/2) {};
    \vertex[dot] (a)  at (0,0) {};

    \vertex (e) at (6/4, -9/8);
    \vertex (d) at (2/4, -3/8);

    \diagram* {
       (a) --  (d) --[fermion] (e) -- (b) --[fermion] (c) --[fermion] (a),
       (e) --[fermion, dashed, half right,looseness=1.7] (d);
    };
  \end{feynman}
\end{tikzpicture} & ~~~~~~~~\begin{tikzpicture}[baseline=(a.base)]
  \begin{feynman}
 \vertex [dot] (c) at (2,3/2) {};
    \vertex[dot] (b) at (2,-3/2) {};
    \vertex[dot] (a)  at (0,0) {};

    \vertex (e) at (4/3, 1);
    \vertex (d) at (4/3, -1);

    \diagram* {
       (a) -- [fermion] (d)  --(b) --[fermion] (c) -- (e) --[fermion] (a),
       (e) --[fermion, dashed] (d);
    };
  \end{feynman}
\end{tikzpicture} \\
  D_0   & D_1 & ~~~~~~~~D_2 & ~~~~~~~~D_3 \\
  & & &  \\
 \begin{tikzpicture}[baseline=(a.base)]
  \begin{feynman}
 \vertex [dot] (c) at (2,3/2) {};
    \vertex[dot] (b) at (2,-3/2) {};
    \vertex[dot] (a)  at (0,0) {};

    \vertex (e) at (1, 3/4);
    \vertex (d) at (2, 0);

    \diagram* {
       (a) --[fermion]  (b) --[fermion] (d) --[fermion] (c) --[fermion] (e) --[fermion] (a),
       (e) --[fermion, dashed] (d);
    };
  \end{feynman}
\end{tikzpicture}& 
 \begin{tikzpicture}[baseline=(a.base)]
  \begin{feynman}
 \vertex [dot] (c) at (2,3/2) {};
    \vertex[dot] (b) at (2,-3/2) {};
    \vertex[dot] (a)  at (0,0) {};

    \vertex (d) at (1, -3/4);
    \vertex (e) at (2, 0);

    \diagram* {
       (a) --[fermion]  (d) --[fermion] (b) --[fermion] (e) --[fermion] (c) --[fermion]  (a),
       (e) --[fermion, dashed] (d);
    };
  \end{feynman}
\end{tikzpicture}
&  ~~~~~~~~\begin{tikzpicture}[baseline=(a.base)]
  \begin{feynman}
 \vertex [dot] (c) at (2,3/2) {};
    \vertex[dot] (b) at (2,-3/2) {};
    \vertex[dot] (a)  at (0,0) {};

    \vertex (e) at (2, 3/4);
    \vertex (d) at (2, -3/4);

    \diagram* {
       (a) -- [fermion] (b)  -- (d) --[fermion] (e) --(c) --[fermion] (a),
       (e) --[fermion, dashed, half right,looseness=1.7] (d);
    };
  \end{feynman}
\end{tikzpicture}  & \\
  D_4 & D_5 & ~~~~~~~~D_6 &
\end{array}
\end{equation*}
    \caption{Feynman diagram for the conformal current 3-point correlator $\langle J^a_\mu J^b_\nu J^c_\rho\rangle$ up to the order $O(1/N)$.}
    \label{fig:JJJ1N}
\end{figure}
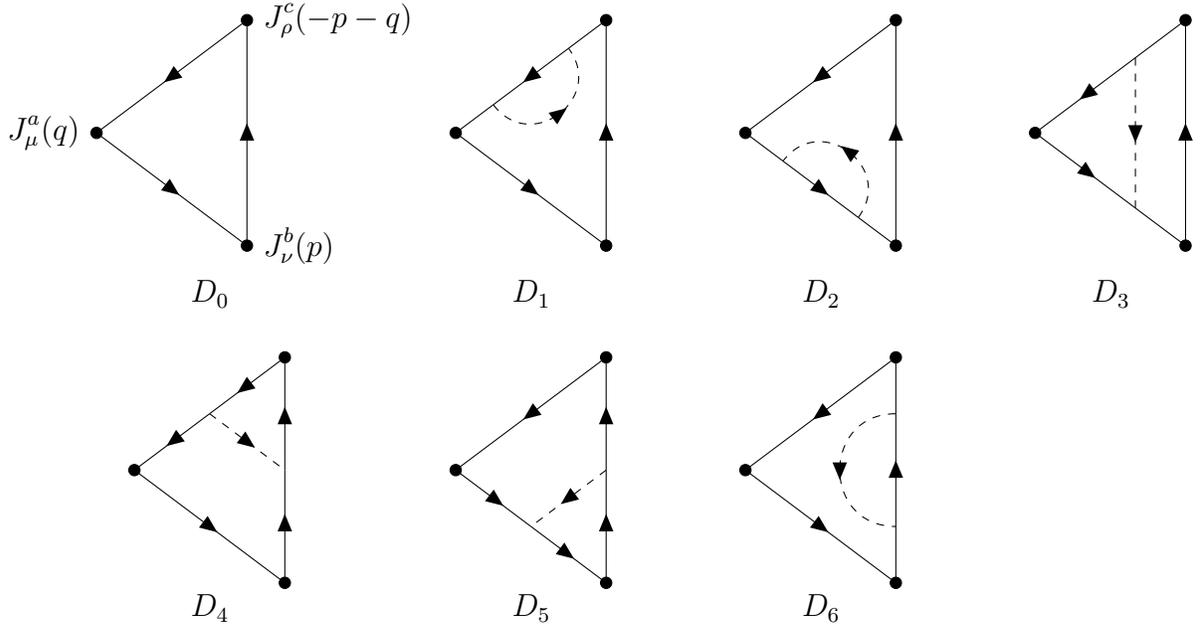

The 3-point correlator $\langle J_\mu^a(q)J_\nu^b(p)J_\rho^c(-p-q)\rangle$ at the order $O(1/N)$ is given by the seven Feynman diagrams in Figure \ref{fig:JJJ1N}
\begin{equation}
    \langle J_\mu^a(q)J_\nu^b(p)J_\rho^c(-p-q)\rangle= D_0+D_1+\cdots+D_6+O(1/N^{2}). \label{3J6D}
\end{equation}
The exact results of the loop integrals in Figure \ref{fig:JJJ1N} are hard to evaluate analytically. Nevertheless, due to the constraints from the conformal symmetry, the conformal 3-point correlator  $\langle J_\mu^a(q)J_\nu^b(p)J_\rho^c(-p-q)\rangle$ is fixed up to two free parameters $C_J$ and $\lambda_{JJJ}$, or equivalently the coefficients $n_s$ and $n_f$ in the linear combination of free scalar and free fermion theories (\ref{ferbos}). The parameter $C_J$ also appears in the 2-point correlator $\langle J_\mu^a(p)J_\nu^b(-p)\rangle$ and its subleading order correction has been computed in \cite{Diab:2016spb}. To evaluate the second parameter $\lambda_{JJJ}$ to the subleading order, we take the zero-momentum limit $q\rightarrow 0$ for the correlator (\ref{3J6D}). The leading term of the correlation function in the zero-momentum limit degenerates to the 2-point correlation function and involves in the parameter $C_J$ only. To fix the second parameter $\lambda_{JJJ}$ it is necessary to expand the correlator $\langle J_\mu^a(q)J_\nu^b(p)J_\rho^c(-p-q)\rangle$ to the subleading order in the small $q$ expansion. In other words, we evaluate the subgraphs of the Feynman diagrams to the subleading order in the small $q$ expansion.

As explained in Section \ref{methodsbg}, the subgraph $\gamma$ of a Feynman diagram is chosen based on the two rules: (a) $\gamma$ contains all the vertices with large external momenta and (b) $\gamma$ is 1PI after identifying all the vertices with large external momenta. We take the diagram $D_4$ as an example to show its expansion by subgraphs. 

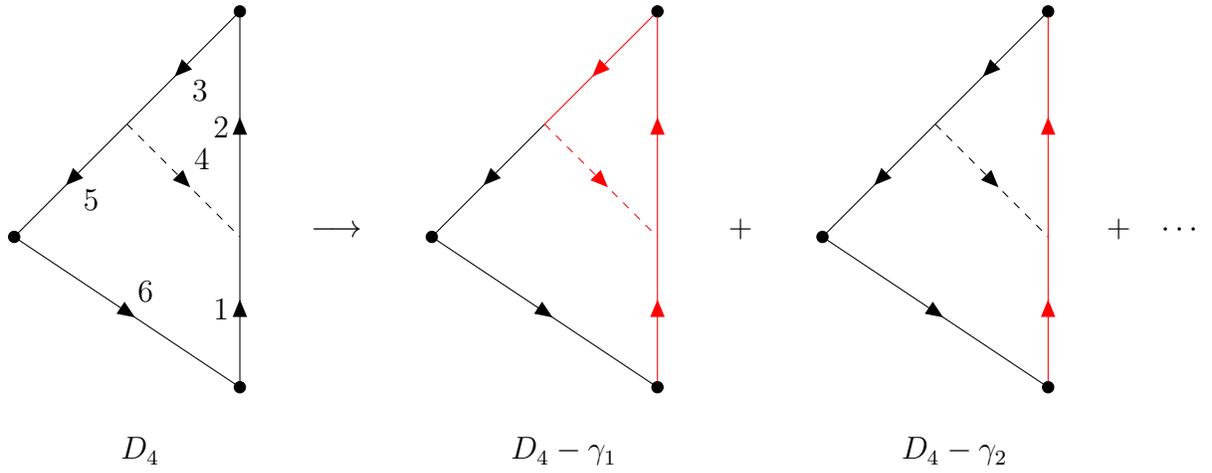
\begin{figure}
    \centering
    \begin{equation*}
    \begin{array}{cccccc}
         \begin{tikzpicture}[baseline=(a.base)]
  \begin{feynman}
 \vertex [dot] (c) at (3,3) {};
    \vertex[dot] (b) at (3,-2) {};
    \vertex[dot] (a)  at (0,0) {};

    \vertex (e) at (3/2, 3/2);
    \vertex (d) at (3, 0);

    \diagram* {
       (a) --[fermion, edge label=\(6\)]  (b) --[fermion, edge label=\(1\)] (d) --[fermion, edge label=\(2\)] (c) --[fermion, edge label=\(3\)] (e) --[fermion, edge label=\(5\)] (a),
       (e) --[fermion, dashed, edge label=\(4\)] (d);
    };
  \end{feynman}
\end{tikzpicture} ~~& ~\longrightarrow ~& ~~ \begin{tikzpicture}[baseline=(a.base)]
  \begin{feynman}
 \vertex [dot] (c) at (3,3) {};
    \vertex[dot] (b) at (3,-2) {};
    \vertex[dot] (a)  at (0,0) {};

    \vertex (e) at (3/2, 3/2);
    \vertex (d) at (3, 0);

    \diagram* {
       (a) --[fermion]  (b) --[fermion,red] (d) --[fermion, red] (c) --[fermion,red] (e) --[fermion] (a),
       (e) --[fermion, dashed,red] (d);
    };
  \end{feynman}
\end{tikzpicture} ~~&~+~& ~~
\begin{tikzpicture}[baseline=(a.base)]
  \begin{feynman}
 \vertex [dot] (c) at (3,3) {};
    \vertex[dot] (b) at (3,-2) {};
    \vertex[dot] (a)  at (0,0) {};

    \vertex (e) at (3/2, 3/2);
    \vertex (d) at (3, 0);

    \diagram* {
       (a) --[fermion]  (b) --[fermion,red] (d) --[fermion,red] (c) --[fermion] (e) --[fermion] (a),
       (e) --[fermion, dashed] (d);
    };
  \end{feynman}
\end{tikzpicture} ~~ &+ ~~\cdots \\
 & & & & & \\
D_4 & & ~~~D_4-\gamma_1 & & ~~~D_4-\gamma_2 & 
\end{array}
    \end{equation*}
    \caption{Subgraphs $\gamma_1$ (1234) and $\gamma_2$ (16) of the Feynman diagram $D_4$ in Figure \ref{fig:JJJ1N}. The diagram $D_4$ itself constructs a trivial subgraph $\Gamma$. In addition, there are extra subgraphs of $D_4$ which relate to scaleless integrals.}
    \label{fig:D3subgraphs}
\end{figure}

The Feynman diagram $D_4$ is given by the 2-scale integral
\begin{equation}
    D_4\propto\int \frac{d^Dk_1}{(2\pi)^D}\frac{d^Dk_2}{(2\pi)^D} \frac{\left(2 k_{1,\nu }+p_{\nu }\right) \left(2 k_{1,\mu }-q_{\mu }\right) \left(2 k_{2,\rho }+p_{\rho }-q_{\rho }\right)}{k_1^2 \left(k_1+p\right)^2 \left(k_2+p\right)^2 \left(k_1-q\right)^2 \left(k_2-q\right)^2 \left(\left(k_1-k_2\right)^2\right)^{\frac{d}{2} -2+\epsilon}}, \label{D3int}
\end{equation}
where we have omitted the overall normalization $\mathcal{N}_\sigma$ (\ref{Lag2}) for $\sigma$ propagator and the algebraic factor $i^3\mathrm{tr}(t^at^bt^c)$ for brevity.
All the subgraphs of $D_4$ contain the vertices $J_\nu^b(p)$ and $J_\rho^c(-p-q)$. There is a trivial subgraph of $D_4$ given the diagram itself $\Gamma=D_4$, and to evaluate the subgraph $\Gamma$, we simply take the Taylor expansion of the whole integrand in (\ref{D3int}) with respect to the small momentum $q$ 
\begin{equation}
    \mathcal{I}_{\Gamma}\propto\int \frac{d^Dk_1}{(2\pi)^D}\frac{d^Dk_2}{(2\pi)^D} \mathcal{T}_{q}\frac{\left(2 k_{1,\nu }+p_{\nu }\right) \left(2 k_{1,\mu }-q_{\mu }\right) \left(2 k_{2,\rho }+p_{\rho }-q_{\rho }\right)}{k_1^2 \left(k_1+p\right)^2 \left(k_2+p\right)^2 \left(k_1-q\right)^2 \left(k_2-q\right)^2 \left(\left(k_1-k_2\right)^2\right)^{\frac{d}{2} -2+\epsilon}}.\label{D3Gamma}
\end{equation}
After the Taylor expansion,  the original 2-scale two-loop integral in (\ref{D3int}) becomes a 1-scale two-loop integral with external momentum $p$. A typical example is given by
\begin{equation}
    K(\Delta,p)\equiv\int \frac{d^Dk_1}{(2\pi)^D}\frac{d^Dk_2}{(2\pi)^D} \frac{1}{k_1^2 \left(k_1+p\right)^2 k_2^2\left(k_2+p\right)^2 \left(\left(k_1-k_2\right)^2\right)^{\Delta}}.\label{Kite}
\end{equation}
This is the so called ``Kite" diagram, which can be analytically evaluated using the Gegenbauer polynomial technique \cite{Chetyrkin:1980pr,Kotikov:1995cw}. The price we pay is that by taking the Taylor expansion $\mathcal{T}_q$, the integrand has higher rank tensor structures. In this work, it suffices to take the Taylor expansion to the second order, which increases the rank of the tensor structures by 1.  The Feynman integrals with tensor structures can be decomposed into scalar integrals using tensor reductions \cite{Davydychev:1991va}. The final integrals in general dimension, even in the zero-momentum limit,  are too cumbersome to show here.  
In 3D, the first few leading terms of the integral is given by
\begin{align}
    \mathcal{I}_\Gamma=&\frac{3 p_{\alpha } p_{\mu } p_{\nu } p_{\rho } q_{\alpha }-p^2 \left(p_{\nu } p_{\rho } q_{\mu }+p_{\mu } \left(p_{\rho } q_{\nu }+p_{\nu } \left(2 p_{\rho }+q_{\rho }\right)\right)\right)}{24 \pi ^2 \left(p^2\right)^{5/2} \epsilon } \nn\\
    &+\frac{p^2 \left(15 p_{\nu } p_{\rho } q_{\mu }+p_{\mu } \left(3 p_{\rho } q_{\nu }+p_{\nu } \left(-6 p_{\rho }-13 q_{\rho }\right)\right)\right)+39 p_{\alpha } p_{\mu } p_{\nu } p_{\rho } q_{\alpha }}{144 \pi ^2 \left(p^2\right)^{5/2}}  \\
    &+\frac{\delta\mathrm{-dependent~terms}}{\epsilon}+\delta\mathrm{-dependent ~terms}. \nn
\end{align}
An important property of the subgraph $\Gamma$ is that its leading order term is independent of the small momentum $q$. 
In contrast, the subgraphs
$\gamma$ will be shown to contain factors of $q^{D-n}$, which makes them radically different from $\mathcal{I}_\Gamma$.

Besides the subgraph $\Gamma$, the Feynman diagram $D_4$ also contains the subgraphs $\gamma_1$ and $\gamma_2$ depicted in red color in Figure \ref{fig:D3subgraphs}. For the subgraph $\gamma_1$, the internal momentum $k_2$ flows in the propagators (234) and is considered large; in contrast, the internal momentum $k_1$ flows in $\Gamma/\gamma_1$ so is considered small. Therefore, the integral for the subgraph $\gamma_1$ is 
\begin{align}
    \mathcal{I}_{\gamma_1}\propto\int \frac{d^Dk_1}{(2\pi)^D}\frac{d^Dk_2}{(2\pi)^D} &\frac{1}{k_1^2 \left(k_1-q\right)^2}\times \nn\\
    & \mathcal{T}_{k_1,q}
    \frac{\left(2 k_{1,\nu }+p_{\nu }\right) \left(2 k_{1,\mu }-q_{\mu }\right) \left(2 k_{2,\rho }+p_{\rho }-q_{\rho }\right)}{ \left(k_1+p\right)^2 \left(k_2+p\right)^2 \left(k_2-q\right)^2 \left(\left(k_1-k_2\right){}^2\right)^{\frac{d}{2} -2+\epsilon}}.\label{D3gamma1}
\end{align}
We take the above Taylor expansion $\mathcal{T}_{k_1,q}$ to second order, after which the new integration the two internal momenta $k_1$ and $k_2$ decouple. The integral $\mathcal{I}_{\gamma_1}$ consists of  two one-loop integrals for $k_{1}$ and $k_2$ which can be evaluated directly. 
\begin{equation}
    \mathcal{I}_{\gamma_1}=\frac{2^{-D} (D-4) \pi ^{-\frac{D}{2}} \left(q^2\right)^{\frac{D}{2}-2} \left(p_{\alpha } p_{\nu } p_{\rho } q_{\alpha } q_{\mu }-p_{\rho } \left(p^2 q_{\mu } q_{\nu }+q^2 p_{\mu } p_{\nu }\right)\right)}{(D-1)  \Gamma \left(\frac{D}{2}+1\right)\left(p^2\right)^2 \epsilon }+\cdots.
\end{equation}
The leading term contributes at the order $O(q^{D-2})$, and the terms with a pole $1/\epsilon$ will be canceled in the final results.

For the subgraph $\gamma_2$, the two internal momenta $k_1$ and $k_2$ are considered small, as both of them flow through the links in $\Gamma/\gamma_2$. The integral of the subgraph $\gamma_2$ is 
\begin{align}
    \mathcal{I}_{\gamma_2}\propto\int \frac{d^Dk_1}{(2\pi)^D}\frac{d^Dk_2}{(2\pi)^D} &
    \frac{\left(2 k_{1,\mu }-q_{\mu }\right) \left(2 k_{1,\nu }+p_{\nu }\right) \left(2 k_{2,\rho }+p_{\rho }-q_{\rho }\right)}{k_1^2 \left(k_1-q\right)^2 \left(k_2-q\right)^2 \left(\left(k_1-k_2\right)^2\right)^{\frac{d}{2} -2+\epsilon}} \nn\\
    & \times \mathcal{T}_{k_1,k_2,q}\frac{1}{\left(k_1+p\right)^2 \left(k_2+p\right)^2 }.\label{D3gamma2}
\end{align}
Here, the Taylor expansion $\mathcal{T}_{k_1,k_2,q}$ is identical to the Taylor expansion with respect to the large external momentum $p$, through which the integral $\mathcal{I}_{\gamma_2}$ becomes a 1-scale (external momentum $q$) two-loop integral, reminiscent to the integral (\ref{D3Gamma}) but with a different scale parameter. At the leading order, the subgraph integral gives
\begin{equation}
    \mathcal{I}_{\gamma_2}=\frac{2^{1-2 D} (D-4) \pi ^{1-D}  \csc \left(\frac{\pi  D}{2}\right) \left(q^2\right)^{\frac{D}{2}-1}p_{\nu } p_{\rho } q_{\mu }}{(D-2) \left(p^2\right)^2 \Gamma (D+1)}+O(q^{D}).
\end{equation}
There is no singular term in $\epsilon$, and the leading term contributes at the order $O(q^{D-1})$, higher than the leading term in $\mathcal{I}_{\gamma_1}$.

In Figure \ref{fig:D3subgraphs}, we have ignored more subgraphs of the diagram $D_4$. Actually, the links (143), (653), (14562), (14563) and (12356) also construct subgraphs of $D_4$. Nevertheless, they all correspond to vanishing integrals. For instance, the subgraph (145632) relates to the integral
\begin{align}
    \mathcal{I}_{145632}=\int \frac{d^Dk_1}{(2\pi)^D}\frac{d^Dk_2}{(2\pi)^D} &\frac{1}{\left(k_2-q\right)^2 } \nn\\
    &\times\mathcal{T}_{k_2,q}
    \frac{\left(2 k_{1,\nu }+p_{\nu }\right) \left(2 k_{1,\mu }-q_{\mu }\right) \left(2 k_{2,\rho }+p_{\rho }-q_{\rho }\right)}{k_1^2 \left(k_1+p\right)^2 \left(k_1-q\right)^2 \left(k_2+p\right)^2  \left(\left(k_1-k_2\right){}^2\right)^{\frac{d}{2} -2+\epsilon}},\label{D3exasb}
\end{align}
in which the integration over $k_2$ becomes scaleless after taking the Taylor expansion
\begin{equation}
    \mathcal{I}_{145632}=\int \frac{d^Dk_2}{(2\pi)^D} \frac{1}{\left(k_2-q\right)^2 }\times \left(\mathrm{polynomials ~of ~ } k_2\right)=0.
\end{equation}
This explains why the subgraphs omitted in Figure \ref{fig:D3subgraphs} have zero contributions to the original loop integral.

\begin{figure}
\begin{equation*}
\begin{array}{cccc}
  \begin{tikzpicture}[baseline=(a.base)]
  \begin{feynman}
    \vertex [dot,label=right:$J^c_{\rho}(-p-    q)$] (c) at (2,3/2) {};
    \vertex[dot,label=right:$J^b_{\nu}(p)$] (b) at (2,-3/2) {};
    \vertex[dot,label=left:$J^a_{\mu}(q)$] (a)  at (0,0) {};

    \diagram* {
       (a) -- [fermion] (b) --[fermion,red] (c) --[fermion] (a)
    };
  \end{feynman}
\end{tikzpicture}   & 
 \begin{tikzpicture}[baseline=(a.base)]
  \begin{feynman}
 \vertex [dot] (c) at (2,3/2) {};
    \vertex[dot] (b) at (2,-3/2) {};
    \vertex[dot] (a)  at (0,0) {};

    \vertex (d) at (6/4, 9/8);
    \vertex (e) at (2/4, 3/8);

    \diagram* {
       (a) -- [fermion] (b) --[fermion,red] (c) -- (d) --[fermion] (e) -- (a),
       (e) --[fermion, dashed, half right,looseness=1.7] (d);
    };
  \end{feynman}
\end{tikzpicture} & ~~~~~~~~\begin{tikzpicture}[baseline=(a.base)]
  \begin{feynman}
 \vertex [dot] (c) at (2,3/2) {};
    \vertex[dot] (b) at (2,-3/2) {};
    \vertex[dot] (a)  at (0,0) {};

    \vertex (e) at (6/4, -9/8);
    \vertex (d) at (2/4, -3/8);

    \diagram* {
       (a) --  (d) --[fermion] (e) -- (b) --[fermion,red] (c) --[fermion] (a),
       (e) --[fermion, dashed, half right,looseness=1.7] (d);
    };
  \end{feynman}
\end{tikzpicture} & ~~~~~~~~\begin{tikzpicture}[baseline=(a.base)]
  \begin{feynman}
 \vertex [dot] (c) at (2,3/2) {};
    \vertex[dot] (b) at (2,-3/2) {};
    \vertex[dot] (a)  at (0,0) {};

    \vertex (e) at (4/3, 1);
    \vertex (d) at (4/3, -1);

    \diagram* {
       (a) -- [fermion] (d)  -- (b) --[fermion,red] (c) -- (e) --[fermion] (a),
       (e) --[fermion, dashed] (d);
    };
  \end{feynman}
\end{tikzpicture} \\
  D_0-\gamma_1   & D_1-\gamma_1 & ~~~~~~~~D_2-\gamma_1 & ~~~~~~~~D_3-\gamma_1 \\
  & & &  \\ \begin{tikzpicture}[baseline=(a.base)]
  \begin{feynman}
 \vertex [dot] (c) at (2,3/2) {};
    \vertex[dot] (b) at (2,-3/2) {};
    \vertex[dot] (a)  at (0,0) {};

    \vertex (e) at (4/3, 1);
    \vertex (d) at (4/3, -1);

    \diagram* {
       (a) -- [fermion] (d)  --[red] (b) --[fermion,red] (c) --[red] (e) --[fermion] (a),
       (e) --[fermion, dashed,red] (d);
    };
  \end{feynman}
\end{tikzpicture}   &
 \begin{tikzpicture}[baseline=(a.base)]
  \begin{feynman}
 \vertex [dot] (c) at (2,3/2) {};
    \vertex[dot] (b) at (2,-3/2) {};
    \vertex[dot] (a)  at (0,0) {};

    \vertex (e) at (1, 3/4);
    \vertex (d) at (2, 0);

    \diagram* {
       (a) --[fermion]  (b) --[fermion,red] (d) --[fermion,red] (c) --[fermion,red] (e) --[fermion] (a),
       (e) --[fermion, dashed,red] (d);
    };
  \end{feynman}
\end{tikzpicture}& 
 ~~~~~~~~\begin{tikzpicture}[baseline=(a.base)]
  \begin{feynman}
 \vertex [dot] (c) at (2,3/2) {};
    \vertex[dot] (b) at (2,-3/2) {};
    \vertex[dot] (a)  at (0,0) {};

    \vertex (d) at (1, -3/4);
    \vertex (e) at (2, 0);

    \diagram* {
       (a) --[fermion]  (d) --[fermion,red] (b) --[fermion,red] (e) --[fermion,red] (c) --[fermion]  (a),
       (e) --[fermion, dashed,red] (d);
    };
  \end{feynman}
\end{tikzpicture}
&  ~~~~~~~~\begin{tikzpicture}[baseline=(a.base)]
  \begin{feynman}
 \vertex [dot] (c) at (2,3/2) {};
    \vertex[dot] (b) at (2,-3/2) {};
    \vertex[dot] (a)  at (0,0) {};

    \vertex (e) at (2, 3/4);
    \vertex (d) at (2, -3/4);

    \diagram* {
       (a) -- [fermion] (b)  --[red] (d) --[fermion,red] (e) --[red] (c) --[fermion] (a),
       (e) --[fermion, dashed, half right,looseness=1.7,red] (d);
    };
  \end{feynman}
\end{tikzpicture}  \\
 D_3-\gamma_2 & D_4-\gamma_1 & ~~~~~~~~D_5-\gamma_1 & ~~~~~~~~D_6-\gamma_1
\end{array}
\end{equation*}
    \caption{The subgraphs $\gamma$ marked in red color give the dominating subgraphs (besides the original diagram $\Gamma$)  of the Feynman diagrams $D_i$ up to the order $O(1/N)$.}
    \label{fig:JJJ1Nsubs}
\end{figure}
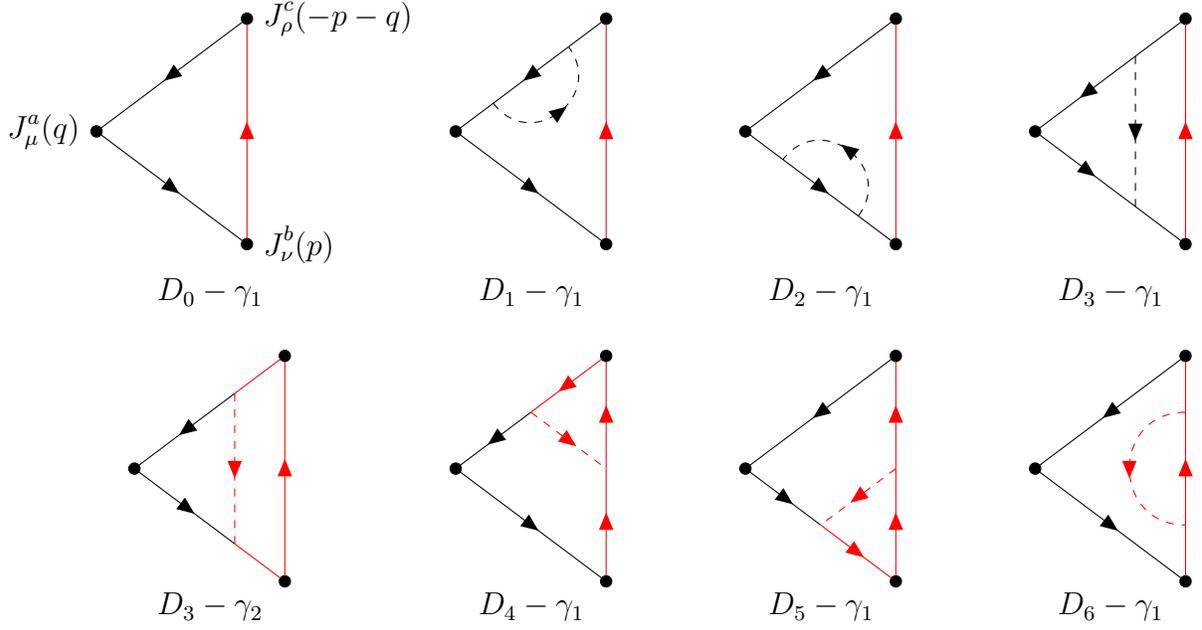

To summarize, besides the subgraph $\Gamma$ constructed by the diagram $D_4$ itself, there are several subgraphs decomposed from $D_4$, and the leading contribution comes from the subgraph $\gamma_1$ in Figure \ref{fig:D3subgraphs}. The rest of the subgraphs either contributes at higher orders or correspond to scaleless integrals.
We have done similar analysis for other Feynman diagrams in Figure \ref{fig:JJJ1N}, and the dominating subgraphs $\gamma$ are shown in Figure \ref{fig:JJJ1Nsubs}. 
Note that the Feynman diagram $D_3$ has two subgraphs $\gamma_1$ and $\gamma_2$, both of which contribute at the order $O(q^{D-2})$:
\begin{align}
    \mathcal{I}_{D_3,\gamma_1}&=\frac{2^{1-D} (4-D) \pi ^{-\frac{D}{2}} \left(q^2\right)^{\frac{D}{2}-2}}{(D-1)   \Gamma \left(\frac{D}{2}+1\right)p^4\epsilon }\left(p_{\rho } \left(p^2 q_{\mu } q_{\nu }+q^2 p_{\mu } p_{\nu }\right)+p^2 p_{\nu } q_{\mu } q_{\rho }-p_{\alpha } p_{\nu } p_{\rho } q_{\alpha } q_{\mu }\right)+\cdots,  \nn\\
    \mathcal{I}_{D_3,\gamma_2}&=-\frac{2^{-D} (4-D) \pi ^{-\frac{D}{2}} \left(q^2\right)^{\frac{D}{2}-2}}{(D-1)  \Gamma \left(\frac{D}{2}+1\right)p^4 \epsilon } \left(p_{\rho } \left(p^2 q_{\mu } q_{\nu }+q^2 p_{\mu } p_{\nu }\right)+p^2 p_{\nu } q_{\mu } q_{\rho }-p_{\alpha } p_{\nu } p_{\rho } q_{\alpha } q_{\mu }\right)+\cdots. \nn 
\end{align}
The two subgraphs have the same tensor structure in the $\epsilon$-divergent terms, but their $\epsilon$-regular terms are different. 

Using the subgraph expansion, the 3-point correlator (\ref{3J6D}) can be expanded as
\begin{equation}
    \langle J_\mu^a(q)J_\nu^b(p)J_\rho^c(-p-q)\rangle = \mathcal{S}_\Gamma+ \mathcal{S}_\gamma, 
\end{equation}
with
\begin{equation}
\mathcal{S}_\Gamma=\sum\limits_{i=0}^6\mathcal{I}_{D_i,\Gamma}, ~~~ \mathcal{S}_\gamma=\sum\limits_{i=0}^6\mathcal{I}_{D_i,\gamma_1} + \mathcal{I}_{D_3,\gamma_2}, \label{subgraphsum}
\end{equation}
where $\mathcal{I}_{D_i,X}$ denotes the integral of the subgraph $X$ of the Feynman diagram $D_i$.
In the small momentum limit $q\rightarrow 0$, the two sectors $\mathcal{S}_\Gamma$ and $\mathcal{S}_\gamma$ scale as:  
\begin{equation}
    \mathcal{S}_\Gamma\sim q^0, ~~~~ \mathcal{S}_\gamma\sim q^{D-2}. 
\end{equation}
As a result, in general dimensions the two sectors do not overlap, and the subleading order corrections to the free parameters $C_J,~\lambda_{JJJ}$ can be extracted from either $\mathcal{S}_\Gamma$ or $\mathcal{S}_\gamma$. We have verified the results from the two different sectors are identical, as they should be.

\subsubsection{Renormalization of the 3-point correlator}
In general, the 3-point correlators obtained from the Feynman diagrams are divergent, and one needs to introduce a renormalization factor $Z_J$ to absorb the infinities
\begin{equation}
    \langle J^{\mathrm{phy},a}_\mu J^{\mathrm{phy},b}_\nu J^{\mathrm{phy},c}_\rho\rangle=Z_J^3  ~\langle J_\mu^a J_\nu^b J_\rho^c\rangle.
\end{equation}
Here, for the spin-1 conserved current $J^a_\mu$, its renormalization factor $Z_J$ has been evaluated in \cite{Diab:2016spb}, and the result is $Z_J=1+O(1/N^2)$. Therefore, at order $O(1/N)$, no renormalization for the correlator $\langle JJJ\rangle$ is needed. The net results $ \mathcal{S}_\Gamma$ and $\mathcal{S}_\gamma$ from the Feynman loop integrals should be finite. This has been confirmed in our loop computations. We have evaluated the Feynman integrals for the subgraphs $\gamma$ in Figure \ref{fig:D3subgraphs} in general dimensions. After combining all the contributions of  $\mathcal{S}_\gamma$ in (\ref{subgraphsum}), the poles in $\epsilon$ and the logarithmic terms $\log{p}, ~\log{q}$ arising from the $\epsilon$-regularization of the $\sigma$ propagator have been cancelled. At the leading order $O(N^0)$,  the result is given by the free boson theory (\ref{JJJg0}). At the next-to-leading order, $O(1/N)$ the result $\mathcal{S}_{\gamma,1/N}$ is
\begin{align}
    \mathcal{S}_{\gamma,1/N}=&\frac{2^{5-D} (3 D-4) \pi ^{-\frac{D}{2}} \left(q^2\right)^{\frac{D}{2}-2} p_{\nu } p_{\rho } \left(q^2 p_{\mu }-p_{\alpha } q_{\alpha } q_{\mu }\right)}{(D-2)^2 (D-1) D^2  \Gamma \left(\frac{D}{2}-2\right)p^4 N} \nn\\
    & +\frac{2^{3-D} (D-4) \pi ^{-\frac{D}{2}} \left(q^2\right)^{\frac{D}{2}-2}}{(D-2) D  \Gamma \left(\frac{D}{2}+1\right) p^2N}\left(p_{\rho } q_{\mu } q_{\nu }+p_{\nu } q_{\mu } q_{\rho }\right) +\delta \mathrm{-dependent~ terms}. \label{Sgamma1N}
\end{align}
The integrals of the subgraphs $\mathcal{I}_{D_i,\Gamma}$ are rather cumbersome in general dimensions; therefore, we evaluate the integrals $\mathcal{I}_{D_i,\Gamma}$ within fixed dimension $D=3$.  At order $O(N^0)$, it is given by the free boson theory (\ref{JJJG0}) and (\ref{JJJG1}). At the subleading order $O(1/N)$, the result is
\begin{align}
    \mathcal{S}_{\Gamma,1/N}|_{D=3}=&\frac{p^2 \left(2 p_{\nu } p_{\rho } q_{\mu }+p_{\mu } \left(3 p_{\rho } q_{\nu }+p_{\nu } \left(4 p_{\rho }+q_{\rho }\right)\right)\right)-6 p_{\alpha } p_{\mu } p_{\nu } p_{\rho } q_{\alpha }}{9 \pi ^2 \left(p^2\right)^{5/2}N} \nn\\&+\delta\mathrm{-dependent~ terms}. \label{SGamma1N}
\end{align}
Like the subgraphs $\gamma$, the poles and logarithmic terms are canceled after collecting all the contributions from $D_i$, and the final result $\mathcal{S}_\Gamma|_{D=3}$ is finite without renormalization. 

\subsubsection{Compare with the correlation functions from conformal symmetry}
Now we are ready to extract the $1/N$ corrections to the free parameters $C_J$ and $\lambda_{JJJ}$ -- or equivalently, the $n_s$ and $n_f$ from the Feynman integrals $\mathcal{S}_\Gamma$ and $\mathcal{S}_\gamma$. Since $\mathcal{S}_\Gamma$ and $\mathcal{S}_\gamma$ represent two distinct sectors of the correlation functions, both of them contain the same information on $n_s$ and $n_f$ at the order $O(1/N)$. 

We start with the sector $\mathcal{S}_\gamma$, which has been evaluated in general dimensions. 
As shown in the seminal work \cite{Osborn:1993cr}, the conformal 3-point correlator $\langle JJJ\rangle$ is strongly restricted by the conformal symmetry. The general conformal 3-point correlator $\langle JJJ\rangle$ consists of functions generated by the free scalar and free fermion theories. The two free parameters in the correlation function can be defined through
\begin{equation}
    \langle JJJ\rangle_{\mathrm{crit.~} O(N)}=n_s\langle JJJ\rangle_{\mathrm{free ~scalar}}+n_f\langle JJJ\rangle_{\mathrm{free ~fermion}}.
\end{equation}
At order $O(N^0)$, we have $n_s=1$ and $n_f=0$. At order $O(1/N)$, the correlation function $\mathcal{S}_{\gamma,1/N}$ is a combination of $\langle JJJ\rangle_{\mathrm{free ~scalar}} $ and $ \langle JJJ\rangle_{\mathrm{free ~fermion}}$. The parameters $n_s$ and $n_f$ can be simply fixed by comparing $\mathcal{S}_{\gamma,1/N}$ in (\ref{Sgamma1N}) with 
$\langle JJJ\rangle_{\mathrm{free ~scalar}} $ in (\ref{JJJg0}) and $ \langle JJJ\rangle_{\mathrm{free ~fermion}}$ in (\ref{JJJfg1}), which are given by
\begin{align}
    n_s &=1-\frac{2^{D+2} (3 D-4) \sin \left(\frac{\pi  D}{2}\right) \Gamma \left(\frac{D-1}{2}\right)}{\pi ^{3/2} (D-2)^2 D^2 \Gamma \left(\frac{D}{2}-2\right)N}, \label{ns}\\
    n_f&=-\frac{2^{D+3} \sin \left(\frac{\pi  D}{2}\right) \Gamma \left(\frac{D-1}{2}\right)}{\pi ^{3/2} (D-2)^2 D^2 \Gamma \left(\frac{D}{2}-2\right)N}. \label{nf}
\end{align}
The above results can also be obtained from the subgraph sector $\mathcal{S}_\Gamma$ (\ref{SGamma1N}). We compare the result $\mathcal{S}_{\Gamma,1/N}|_{D=3}$ with correlation functions from the free scalar  and free fermion theories in Section \ref{JT3pt}. The 3D results are \begin{equation}
    n_s|_{D=3}=-\frac{80}{9\pi^2N},~~~~n_f|_{D=3}=-\frac{32}{9\pi^2N},
\end{equation}
which agree with the solutions of the subgraph sector $\mathcal{S}_\gamma$ in (\ref{ns}) and (\ref{nf}). This provides a consistency check for the method of subgraphs.

\subsection{$\langle JJJ\rangle$ in the GNY model}
We use the same method to compute the $1/N$ corrections to the free parameters in $\langle JJJ\rangle$ in the GNY model.\footnote{See the recent works \cite{Erramilli:2022kgp,Mitchell:2024hix} for bootstrap studies of the GNY model.} 
The effective Lagrangian for the IR fixed point reads
\begin{equation}
    \mathcal{L}_{IR,f}=-\bar{\psi}_{0i}\slashed{\partial}\psi_0^i +\frac{1}{\sqrt{N}}\sigma_0 \bar{\psi}_{0i}\psi_0^i. \label{LagGNY}
\end{equation}
The propagator of the bare fermion operator is provided in (\ref{prop}). The critical scalar $\sigma_0$ has an effective propagator: 
\begin{equation}
    \langle \sigma_0(p)\sigma_0(-p)\rangle= \frac{\mathcal{M}_\sigma}{\left(p^2\right)^{\frac{d}{2}-1+\epsilon}}, ~~~\mathcal{M}_\sigma=-2^{D+1} (4\pi)^{\frac{D-3}{2}}\Gamma\left(\frac{D-1}{2}\right)\sin(\frac{\pi D}{2}), \label{psiprop}
\end{equation}
where the parameter $\epsilon$ is introduced to regularize the divergences in the loop integrals, like we have done in the critical $O(N)$ vector model. The vertices of the spin-1 conserved current $J^a_\mu$ and stress tensor $T_{\mu\nu}$ have been presented in (\ref{Jopf}) and (\ref{Topf}).
The normalization of the bare operators $\psi_0^i,~\sigma_0$, the spin-1 conserved current $J^a_\mu$, and the stress tensor $T_{\mu\nu}$ in GNY model have been studied in \cite{Diab:2016spb}. An interesting property is that the renormalization factor for the operator $J_\mu^a$ is $Z_J=1+O(1/N^2)$; therefore, the renormalization for the 3-point correlator $\langle JJJ\rangle$ is trivial at the order $O(1/N)$.

We compute the 3-point correlator $\langle J_\mu^a(q)J_\nu^b(p)J_\rho^c(-p-q)\rangle$ in GNY model using the Feynman diagrams to the order $O(1/N)$. Actually, the Feynman diagrams for this correlator have the same geometry as those in Figure \ref{fig:JJJ1N} for the critical $O(N)$ vector model. The reason is that 
the propagators and vertices defined by the Lagrangian (\ref{LagGNY}) have the same structures as the critical $O(N)$ vector model. Differences appear in the vertices of $J^a$; however, they only affect the elements in the numerators of the integrands.

We computed the Feynman diagrams using subgraph expansions, and the leading subgraphs $\gamma$ for the Feynman diagrams are also given by those in Figure \ref{fig:JJJ1Nsubs}, but with propagators and vertices replaced by those of the GNY model. We focus on the subgraph sector $\mathcal{S}_\gamma$ for simplicity. 
The leading term at the order $O(N^0)$ is given by the free fermion theory (\ref{JJJfg1}). The subleading terms at the order $O(1/N)$ are given by the subgraphs in Figure \ref{fig:JJJ1Nsubs}.
After collecting all the contributions from each subgraph, the net result is
\begin{align}
\mathcal{S}_\gamma=&\frac{2^{4-D} \pi ^{-\frac{D}{2}} \left(q^2\right)^{\frac{D}{2}-2} \left(q_{\mu } \left((D-1) p^2 \left(p_{\rho } q_{\nu }+p_{\nu } q_{\rho }\right)+2 p_{\alpha } p_{\nu } p_{\rho } q_{\alpha }\right)-2 q^2 p_{\mu } p_{\nu } p_{\rho }\right)}{(D-1) D^2 \Gamma \left(\frac{D}{2}-1\right) p^4  N}  \nn \\
&+ \delta\mathrm{-dependent~terms}.  
\end{align}
We decompose the correlation function into free scalar and free fermion parts:
\begin{equation}
    \langle JJJ\rangle_{\mathrm{GNY}}=n_s\langle JJJ\rangle_{\mathrm{free ~scalar}}+n_f\langle JJJ\rangle_{\mathrm{free ~fermion}}.
\end{equation}
By matching the subgraph integrals on both sides, we obtain the free parameters
\begin{align}
    n_s &=\frac{2^{D+2} \sin \left(\frac{\pi  D}{2}\right) \Gamma \left(\frac{D-1}{2}\right)}{\pi ^{3/2} D^2 \Gamma \left(\frac{D}{2}-1\right)N}, \label{nsf}\\
    n_f&=1+\frac{2 (D+1) \sin \left(\frac{\pi  D}{2}\right) \Gamma (D-1)}{\pi  \Gamma \left(\frac{D}{2}+1\right)^2 N}. \label{nff}
\end{align}
The two parameters can also be evaluated from the subgraph sector $\mathcal{S}_\Gamma$. For simplicity, we compute the integrations of the subgraph $\Gamma$ in 3D. The result is 
\begin{align}
    \mathcal{S}_{\Gamma,1/N}|_{D=3}=&\frac{3 p_{\alpha } p_{\mu } p_{\nu } p_{\rho } q_{\alpha }-p^2 \left(p_{\nu } p_{\rho } q_{\mu }+p_{\mu } \left(p_{\nu } \left(2 p_{\rho }+5 q_{\rho }\right)-3 p_{\rho } q_{\nu }\right)\right)}{9 \pi ^2p^{5}N} \nn\\&+\delta\mathrm{-dependent~ terms}. 
\end{align}
Comparing with the correlation functions from the free scalar  and free fermion theories, we obtain
\begin{equation}
    n_s|_{D=3}=-\frac{32}{9 \pi ^2 N},~~~~n_f|_{D=3}=-\frac{128}{9 \pi ^2 N},
\end{equation}
which are well consistent with the results (\ref{nsf}) and (\ref{nff}) obtained from $\mathcal{S}_\gamma$.

\section{$1/N$ expansions of $\langle JJ\cO\rangle$ and conductivity at finite temperature} \label{sec5}
Near the IR fixed point, some fundamental physical parameters can be computed using conformal perturbation theory. In particular, the conductivity near the quantum critical point can be estimated using the CFT data related to the spin-1 conserved current operators. In this section, we briefly review the conformal perturbation studies of the conductivity near the IR fixed point \cite{Katz:2014rla,Witczak-Krempa:2015pia,Lucas:2016fju,Lucas:2017dqa}, which motivates our perturbative study of the 3-point correlator $\langle JJ\cO \rangle$. Then we apply the method of subgraphs to compute the $1/N$ corrections to the correlators $\langle JJ\cO \rangle$ in the critical $O(N)$ vector model.

\subsection{Conformal correlator $\langle JJ\cO\rangle$ and conductivity at finite temperature}
Consider a general CFT in $2+1$ dimensions with coordinates $x_\mu=(t,\mathbf{x})$. The imaginary frequency conductivity, $\omega(ik)$, is defined by the 2-point correlator of the spin-1 conserved current $J_\mu$ in the space-like direction, e.g.,  $\mu=2$:
\begin{equation}
    \frac{\omega(ik)}{\omega_Q}=-\frac{1}{k}\langle J_{\mu=2}(\mathbf{k}) J_{\nu=2}(-{\mathbf{k}})\rangle_T+ \mathrm{contact~ terms}, \label{conductivity}
\end{equation}
where $\omega_Q=e^2/\hbar$ is the quantum unit of
conductance, and the external momentum $\mathbf{k}$ is taken to be in the time direction, $\mathbf{k}=(k,0,0)$, where $k$ is the analytic continuation of the  Matsubara frequencies $k_n=2\pi n T$. The 2-point correlator is evaluated at a finite temperature $T$. In the large momentum limit $k\gg T$, the thermal 2-point correlator $\langle J_{2}(\mathbf{k}) J_{2}(-{\mathbf{k}})\rangle_T$ can be evaluated using the OPE 
\begin{equation}
    \lim\limits_{k\gg |\mathbf{q}|}J_{\mu=2}(\mathbf{k})J_{\nu=2}(\mathbf{-k+q})=-c_1 |k| \delta^3(\mathbf{q})-\frac{c_2}{|k|^{\Delta-1}}\cO(\mathbf{q})+ \cdots, \label{JJOPE}
\end{equation}
where $c_1$ and $c_2$ are the OPE coefficients in momentum space, and we have ignored the contributions from operators with higher scaling dimensions—in particular, the stress tensor $T_{\mu\nu}$. Here we have implicitly assumed that the lowest scalar operator $\cO$ is relevant; therefore, in the OPE limit, it provides the subleading order contribution to the thermal 2-point function. 
Applying the above OPE in the thermal 2-point correlator in (\ref{conductivity}), it gives
\begin{equation}
    \frac{\omega(ik)}{\omega_Q}=\frac{C_J}{32}+\lambda_{JJ\cO}\frac{C_J }{4\pi}\frac{\Gamma(\Delta_\cO+1)\sin\left(\frac{\pi}{2}\Delta_\cO\right)}{2-\Delta_\cO} f(T,k)+\cdots
\end{equation}
in which the factor $f(T,k)$ is related to the thermal one-point function of the scalar $\langle \cO\rangle_T$; see \cite{Katz:2014rla} for more details. The parameters $C_J$ and $\lambda_{JJ\cO}$ are the usual OPE coefficients defined in position space; they relate to the parameters $c_1$ and $c_2$ in (\ref{JJOPE}) up to the constants from Fourier transformations. 

The conductivity at finite temperature (\ref{conductivity}) can be generalized to a CFT deformed by a relevant operator $\cO$
\begin{equation}
    S_h=S_{CFT}-h\int d^Dx\cO.
\end{equation}
The conductivity in the theory $S_h$ with $h\ll k$ can also be estimated using the conformal perturbation theory \cite{Lucas:2017dqa}. Again, the leading two terms are related to the correlator $\langle J_2(k)J_2(-k)\rangle$  and  the 3-point correlator $\langle \cO(q)J_2(k)J_2(-k-q)\rangle$.

We compute the conformal 3-point correlator $\langle \cO(q)J_\mu(p)J_\nu(-p-q)\rangle$ in momentum space. We follow the normalization of the 3-point correlator in the bootstrap work \cite{Reehorst:2019pzi}, which has obtained a nonperturbative estimation of the OPE coefficient $\lambda_{JJO}$ for the critical $O(2)$ vector model. Assume that the scalar $\cO$ with scaling dimension $\Delta$ is normalized as
\begin{equation}
    \langle \cO(0)\cO(x)\rangle=\frac{C_\cO}{\left(x^2\right)^{\Delta}}.
\end{equation}
The correlator $\langle \cO J_\mu J_\nu\rangle$ in position space  is given by \cite{Reehorst:2019pzi,Dymarsky:2017xzb, Fortin:2020des}
\begin{equation}
    \langle \cO(x_1) J_\mu(x_2)J_\nu(x_3) \rangle=\frac{C_J  C_\cO^{1/2}}{(4\pi)^2}\lambda_{JJ\cO}\frac{(\Delta-D+1)H_{\mu\nu}+\Delta V_{1,\mu}V_{2,\nu}}{\Delta\, x_{12}^{\Delta}x_{13}^{\Delta}x_{23}^{2D-2-\Delta}}, \label{JJO}
\end{equation}
where we have ignored the global symmetry indices. The correlation function contains only one independent parameter, $\lambda_{JJ\cO}$, while the tensor structures in (\ref{JJO}) are defined as follows
\begin{align}
    H_{\mu\nu}&=\delta (\mu ,\nu )-2\frac{x_{23,\mu } x_{23,\nu }}{x_{23}^2}, \\
    V_{1,\mu}&= \frac{x_{23} }{x_{12} x_{13}}x_{12,\mu } +\frac{x_{12} }{x_{23} x_{13}} x_{23,\mu }, \\
     V_{2,\nu}&=\frac{x_{23} }{x_{12} x_{13}}x_{13,\nu }-\frac{x_{13} }{x_{23} x_{12}} x_{23,\nu }, 
\end{align}
where $x_{ij,\mu}=x_{i,\mu}-x_{j,\mu}$ and $x_{ij}=\sqrt{x_{ij}^2}$.

\begin{figure}
   \begin{equation*}
   \begin{array}{ccccc}
    \begin{tikzpicture}[baseline=(a.base)]
  \begin{feynman}
    \vertex [dot,label=right:$J_{\nu}(-p-    q)$] (c) at (2,3/2) {};
    \vertex[dot,label=right:$J_\mu(p)$] (b) at (2,-3/2) {};
    \vertex[dot,label=left:$\cO(q)$] (a)  at (0,0) {};

    \diagram* {
       (a) -- [fermion, edge label=\(k\)] (b) --[fermion] (c) --[fermion] (a)
    };
  \end{feynman}
\end{tikzpicture} &=&
\begin{tikzpicture}[baseline=(a.base)]
  \begin{feynman}
    \vertex [dot,label=right:$J_\nu(-p-    q)$] (c) at (2,3/2) {};
    \vertex[dot,label=right:$J_\mu(p)$] (b) at (2,-3/2) {};
    \vertex[dot,label=left:$\cO(q)$] (a)  at (0,0) {};

    \diagram* {
       (a) -- [fermion,red, edge label=\({\color{black}k}\)] (b) --[fermion,red] (c) --[fermion,red] (a)
    };
  \end{feynman}
\end{tikzpicture}
&+&
\begin{tikzpicture}[baseline=(a.base)]
  \begin{feynman}
    \vertex [dot] (c) at (2,3/2) {};
    \vertex[dot] (b) at (2,-3/2) {};
    \vertex[dot] (a)  at (0,0) {};

    \diagram* {
       (a) -- [fermion, edge label=\(k\)] (b) --[fermion,red] (c) --[fermion] (a)
    };
  \end{feynman}
\end{tikzpicture} \\
& & & & 
\\ \mathrm{Diagram~ }\displaystyle{\Gamma} & & \mathrm{Subgraph ~ } {\textstyle\color{red}\Gamma} & & \mathrm{Subgraph ~ } {\color{red}\gamma}
\end{array}
\end{equation*}
    \caption{Diagrammatic representation of $\mathcal{I}(\nu_1,\nu_2,\nu_3)$ and the two subgraphs in the expansion: $\Gamma$ and $\gamma$ marked in red color.}
    \label{fig:JJO}
\end{figure}
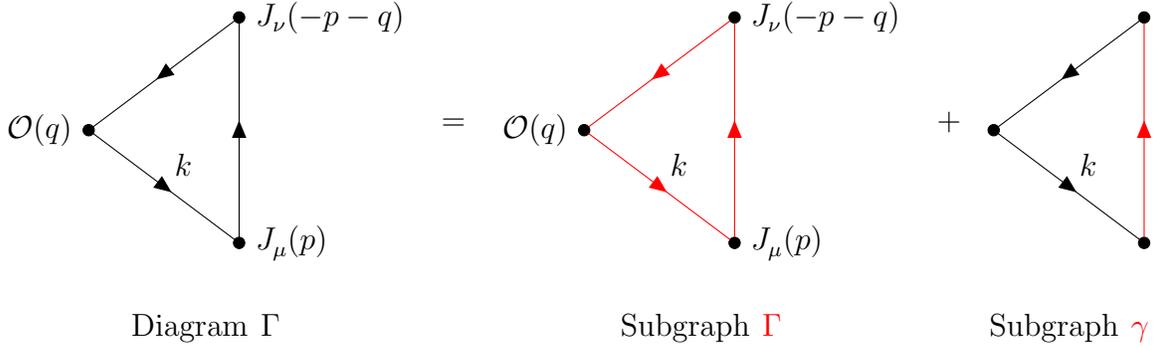

The correlator $\langle \cO(q)J_\mu(p)J_\nu(-p-q)\rangle$ in momentum space can be obtained by Fourier transformation of the above formula. There could be extra contributions from contact terms in position space, whose Fourier transformations only give terms analytical in the external momenta. Like the scalar 3-point correlator, the Fourier transformation of (\ref{JJO}) leads to one-loop integrals (\ref{3DI}) associated with tensor structures for the spinning operators. We take subgraph expansions in the small $q$ limit for these integrals. The results consist of two terms which are regular/singular in $q$
\begin{align}
    &\langle \cO(q) J_\mu(p)J_\nu(-p-q)\rangle = \frac{C_J  C_\cO^{1/2}}{(4\pi)^2}\lambda_{JJ\cO}\times \label{JJOinP}\\
    &\left(\frac{2^{3-D} \pi ^{D+\frac{1}{2}}   \Gamma \left(2-\frac{\Delta }{2}\right) \left(p^2\right)^{\frac{\Delta }{2}-2} \Gamma \left(\frac{1}{2} (D-\Delta +2)\right) \Gamma \left(\Delta -\frac{D}{2}\right) \left(p^2 \delta (\mu ,\nu )-p_{\mu } p_{\nu }\right)}{\Gamma \left(\frac{\Delta }{2}+1\right)^2 \Gamma \left(\frac{1}{2} (D-\Delta +1)\right) \Gamma \left(\frac{1}{2} (D+\Delta -2)\right)}+\right. \nn \\
    &\left.\frac{2^{2-\Delta } \pi ^D (-D+\Delta +1) \Gamma \left(\frac{D}{2}-\Delta \right) \Gamma \left(\frac{1}{2} (-D+\Delta +2)\right) \left(p^2\right)^{\frac{1}{2} (D-\Delta -2)} \left(q^2\right)^{\Delta -\frac{D}{2}} \delta (\mu ,\nu )}{\Gamma (\Delta+1 ) \Gamma \left(D-\frac{\Delta }{2}-1\right)}+\cdots\right),\nn
\end{align}
where we only keep the leading terms in the subgraph expansions.
The $q$-singular ($q$-regular) term is from the Subgraph $\gamma$ ($\Gamma$). For the scalars with scaling dimension $\Delta<d/2$, the $q$-singular term dominates in the small $q$ limit.

In this work, we consider the conductivity near the critical $O(N)$ vector model. 
The leading order result of $\omega(ik)$ in the large $k$ limit is given by the spin-1 conserved current central charge $C_J$.
The parameter $C_J$ in the critical $O(N)$ vector model has been computed to the order $O(1/N)$ in \cite{Petkou:1995vu,Huh:2013vga,Huh:2014eea,Diab:2016spb}. The subleading order contribution to $\omega(ik)$ depends on the OPE coefficient $\lambda_{JJ\sigma}$. The leading order result of $\lambda_{JJ\sigma}$ has been computed in \cite{Katz:2014rla}. We will compute its next-to-leading order correction using the method of subgraphs. 

\subsection{$\langle JJ\sigma_T\rangle$ in the $O(N)$ vector model}
In the critical $O(N)$ vector model, there are two relevant scalars in the $J\times J$ OPE: the $O(N)$ traceless symmetric scalar $\sigma_T=\phi_i\phi_j-\frac{1}{N} \delta_{ij}\phi_k\phi_k$ and the $O(N)$ singlet $\sigma=\phi_i\phi_i$.\footnote{The expressions $\phi_i\phi_j-\frac{1}{N} \delta_{ij}\phi_k\phi_k$ and $\phi_i\phi_i$ give the physical operators only for the free scalar theory or in the large $N$ limit. At finite $N$, these operators are modified by the renormalization factors. }  It is the coefficient of the correlator $\langle JJ\sigma\rangle$ that appears in the conductivity formula (\ref{conductivity}). The coefficient $\langle JJ\sigma_T\rangle$ is expected to play a significant role in bootstrap studies involving conserved currents. It turns out that subgraph expansions of the two 3-point correlators show interesting differences.

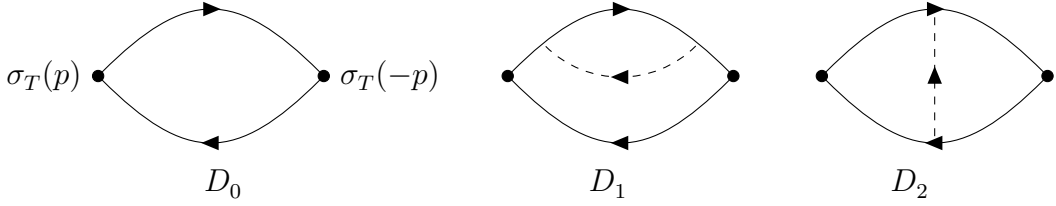
\begin{figure}
   \begin{equation*}
\begin{array}{ccc}
  \begin{tikzpicture}[baseline=(a.base)]
  \begin{feynman}
    \vertex [dot,label=right:$\sigma_T(-p)$] (b) at (3,0) {};
    \vertex[dot,label=left:$\sigma_T(p)$] (a)  at (0,0) {};

    \diagram* {
       (a) -- [fermion, quarter left,looseness=1.4] (b) --[fermion, quarter left,looseness=1.4] (a),
    };
  \end{feynman}
\end{tikzpicture}   & 
 ~~\begin{tikzpicture}[baseline=(a.base)]
  \begin{feynman}
    \vertex [dot] (b) at (3,0) {};
    \vertex[dot] (a)  at (0,0) {};
    \vertex (d) at (1/2, 2/5);
    \vertex (c) at (5/2, 2/5);

    \diagram* {
       (a) -- [fermion, quarter left,looseness=1.4] (b) --[fermion, quarter left,looseness=1.4] (a),
       (c) --[fermion, dashed, quarter left](d),
    };
  \end{feynman}
\end{tikzpicture} & ~~~~
 \begin{tikzpicture}[baseline=(a.base)]
  \begin{feynman}
    \vertex [dot] (b) at (3,0) {};
    \vertex[dot] (a)  at (0,0) {};
    \vertex (d) at (3/2, 4/5);
    \vertex (c) at (3/2, -4/5);

    \diagram* {
       (a) -- [fermion, quarter left,looseness=1.4] (b) --[fermion, quarter left,looseness=1.4] (a),
       (c) --[fermion, dashed](d),
    };
  \end{feynman}
\end{tikzpicture}  \\
D_0 & D_1 & D_2 
\end{array}
\end{equation*}
    \caption{Feynman diagrams for the $\langle \sigma_T(p) \sigma_T(-p)\rangle$ propagator.}
    \label{fig:JJSt}
\end{figure}

We first study the 3-point correlator with a non-singlet scalar $\langle JJ\sigma_T\rangle$. The scalar operator $\sigma_T$ is associated with a renormalization factor $Z_{\sigma_T}$:
\begin{equation}
    \sigma_T^{\mathrm{phy}}= Z_{\sigma_T}  \sigma_T, ~~~~ Z_{\sigma_T}=1+\frac{1}{N}\frac{r_{\sigma_T}}{\epsilon}.
\end{equation}
The renormalization factor $Z_{\sigma_T}$ is fixed by the cancellation of the divergences in the propagator $\langle \sigma_T(p) \sigma_T(-p)\rangle$, after which the correlation function $\langle \sigma_T(p) \sigma_T(-p)\rangle$ takes the standard form fixed by conformal symmetry
\begin{equation}
    \langle \sigma_T^{\mathrm{phy}}(p) \sigma_T^{\mathrm{phy}}(-p)\rangle = Z_{\sigma_T}^2 \langle \sigma_T(p) \sigma_T(-p)\rangle= \frac{C_{\sigma_T}}{\left(p^2\right)^{\frac{D}{2}-\Delta_{\sigma_T}}}.
\end{equation}
At the order $O(1/N)$, the propagator is given by the Feynman diagrams in Figure \ref{fig:JJSt}, from which the factors $r_{\sigma_T}, C_{\sigma_T}$ and $\Delta_{\sigma_T}$ can be solved to the order $O(1/N)$. The results read
\begin{equation}
    r_{\sigma_T}=-\frac{2^D \sin \left(\frac{\pi  D}{2}\right) \Gamma \left(\frac{D-1}{2}\right)}{\pi ^{3/2} D \Gamma \left(\frac{D}{2}\right)},~~~ \Delta_{\sigma_T}=D-2-\frac{2^D \sin \left(\frac{\pi  D}{2}\right) \Gamma \left(\frac{D-1}{2}\right)}{\pi ^{3/2} \Gamma \left(\frac{D}{2}+1\right)N}.
\end{equation}
The scaling dimension $\Delta_{\sigma_T}$ agrees with previous results \cite{Ma:1974qh,Gracey:2002qa}. The coefficients $C_{\sigma_T}$ is given by
\begin{equation}
    C_{\sigma_T}=C_{\sigma_T}^0\left(1+\frac{C_{\sigma_T}^1}{N}+O(\frac{1}{N^2})\right),
\end{equation}
in which
\begin{align}
    C_{\sigma_T}^0 &=-\frac{2^{3-2 D} \pi ^{\frac{3}{2}-\frac{D}{2}} \csc \left(\frac{\pi  D}{2}\right)}{\Gamma \left(\frac{D-1}{2}\right)}, \\
    C_{\sigma_T}^1 &=\left(\frac{8 \mathcal{C}_{\mathrm{O(N)}}}{D-4}+\frac{128-12 (D-6) (D-4) D}{(D-4)^2 (D-2) D}\right)\frac{2 \sin \left(\frac{\pi  D}{2}\right) \Gamma (D-2)}{\pi  \Gamma \left(\frac{D}{2}-2\right) \Gamma \left(\frac{D}{2}+1\right)N},
\end{align}
where $\mathcal{C}_{\mathrm{O(N)}}=\psi\left(3-\frac{D}{2}\right)+\psi(D-1)-\psi\left(\frac{D}{2}\right)+\psi(1)$, and $\psi(x)=\Gamma'(x)/\Gamma(x)$.

The Feynman diagrams for the 3-point correlator $\langle \sigma_T(q)J_\mu^a(p)J_\nu^a(-p-q) \rangle$ have the same geometry as the seven Feynman diagrams in Figure \ref{fig:JJJ1N}, but with the first vertex $J_\mu^a(q)$ replaced by the scalar vertex $\sigma_T(q)$. We use the subgraph expansions to evaluate the integrals—in particular the subgraphs $\Gamma$ of the original diagrams, which are regular in $q$ in the small $q$ expansion. Since this 3-point correlator has one independent tensor structure, it suffices to keep the leading order in $q$ in the small $q$ expansion. The integrals of these subgraphs are too cumbersome to be shown here or in appendices; instead, we present the integrals in a {\it Mathematica} file attached to this submission. 

Combining the renormalization factor of $\sigma_T$ and the loop integrals to the order $O(1/N)$, we extract the coefficient $\lambda_{JJ\sigma_T}$ of the 3-point correlator $\langle \sigma_T(q)J_\mu^a(p)J_\nu^a(-p-q) \rangle$. Using the normalization (\ref{JJOinP}), the coefficient reads
\begin{align}
    \lambda_{JJ\sigma_T}=D-2+\frac{((D-12) D+16) (\cos (\pi  D)-1) \Gamma \left(2-\frac{D}{2}\right) \Gamma (D-2)}{\pi ^2 D \Gamma \left(\frac{D}{2}+1\right)N} + O(1/N^2).
\end{align}
We emphasize that for the above 3-point correlator, it is the subgraphs $\Gamma$ which produce the $q$-regular term in (\ref{JJOinP}). The integrals of the subgraphs $\gamma$ are not mentioned here, but they generate the $q$-singular term in $\langle JJ\sigma_T\rangle$.
In contrast, the roles of the subgraphs $\Gamma$ and $\gamma$ are switched for the 3-point correlator $\langle \sigma_T(q)J_\mu^a(p)J_\nu^a(-p-q) \rangle$.

\subsection{$\langle JJ\sigma\rangle$ in the $O(N)$ vector model} \label{section:JJSig}
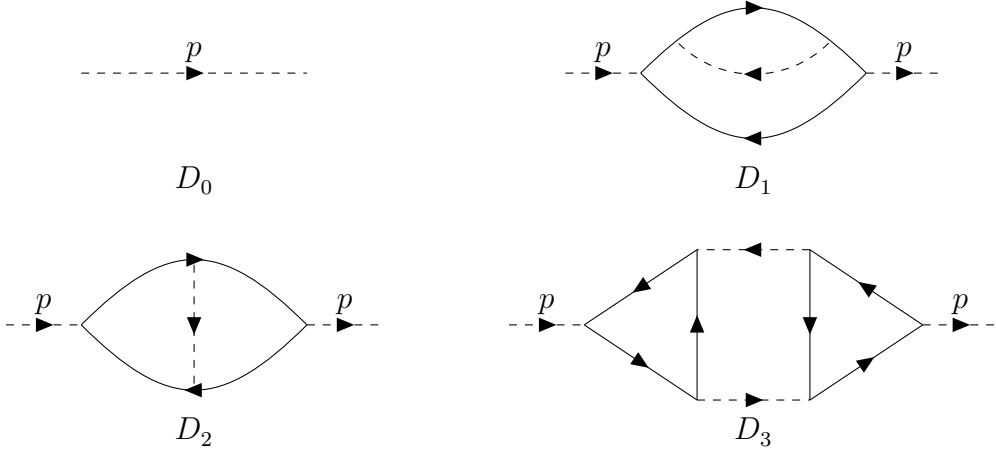
\begin{figure}
    \begin{equation*}
\begin{array}{cc}
  \begin{tikzpicture}[baseline=(a.base)]
  \begin{feynman}
   
    \vertex (a) at (0,0);
    \vertex (b) at (3, 0);

    \diagram* {
       (a) --[fermion, dashed,edge label=\(p\)] (b)
    };
  \end{feynman}
\end{tikzpicture} ~~~~ &~~~~ 
 \begin{tikzpicture}[baseline=(a.base)]
  \begin{feynman}
    \vertex (d) at (5,0);
    \vertex (a) at (0,0);
    \vertex (b) at (1, 0);
    \vertex (c) at (4, 0);
    \vertex (e) at (3/2, 2/5);
    \vertex (f) at (7/2, 2/5);

    \diagram* {
       (a) --[fermion, dashed,edge label=\(p\)] (b) -- [fermion, quarter left,looseness=1.4] (c) -- [fermion, dashed,edge label=\(p\)](d),
       (c)--[fermion, quarter left,looseness=1.4] (b),
       (f) --[fermion, dashed, quarter left](e),
    };
  \end{feynman}
\end{tikzpicture} \\
D_0 ~~~~&~~~~ D_1 \\ &  \\
 \begin{tikzpicture}[baseline=(a.base)]
  \begin{feynman}
      \vertex (d) at (5,0);
    \vertex (a) at (0,0);
    \vertex (b) at (1, 0);
    \vertex (c) at (4, 0);
    \vertex (f) at (5/2, 4/5);
    \vertex (e) at (5/2, -4/5);

    \diagram* {
       (a) --[fermion, dashed,edge label=\(p\)] (b) -- [fermion, quarter left,looseness=1.4] (c) -- [fermion, dashed,edge label=\(p\)](d),
       (c)--[fermion, quarter left,looseness=1.4] (b),
       (f) --[fermion, dashed](e),
    };
  \end{feynman}
\end{tikzpicture} ~~~~&~~~~
 \begin{tikzpicture}[baseline=(a.base)]
  \begin{feynman}
    \vertex (d) at (5/2,1);
    \vertex (a) at (0,0);
    \vertex (b) at (1, 0);
    \vertex (c) at (5/2, -1);
    \vertex (e) at (11/2, 0);
    \vertex (f) at (8/2, 1);
    \vertex (g) at (8/2, -1);
    \vertex (h) at (13/2, 0);
    \diagram* {
        (a) --[fermion, dashed,edge label=\(p\)] (b) -- [fermion] (c)-- [fermion] (d)-- [fermion] (b),
        (e) --[fermion] (f) -- [fermion] (g)-- [fermion] (e)--[fermion, dashed,edge label=\(p\)] (h),
       (c) --[fermion, dashed](g),
       (f) --[fermion, dashed](d),
    };
  \end{feynman}
\end{tikzpicture} \\
D_2 ~~~~&~~~~ D_3
\end{array}
\end{equation*}
\caption{Feynman diagrams  for the propagator of the singlet scalar $\langle \sigma(p)\sigma(-p) \rangle$ up to the order $O(1/N)$.}
    \label{fig:SigSig}
\end{figure}

The critical scalar $\sigma$ has a propagator (\ref{phisigpropagator}), and its renormalization factor $Z_\sigma$ is defined in (\ref{renormalizations}). Up to the order $O(1/N)$, the 2-point correlator $\langle \sigma(p)\sigma(-p)\rangle$ is given by the four Feynman diagrams in Figure \ref{fig:SigSig}. The Feynman diagrams can be evaluated using the formulas shown before, except for the three-loop diagram $D_3$, which is of Aslamazov-Larkin type \cite{ASLAMASOV1968238}. Details of these integrals are presented in an attached {\it Mathematica} file.

Based on the Feynman integrals in Figure \ref{fig:SigSig}, the renormalization factor of the field $\sigma$ can be solved to the subleading order
\begin{equation}
  Z_\sigma=1+\frac{r_\sigma}{\Delta}\frac{1}{N}+O(1/N^2),~~~  r_\sigma=\frac{2^D \sin \left(\frac{\pi  D}{2}\right) \Gamma \left(\frac{D-1}{2}\right)}{\pi ^{3/2} \Gamma \left(\frac{D}{2}+1\right)}.
\end{equation}
The scaling dimension $\Delta_\sigma$ and the coefficient $C_\sigma$ are given by
\begin{equation}
    \Delta_\sigma=2+\frac{4 (D-2) (D-1) \eta _{\text{ON}}}{(D-4) }\frac{1}{N}+O(1/N^2), \label{deltasig}
\end{equation}
and
\begin{align}
    C_\sigma &=C_\sigma^0 \left(1+\frac{C_\sigma^1}{N}+O(1/N^2)\right), \\
    C_\sigma^1 &= \frac{\eta_{\mathrm{ON}}}{N}\left(\frac{2 \mathcal{C}_{\mathrm{ON}} ((D-3) D+4)}{4-D}-\frac{2 \left((D-4) D \left(2 (D-4) D^2+D+36\right)+64\right)}{(D-4)^2 (D-2) D}\right), \label{si}
\end{align}
where 
\begin{equation}
    \eta_{\mathrm{ON}}=\frac{2 \sin \left(\frac{\pi  D}{2}\right) \Gamma (D-2)}{\pi  \Gamma \left(\frac{D}{2}-2\right) \Gamma \left(\frac{D}{2}+1\right)}.
\end{equation}
The anomalous dimension of $\sigma$ agrees with previous result \cite{Okabe:1978mp,Vasilev19811nEC}.

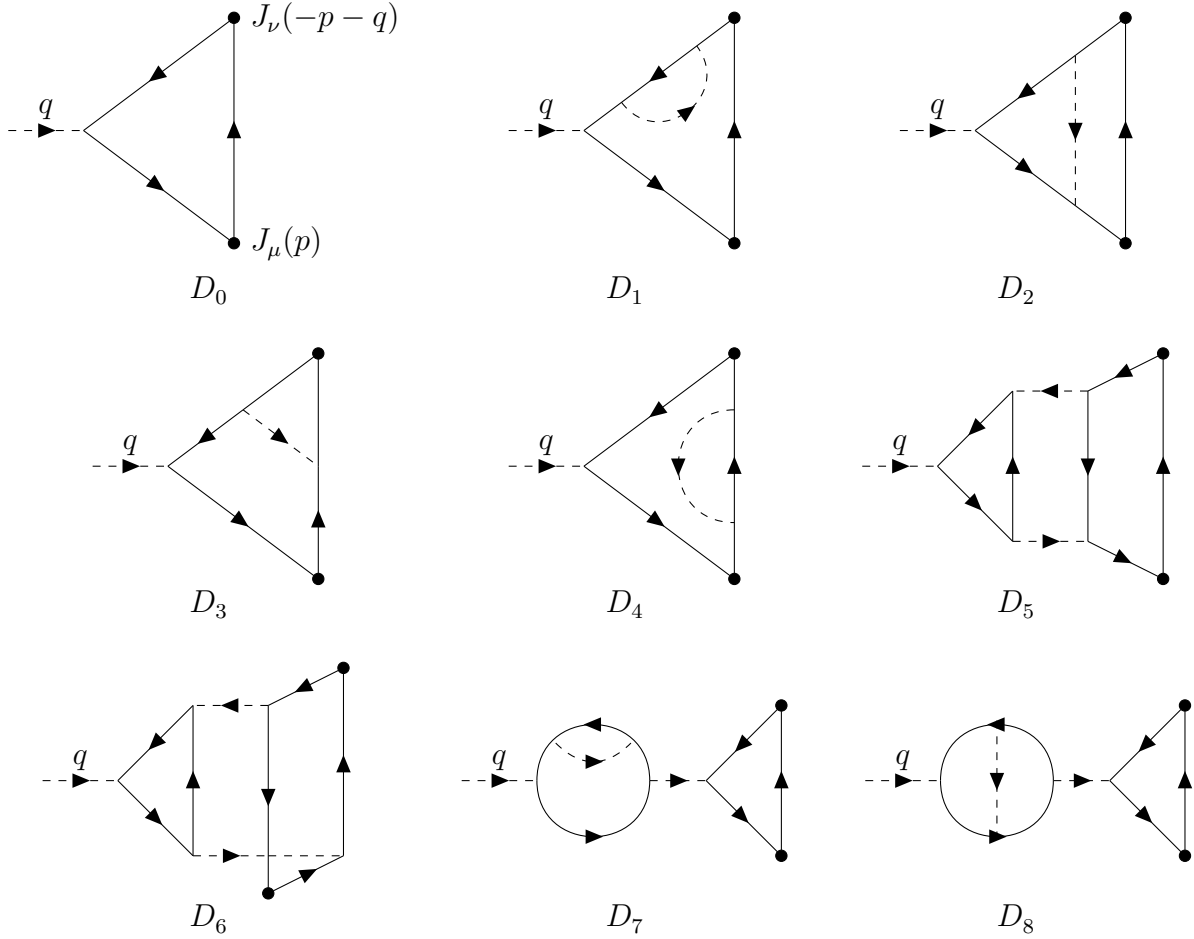
\begin{figure}
    \centering
    \begin{equation*}
\begin{array}{ccc}
  \begin{tikzpicture}[baseline=(a.base)]
  \begin{feynman}
    \vertex [dot,label=right:$J_{\nu}(-p-    q)$] (d) at (3,3/2) {};
    \vertex[dot,label=right:$J_{\mu}(p)$] (c) at (3,-3/2) {};
    \vertex (a)  at (0,0);
    \vertex (b)  at (1,0);

    \diagram* {
        (a) --[fermion, dashed,edge label=\(q\)] (b) -- [fermion] (c)-- [fermion] (d)-- [fermion] (b),
    };
  \end{feynman}
\end{tikzpicture}   & 
 \begin{tikzpicture}[baseline=(a.base)]
  \begin{feynman}
 \vertex [dot] (d) at (3,3/2) {};
    \vertex[dot] (c) at (3,-3/2) {};
    \vertex (a)  at (0,0);
    \vertex (b)  at (1,0);

    \vertex (e) at (5/2, 9/8);
    \vertex (f) at (3/2, 3/8);

    \diagram* {
        (a) --[fermion, dashed,edge label=\(q\)] (b) -- [fermion] (c)-- [fermion] (d)-- [fermion] (b),
       (f) --[fermion, dashed, half right,looseness=1.7] (e);
    };
  \end{feynman}
\end{tikzpicture} &  \begin{tikzpicture}[baseline=(a.base)]
  \begin{feynman}
 \vertex [dot] (d) at (3,3/2) {};
    \vertex[dot] (c) at (3,-3/2) {};
    \vertex (a)  at (0,0);
    \vertex (b)  at (1,0);

    \vertex (f) at (7/3, 1);
    \vertex (e) at (7/3, -1);

    \diagram* {
       (a) --[fermion, dashed,edge label=\(q\)] (b) -- [fermion] (e) --(c)-- [fermion] (d)--(f) -- [fermion] (b),
       (f) --[fermion, dashed] (e);
    };
  \end{feynman}
\end{tikzpicture} \\
  D_0   & D_1  & D_2 \\ & & \\
    \begin{tikzpicture}[baseline=(a.base)]
  \begin{feynman}
    \vertex [dot] (d) at (3,3/2) {};
    \vertex [dot] (c) at (3,-3/2) {};
    \vertex (a)  at (0,0);
    \vertex (b)  at (1,0);

    \vertex (f) at (2, 3/4);
    \vertex (e) at (3, 0);

    \diagram* {
       (a) --[fermion, dashed,edge label=\(q\)] (b) -- [fermion](c)-- [fermion] (e) -- (d)--(f) -- [fermion] (b),
       (f) --[fermion, dashed] (e);
    };
  \end{feynman}
\end{tikzpicture}& 
\begin{tikzpicture}[baseline=(a.base)]
  \begin{feynman}
    \vertex [dot] (d) at (3,3/2) {};
    \vertex [dot] (c) at (3,-3/2) {};
    \vertex (a)  at (0,0);
    \vertex (b)  at (1,0);
    
    \vertex (f) at (3, 3/4);
    \vertex (e) at (3, -3/4);

    \diagram* {
      (a) --[fermion, dashed,edge label=\(q\)] (b) -- [fermion](c)-- [fermion] (d) -- [fermion] (b),
       (f) --[fermion, dashed, half right,looseness=1.7] (e);
    };
  \end{feynman}
\end{tikzpicture} & \begin{tikzpicture}[baseline=(a.base)]
  \begin{feynman}
    \vertex (d) at (4/2,1);
    \vertex (a) at (0,0);
    \vertex (b) at (1, 0);
    \vertex (c) at (4/2, -1);
    
    \vertex [dot] (e) at (8/2, 6/4) {};
    \vertex (f) at (6/2, 1);
    \vertex (g) at (6/2, -1);
    \vertex [dot] (h) at (8/2, -6/4) {};
    \diagram* {
        (a) --[fermion, dashed,edge label=\(q\)] (b) -- [fermion] (c)-- [fermion] (d)-- [fermion] (b),
        (e) --[fermion] (f) -- [fermion] (g)-- [fermion] (h)--[fermion] (e),
       (c) --[fermion, dashed](g),
       (f) --[fermion, dashed](d),
    };
  \end{feynman}
\end{tikzpicture} 
 \\
 D_3 & D_4 &  D_5  \\ & & \\
\begin{tikzpicture}[baseline=(a.base)]
  \begin{feynman}
    \vertex (d) at (4/2,1);
    \vertex (a) at (0,0);
    \vertex (b) at (1, 0);
    \vertex (c) at (4/2, -1);
    
    \vertex [dot] (e) at (8/2, 6/4) {};
    \vertex (f) at (6/2, 1);
    \vertex (g) at (8/2, -1);
    \vertex [dot] (h) at (6/2, -6/4) {};
    \vertex (i) at (6/2,-1);
    \diagram* {
        (a) --[fermion, dashed,edge label=\(q\)] (b) -- [fermion] (c)-- [fermion] (d)-- [fermion] (b),
        (e) --[fermion] (f) -- [fermion] (h)-- [fermion] (g)--[fermion] (e),
       (c) --[fermion, dashed](i) --[dashed] (g),
       (f) --[fermion, dashed](d),
    }; 
  \end{feynman}
\end{tikzpicture} ~~&~~ 
\begin{tikzpicture}[baseline=(a.base)]
  \begin{feynman}
    \vertex [dot] (f) at (17/4,3/2) {};
    \vertex[dot] (e) at (17/4,-3/2) {};
    \vertex (a)  at (0,0);
    \vertex (b)  at (1,0);
    \vertex (c)  at (5/2,0);
    \vertex (d)  at (13/4,0);
    \vertex (g)  at (5/4,1/2);
    \vertex (h)  at (9/4,1/2);

    \diagram* {
        (a) --[fermion, dashed,edge label=\(q\)] (b) -- [fermion, half right,looseness=1.7] (c)-- [fermion,dashed] (d),
        (d)--[fermion] (e) --[fermion] (f) --[fermion] (d),
        (c) -- [fermion, half right,looseness=1.7] (b),
        (g)--[fermion, dashed, quarter right,looseness=1.2] (h)
    };
  \end{feynman}
\end{tikzpicture}   ~~&~~ \begin{tikzpicture}[baseline=(a.base)]
  \begin{feynman}
    \vertex [dot] (f) at (17/4,3/2) {};
    \vertex[dot] (e) at (17/4,-3/2) {};
    \vertex (a)  at (0,0);
    \vertex (b)  at (1,0);
    \vertex (c)  at (5/2,0);
    \vertex (d)  at (13/4,0);
    \vertex (g)  at (7/4,4/5);
    \vertex (h)  at (7/4,-4/5);

    \diagram* {
        (a) --[fermion, dashed,edge label=\(q\)] (b) -- [fermion, half right,looseness=1.7] (c)-- [fermion,dashed] (d),
        (d)--[fermion] (e) --[fermion] (f) --[fermion] (d),
        (c) -- [fermion, half right,looseness=1.7] (b),
        (g)--[fermion, dashed] (h)
    };
  \end{feynman}
\end{tikzpicture} \\
D_6 & D_7 & D_8 \\ & & \\
\begin{tikzpicture}[baseline=(a.base)]
  \begin{feynman}

    \vertex [dot] (i) at (22/4,3/2) {};
    \vertex[dot] (j) at (22/4,-3/2) {};
    \vertex (a) at (0,0);
    \vertex (b) at (3.5/4, 0);
    \vertex (c) at (7/4, -5/5);
    \vertex (d) at (7/4,5/5);
    \vertex (e) at (11/4, 5/5);
    \vertex (f) at (11/4, -5/5);
    \vertex (g) at (14.5/4, 0);
    \vertex (h) at (18/4, 0);
    \diagram* {
        (a) --[fermion, dashed,edge label=\(q\)] (b) -- [fermion] (c)-- [fermion] (d) -- [fermion]  (b), (d) --[fermion, dashed]  (e)-- [fermion] (f)-- [fermion] (g) -- [fermion] (e), (c) --[fermion, dashed] (f), (g) --[fermion, dashed]  (h) -- [fermion] 
        (j)--[fermion] (i) --[fermion] (h)
    };
  \end{feynman}
\end{tikzpicture}  & & \\
D_9 & &
\end{array}
\end{equation*}
    \caption{Feynman diagrams for the 3-point correlator $\langle \sigma(q)J_\mu(p)J_\nu(-p-q)\rangle$ up to the order $O(1/N)$. There are extra Feynman diagrams similar to $D_1$ and $D_3$ but with the $\sigma$ propagator connecting with different solid lines. However, since we only keep the leading terms in the small $q$ expansion,  these diagrams are identical to $D_1$ or $D_3$. }
    \label{fig:JJsig}
\end{figure}

To the subleading order, the 3-point correlator $\langle \sigma(q)J_\mu(p)J_\nu(-p-q)\rangle$ can be expanded by the Feynman diagrams in Figure \ref{fig:JJsig}. The Feynman diagrams $D_i, ~i=0,1,\dots,4$ are the same as the Feynman diagrams in Figure \ref{fig:JJJ1N}, attached with an extra $\sigma$ propagator. The Feynman diagrams $D_7$, $D_8$, and $D_9$ are the combinations of the 1PI loop corrections of the $\sigma$ propagator and the one-loop diagram of the correlator $\langle JJ \phi^2\rangle$.
The three-loop integrals $D_5$
and $D_6$ are the generalizations of the Aslamazov-Larkin diagram  to three external momenta, which  are hard to evaluate analytically. We use the method of subgraphs to compute the two integrals in the small $q$ limit. Since the 3-point correlator contains only one independent coefficient $\lambda_{JJ\sigma}$, to fix its $1/N$ correction we only need to keep the leading order term in the small $q$ expansion of each subgraph.

We focus on the subgraphs of the Feynman diagrams which generate the $q$-regular terms in the small $q$ expansion (\ref{JJOinP}) of the correlator $\langle JJ\sigma\rangle$. 
We separate the Feynman integrals $\mathcal{I}_{\langle \sigma JJ\rangle}$ in Figure \ref{fig:JJsig}  into two components:  
\begin{equation}
    \mathcal{I}_{\langle \sigma JJ\rangle}=\mathcal{P}_\sigma \cdot \mathcal{I}_{\langle JJ\phi^2\rangle} \propto \left(q^2\right)^{2-\frac{D}{2}} \cdot \mathcal{I}_{\langle JJ\phi^2\rangle}, \label{JJsigma2cmp}
\end{equation}
where $\mathcal{P}_\sigma$ gives the $\sigma$ propagator or its 1PI loop corrections, while  $\mathcal{I}_{\langle JJ\phi^2\rangle}$ denotes the contributions from the amputated diagrams in Figure \ref{fig:JJsig}  without the $\sigma$ propagators. The subgraph expansion will be taken with respect to these amputated diagrams.
The subgraphs consist of two sectors $\Gamma$ and $\gamma$, which depend on the external momenta as follows
\begin{align}
    \mathcal{I}_{\langle JJ\phi^2\rangle,\Gamma} &\propto \left(p^2\right)^{\frac{D}{2}-3} p_\mu p_\nu  +\cdots, \\
    \mathcal{I}_{\langle JJ\phi^2\rangle,\gamma}&\propto \left(p^2\right)^{-1}\left(q^2\right)^{\frac{D}{2}-2}p_\mu p_\nu +\cdots.
\end{align}
Combining with the $\sigma$ propagator factor $\mathcal{P}_\sigma$ in (\ref{JJsigma2cmp}), it is the subgraphs $\gamma$ of the amputated diagram which generate the $q$-regular term in the 3-point correlator $\langle \sigma JJ\rangle$. This is  opposite to the 3-point correlator $\langle JJ \sigma_T\rangle $.

\begin{figure}
    \begin{equation*}
    \begin{array}{ccc}
       \begin{tikzpicture}[baseline=(a.base)]
  \begin{feynman}
    \vertex (d) at (4/2,1);
    \vertex (a) at (0,0);
    \vertex (b) at (1, 0);
    \vertex (c) at (4/2, -1);
    
    \vertex [dot,label=right:$J_{\nu}(-p-    q)$] (e) at (8/2, 6/4) {};
    \vertex (f) at (6/2, 1);
    \vertex (g) at (6/2, -1);
    \vertex [dot, label=right:$J_{\mu}(p)$] (h) at (8/2, -6/4) {};
    \diagram* {
        (a) --[fermion, dashed,edge label=\(q\)] (b) -- [fermion] (c)-- [fermion] (d)-- [fermion] (b),
        (e) --[fermion] (f) -- [fermion] (g)-- [fermion] (h)--[fermion,red] (e),
       (c) --[fermion, dashed](g),
       (f) --[fermion, dashed](d),
    };
  \end{feynman}
\end{tikzpicture}   & \begin{tikzpicture}[baseline=(a.base)]
  \begin{feynman}
    \vertex (d) at (4/2,1);
    \vertex (a) at (0,0);
    \vertex (b) at (1, 0);
    \vertex (c) at (4/2, -1);
    
    \vertex [dot] (e) at (8/2, 6/4) {};
    \vertex (f) at (6/2, 1);
    \vertex (g) at (6/2, -1);
    \vertex [dot] (h) at (8/2, -6/4) {};
    \diagram* {
        (a) --[fermion, dashed,edge label=\(q\)] (b) -- [fermion] (c)-- [fermion] (d)-- [fermion] (b),
        (e) --[fermion,red] (f) -- [fermion,red] (g)-- [fermion,red] (h)--[fermion,red] (e),
       (c) --[fermion, dashed](g),
       (f) --[fermion, dashed](d),
    };
  \end{feynman}
\end{tikzpicture}  ~~&~~
\begin{tikzpicture}[baseline=(a.base)]
  \begin{feynman}
    \vertex (d) at (4/2,1);
    \vertex (a) at (0,0);
    \vertex (b) at (1, 0);
    \vertex (c) at (4/2, -1);
    
    \vertex [dot] (e) at (8/2, 6/4) {};
    \vertex (f) at (6/2, 1);
    \vertex (g) at (6/2, -1);
    \vertex [dot] (h) at (8/2, -6/4) {};
    \diagram* {
        (a) --[fermion, dashed,edge label=\(q\)] (b) -- [fermion] (c)-- [fermion,red] (d)-- [fermion] (b),
        (e) --[fermion,red] (f) -- [fermion,red] (g)-- [fermion,red] (h)--[fermion,red] (e),
       (c) --[fermion, dashed,red](g),
       (f) --[fermion, dashed,red](d),
    };
  \end{feynman}
\end{tikzpicture}  \\
D_{5,\gamma_1} & ~~~~~~D_{5,\gamma_2}  & ~~~~~~D_{5,\gamma_3}  \\ 
& & \\
\begin{tikzpicture}[baseline=(a.base)]
  \begin{feynman}
    \vertex (d) at (4/2,1);
    \vertex (a) at (0,0);
    \vertex (b) at (1, 0);
    \vertex (c) at (4/2, -1);
    
    \vertex [dot, label=right:$J_{\nu}(-p-q)$] (e) at (8/2, 6/4) {};
    \vertex (f) at (6/2, 1);
    \vertex (g) at (8/2, -1);
    \vertex [dot, label=below:$J_{\mu}(p)$] (h) at (6/2, -6/4) {};
    \vertex (i) at (6/2,-1);
    \diagram* {
        (a) --[fermion, dashed,edge label=\(q\)] (b) -- [fermion] (c)-- [fermion] (d)-- [fermion] (b),
        (e) --[fermion] (f) -- [fermion] (h)-- [fermion,red] (g)--[fermion,red] (e),
       (c) --[fermion, dashed](i) --[dashed] (g),
       (f) --[fermion, dashed](d),
    }; 
  \end{feynman}
\end{tikzpicture} & 
\begin{tikzpicture}[baseline=(a.base)]
  \begin{feynman}
    \vertex (d) at (4/2,1);
    \vertex (a) at (0,0);
    \vertex (b) at (1, 0);
    \vertex (c) at (4/2, -1);
    
    \vertex [dot] (e) at (8/2, 6/4) {};
    \vertex (f) at (6/2, 1);
    \vertex (g) at (8/2, -1);
    \vertex [dot] (h) at (6/2, -6/4) {};
    \vertex (i) at (6/2,-1);
    \diagram* {
        (a) --[fermion, dashed,edge label=\(q\)] (b) -- [fermion] (c)-- [fermion] (d)-- [fermion] (b),
        (e) --[fermion,red] (f) -- [fermion,red] (h)-- [fermion,red] (g)--[fermion,red] (e),
       (c) --[fermion, dashed](i) --[dashed] (g),
       (f) --[fermion, dashed](d),
    }; 
  \end{feynman}
\end{tikzpicture} & ~~
\begin{tikzpicture}[baseline=(a.base)]
  \begin{feynman}
    \vertex (d) at (4/2,1);
    \vertex (a) at (0,0);
    \vertex (b) at (1, 0);
    \vertex (c) at (4/2, -1);
    
    \vertex [dot] (e) at (8/2, 6/4) {};
    \vertex (f) at (6/2, 1);
    \vertex (g) at (8/2, -1);
    \vertex [dot] (h) at (6/2, -6/4) {};
    \vertex (i) at (6/2,-1);
    \diagram* {
        (a) --[fermion, dashed,edge label=\(q\)] (b) -- [fermion] (c)-- [fermion,red] (d)-- [fermion] (b),
        (e) --[fermion,red] (f) -- [fermion,red] (h)-- [fermion,red] (g)--[fermion,red] (e),
       (c) --[fermion, dashed,red](i) --[dashed,red] (g),
       (f) --[fermion, dashed,red](d),
    }; 
  \end{feynman}
\end{tikzpicture} \\
D_{6,\gamma_1} & ~~~~~~D_{6,\gamma_2}  & ~~~~~~D_{6,\gamma_3} 
    \end{array}
    \end{equation*}
    \caption{Subgraphs (in red color) of the Feynman diagrams $D_5$ and $D_6$ in Figure \ref{fig:JJsig}. }
    \label{fig:D5subgraphs}
\end{figure}
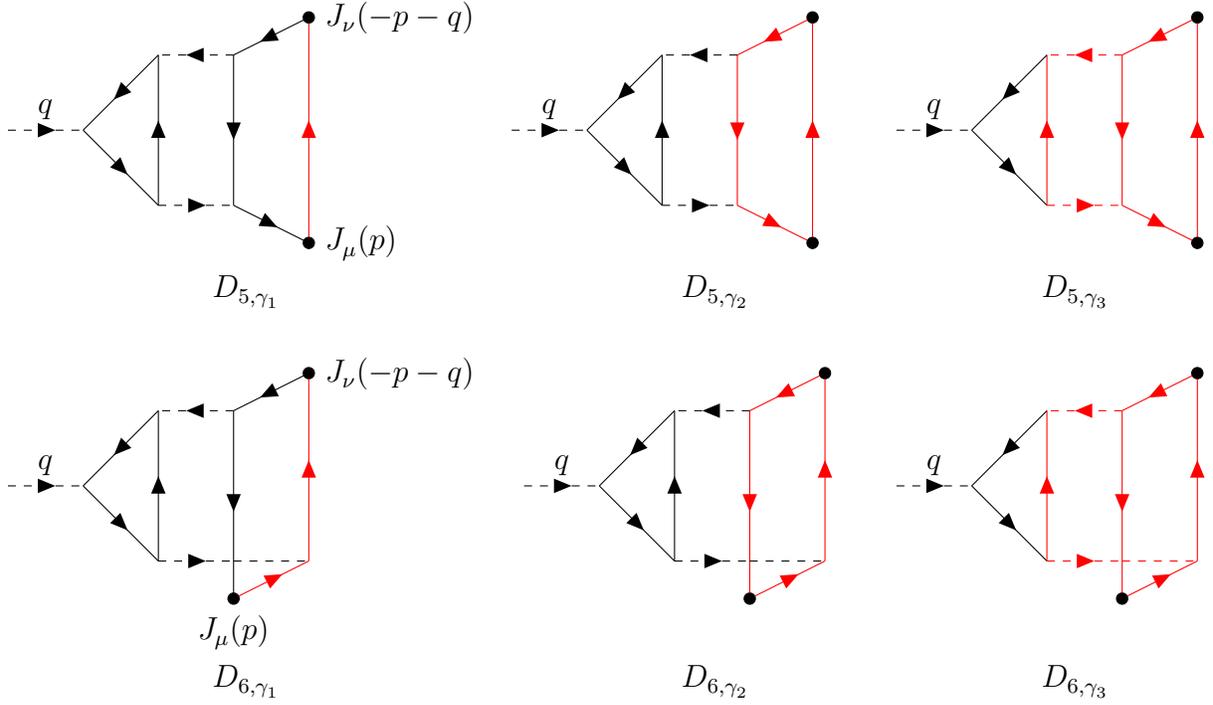

The amputated diagrams for the Feynman diagrams $D_i, ~i=0,1,2,3,4, 7,8,9$ have the same geometry as those in Figure \ref{fig:JJJ1N}, and their subgraph expansions are given by those in Figure \ref{fig:JJJ1Nsubs}. The three-loop Feynman diagrams $D_5$ and $D_6$ contain more complicated subgraphs. In Figure \ref{fig:D5subgraphs}, we show the subgraphs for the Feynman diagram $D_5$ and $D_6$ in Figure \ref{fig:JJsig}. There are more subgraphs for the diagram $D_5$, but they lead to scaleless integrals and are ignored in Figure \ref{fig:D5subgraphs}. The integrals for the subgraphs $\gamma$ are significantly simplified compared with the original Feynman integrals. In particular, the original 2-scale three-loop Feynman diagrams $D_{5}$ and $D_6$ reduce to the Aslamazov-Larkin diagrams in the subgraphs $D_{5,\gamma_1}$ and $D_{6,\gamma_1}$, and to the Kite diagrams in the subgraphs $D_{5,\gamma_2}$, $D_{6,\gamma_2}$ and $D_{6,\gamma_3}$. To evaluate the subgraph $D_{6,\gamma_3}$, we have employed the following identity between two Kite diagrams in 3D:
\begin{equation}
    \left. K(D-3+2\epsilon,p)p^2+2K(D-4+2\epsilon,p)\right|_{D=3}=-\frac{3}{16\pi^2}+O(\epsilon),
\end{equation}
which we have  verified numerically. Its generalization to general $D$ is not known yet.

The integrals of the subgraphs are presented in the attached 
{\it Mathematica} file. For simplicity, we focus on the leading order results in 3D, and ignore the terms with $\delta$-functions.
Collecting all subgraph contributions, the final result is given by
\begin{equation}
    \langle \sigma(q)J_\mu(p)J_\nu(-p-q)\rangle|_{D=3}=\frac{2 p_{\rho } p_{\sigma }}{p^2}-\frac{32 \left(3 \log \left(p^2\right)-2\right)p_{\rho } p_{\sigma }}{9 \pi ^2 p^2 N}+\cdots.
\end{equation} 
Comparing with the same conformal 3-point correlation function (\ref{JJOinP}) solved from conformal symmetry, we obtain
\begin{equation}
    \left. \lambda_{JJ\sigma}\right|_{D=3}=\frac{4}{\pi \sqrt{N}}-\frac{80}{9 \pi ^3\sqrt{N}N}+O(1/N^{5/2}). \label{JJO2nd}
\end{equation}
For the $O(2)$ vector model, the leading order result is $\lambda_{JJ\sigma}\simeq 0.900$ \cite{Reehorst:2019pzi}, and the subleading order correction leads to $\lambda_{JJ\sigma}\simeq 0.799$, while the nonperturbative bootstrap result gives $\lambda_{JJ\sigma}=0.645(4)$ \cite{Reehorst:2019pzi}. There is a notable difference between the large $N$ expansion and the nonperturbative result, which is expected for small $N=2$.

The $1/N$ correction  to the parameter $\lambda_{JJ\sigma}$ in (\ref{JJO2nd}) provides necessary ingredient to compute the conductivity near the $O(N)$ critical points at finite temperature. The next-to-next-to-leading-order term of the conductivity $\omega(ik)$ arises from the stress tensor, whose contribution depends on the 3-point correlator $\langle TJJ\rangle$, and we leave its $1/N$ correction for future work.

\section{Conclusion and discussion}\label{sec6}
We have developed a new approach to compute the subleading order corrections to the conformal current 3-point correlators in the large $N$ expansion. The main technical challenge is that both the conformal 3-point correlators in momentum space and the Feynman loop integrals with three external momenta are hard to evaluate analytically. We introduced the method of subgraphs to resolve this problem, which can generate graph-theoretical expansions for the 3-point loop integrals. Specifically, we diagrammatically expanded the original 2-scale loop integrals in terms of the 1-scale loop integrals, which are much easier to evaluate analytically. Such diagrammatic expansions of the Feynman integrals are subtle in Lorentzian spacetime, while the Feynman loop integrals for CFT data are given in Euclidean spacetime and can be expanded in terms of their subgraphs. This provides a powerful approach to compute the multi-point correlation functions in general CFTs. We have applied the method of subgraphs for the following problems:
\begin{itemize}
    \item The small momentum expansions of the conformal conserved current 3-point correlators $\langle JJJ\rangle$, $\langle TJJ\rangle$, and $\langle TTT\rangle$. 
    
    The conformal 3-point correlation functions can be solved from the conformal Ward identities in momentum space, while the solutions in general dimensions are given by complicated integrals and are hard to apply for the $1/N$ perturbative computations. Using the method of subgraphs, these correlators can be expanded in a graph-theoretical way. The subgraph expansions provide diagrammatic representations for the behaviors of the correlation functions in the zero-momentum limit. In particular, the method of subgraphs captures the singularities of the correlator in the OPE limit in momentum space.

    \item The $1/N$ corrections to the conformal 3-point correlators $\langle JJJ\rangle$ in the critical $O(N)$ vector model and GNY model. 

    The conformal conserved current 3-point correlators play important roles in various directions, while they are only known at the leading order. We use the correlator $\langle JJJ\rangle$ as an example to demonstrate that the method of subgraphs provides a powerful approach to compute the $1/N$ corrections. Different from the subgraph expansions in general QFTs, for CFTs, one needs to keep only the first few terms in the subgraph expansions to obtain the $1/N$ corrections.

    \item The $1/N$ corrections to the conformal 3-point correlators $\langle JJ \cO\rangle$ in the critical $O(N)$ vector model.

    This problem is motivated by the fundamental role of the correlator $\langle JJ\sigma\rangle$ in evaluating the conductivity of quantum critical systems  at finite temperature. We computed the coefficients for both the lowest $O(N)$ traceless symmetric scalar $\sigma_T$ and the singlet scalar $\sigma$, which can be used to compute the subleading term of the conductivity. 
    An interesting observation is that,  due to the attachment of the $\sigma$ propagator in the Feynman diagrams for $\langle JJ\sigma\rangle$, the subgraphs $\Gamma$ and $\gamma$ switch their roles in reproducing the asymptotic behaviors of the correlators $\langle JJ\cO\rangle$ in the subgraph expansion.  
    
\end{itemize}

We expect this work to pave the way to perturbatively study the conformal current 3-point correlators in general CFTs and their applications in various directions. We elaborate on part of the problems below.

It is straightforward to apply the method of subgraphs to compute the subleading order corrections to the conformal current 3-point correlators $\langle JJ\cO\rangle$ and $\langle JJT\rangle$ in the conformal gauge theories, e.g., the IR fixed points of the QED$_3$ coupled with massless fermions \cite{Appelquist86,Appelquist88}. The theories are strongly coupled in the IR, and they provide ideal candidates for the modern conformal bootstrap studies \cite{Rattazzi:2008pe,Poland:2018epd}. The conserved currents are important elements in the bootstrap equations, and the large $N$ perturbative estimations on the conserved current 3-point correlators can provide necessary input data for the bootstrap computations. 

We have computed the conserved current 3-point correlator $\langle JJ\cO\rangle$ in the critical $O(N)$ vector model, motivated by its application in the conductivity at finite temperature \cite{Chowdhury:2012km,Witczak-Krempa:2013nua,Katz:2014rla,Witczak-Krempa:2015pia,Lucas:2016fju,Lucas:2017dqa}. The next dominant contribution comes from the stress tensor $T_{\mu\nu}$. For the CFTs without relevant singlet scalars, e.g., the conformal QED$_3$ coupled with many-flavor massless fermions, the stress tensor provides the dominating contribution to the thermal conductivity in addition to the vacuum state. Therefore, it is important to evaluate the conformal current 3-point correlators $\langle JJT\rangle$ in these theories to provide better estimations of the conductivities near the IR fixed points. The leading order results of the correlator $\langle JJT\rangle$ are given by the free fermion and free scalar theories, which have been computed in \cite{Chowdhury:2012km}. The Feynman diagrams for the correlator $\langle JJT\rangle$ to order $O(1/N)$ are similar to those for $\langle JJ\sigma\rangle$ in Figure \ref{fig:JJsig}, but with a new vertex for the stress tensor $T_{\mu\nu}$. We expect the subleading order correction to $\langle JJT\rangle$ can be solved using the method developed in this work, and it would be interesting to compare the perturbative results with the Monte Carlo simulations \cite{Witczak-Krempa:2013nua,Katz:2014rla,Lucas:2016fju,Chen:2013ppa,Gazit_2014}.

It is tempting to generalize the studies of conformal current 3-point correlators to the parity-breaking CFTs. A prominent family of such theories is the Chern-Simons matter field theories in 3D, which exhibit interesting conductivity properties at finite temperature; see \cite{Maity:2024nol} for a recent study. Prior to the Feynman loop computations, it is necessary to know the conformal symmetry constraints on the parity-odd 3-point correlation functions \cite{Jain:2021wyn, Coriano:2023hts,Coriano:2023gxa,Coriano:2023cvf}. To compute the $1/N$ corrections, it would be helpful to get the diagrammatic representations of the parity-odd  correlation functions. A remarkable property of the 3D Chern-Simons matter field theories is the so-called bosonization, namely the duality between the Chern-Simons theories coupled with fermions and scalars. A classical example of the 3D bosonization is provided by the $U(N)_k$ Chern-Simons theory coupled to a fundamental boson or fermion \cite{Giombi:2011kc,Aharony:2011jz,Aharony:2012nh}. In the planar limit $N\rightarrow \infty$ with fixed 't Hooft coupling $\lambda=N/k$, all the current 3-point correlators can be solved through a bootstrap approach \cite{Maldacena:2011jn,Maldacena:2012sf}, and the results confirm the bosonization duality at large $N$. The duality is conjectured to be true even at finite $N$, with more details on the relations between the gauge groups and matter fields \cite{Aharony:2015mjs,Aharony:2016jvv}. It would be interesting to study this duality at finite $N$ by computing the $1/N$ corrections of the conformal current 3-point correlators. 
The 3D bosonization can also be realized by the Abelian Chern-Simons theories coupled to matter fields \cite{Seiberg:2016gmd,Karch:2016sxi,Kachru:2016rui,Wang:2017txt}, in which different theories are connected through the $SL(2,Z)$ actions and form a duality web. The duality web has a close relation to the deconfined quantum critical points \cite{Senthil_2004}. The current 3-point correlators with $1/N$ corrections can help to decode the underlying physics of these theories.

We have shown that the subgraph expansion is closely related to the operator product expansion in momentum space. Specifically, it provides a diagrammatic classification of the contributions from different regions in the integration domain. 
This approach can be straightforwardly generalized to the conformal 4-point correlators. The conformal 4-point correlator and its OPE are the key ingredients for the bootstrap studies, which are usually formulated in position space \cite{Rattazzi:2008pe,Poland:2018epd}. There have been attempts to implement bootstrap methods in momentum space \cite{Isono:2018rrb,Isono:2019wex,Gillioz:2019iye,Gillioz:2025yfb}. The challenge is that the OPE for conformal 4-point correlator is quite subtle in momentum space, as the external operators have complicated coincident limits, which requires careful classification of the integration domains. We expect the method of subgraphs can help to resolve this problem in a graph-theoretical approach. In the conformal 4-point correlator, there are three independent external momenta and the correlation functions consist of 3-scale loop integrals, for which it needs to take two subgraph expansions sequentially to realize the OPE of the 4-point correlator. It would be interesting to understand how the associativity of the OPEs and also the crossing symmetry are realized in the final results. Moreover, there are constraints from the consistency conditions, including unitarity and locality \cite{Polyakov:1974gs}. These constraints have been explored in Mellin space and lead to remarkable results for the CFT data \cite{Gopakumar:2016wkt, Gopakumar:2016cpb} . It would be intriguing to explore these constraints further in momentum space. We leave this problem for future study.

\section*{Acknowledgements}
The author would like to thank Rajeev Erramilli and Ke Ren for the helpful discussions and collaborations on related projects. The author is in debt to Christoph Dlapa and Zhengwen Liu for the instructive discussions on the method of regions and method of subgraphs. The author is grateful to Matthew Mitchell, David Poland, Kostas Skenderis, Yuan Xin and Gang Yang for valuable discussions. The author would like to thank David Poland and the Yale Physics Department for the support during early stage of this work. The author thanks the organizers of the ICBS 2025 and Bootstrap 2025 for the support during the course of this work.
This research was supported by the Startup Funding  4007022314 of the Southeast University, the Southeast University Global Engagement of Excellence Fund No. 3360682401E, and the National Natural Science Foundation of China funding No. 12375061.

\bibliographystyle{utphys.bst}
\bibliography{CurrentLargeN}% Produces the bibliography via BibTeX.
\end{document}